\documentclass[twocolumn]{emulateapj}

\usepackage{amsmath}
\usepackage{natbib}
\usepackage{graphicx}
\usepackage{mathrsfs}

\newcommand{\numberkok}{9355}

\newcommand{\hi}{H\textsc{i}}

\def\arcdeg{\hbox{$^\circ$}}
\def\slantfrac#1#2{\hbox{$\,^#1\!/_#2$}}

\def\nl{\cr}

\def\ga{\mathrel{\hbox{\rlap{\hbox{\lower4pt\hbox{$\sim$}}}\hbox{$>$}}}}
\def\la{\mathrel{\hbox{\rlap{\hbox{\lower4pt\hbox{$\sim$}}}\hbox{$<$}}}}
\newcommand{\numberabs}{280}
\newcommand{\numberabscut}{222}
\newcommand{\numberabshigh}{134}
\newcommand{\numberem}{278}
\newcommand{\numberkokabs}{14954}

\newcommand\Tau{\mathcal{T}}

\newcommand\cnmfrac{$0.56 \pm 0.10$ }
\newcommand\unmfrac{$0.41 \pm 0.10$}
\newcommand\wnmfrac{$0.03 \pm 0.05$}

\newcommand\cnmfracperct{$56 \pm 10\%$}
\newcommand\cnmfracpercttot{$28\%$}
\newcommand\unmfracperct{$41 \pm 10\%$}
\newcommand\unmfracpercttot{$20\%$}
\newcommand\wnmfracperct{$3 \pm 5\%$}
\newcommand\wnmfracpercttot{$52\%$}
\submitted{Submitted June 2, 2018}
\begin{document}

\title{The 21-SPONGE H{\sc i} Absorption Line Survey II: The temperature of Galactic H{\sc i}}

\author{Claire E. Murray\altaffilmark{1,2$\dagger$}, Sne\v{z}ana Stanimirovi\'c\altaffilmark{1},  W.\,M. Goss\altaffilmark{3}, Carl Heiles\altaffilmark{4}, John M. Dickey\altaffilmark{5}, Brian Babler\altaffilmark{1}, Chang-Goo Kim\altaffilmark{6,7} }
\altaffiltext{1}{Department of Astronomy, 
                 University of Wisconsin, 
                 Madison, WI 53706, USA}
 \altaffiltext{2}{Space Telescope Science Institute, 
                3700 San Martin Drive, 
                Baltimore, MD, 21218, USA}
\altaffiltext{3}{National Radio Astronomy Observatory, 
		P.O. Box O, 1003 Lopezville, Socorro, NM 87801, USA}
\altaffiltext{4}{Radio Astronomy Lab, UC Berkeley,
		601 Campbell Hall, Berkeley CA 94720, USA}
\altaffiltext{5}{University of Tasmania, 
		School of Maths and Physics, 
		Hobart, TAS 7001, Australia}
\altaffiltext{6}{Department of Astrophysical Sciences, 
		Princeton University, 
		Princeton, NJ 08544, USA}
\altaffiltext{7}{ Center for Computational Astrophysics, 
		Flatiron Institute, 
		New York, NY 10010, USA}
\altaffiltext{$\dagger$}{clairemurray56@gmail.com}

%%%%%%%%%%%%%%%%%%%%%%%%%%%%%%%%%%%%%%%%%%%%%%%%%%%%%%%%%%%%% 
\begin{abstract}
 
We present 21-cm Spectral Line Observations of Neutral Gas with the VLA (21-SPONGE), a Karl G. Jansky Very Large Array (VLA) large project ($\sim 600\rm\,hours$) for measuring the physical properties of Galactic neutral hydrogen (\hi). 21-SPONGE is distinguished among previous Galactic \hi\ studies as a result of: (1) exceptional optical depth sensitivity ($\sigma_{\tau} < 10^{-3}$ per $0.42\rm\,km\,s^{-1}$ channels over 57 lines of sight); (2) matching $21\rm\,cm$ emission spectra with highest-possible angular resolution ($\sim 4'$) from the Arecibo Observatory; (3) detailed comparisons with numerical simulations for assessing observational biases. We autonomously decompose $21\rm\,cm$ spectra and derive the physical properties (i.e., spin temperature, $T_s$, column density) of the cold neutral medium (CNM; $T_s<250\rm\,K$), thermally unstable medium (UNM; $250< T_s < 1000\rm\,K$) and warm neutral medium (WNM; $T_s > 1000\rm\,K$) simultaneously. We detect $50\%$ of the total \hi\ mass in absorption, the majority of which is CNM (\cnmfracperct{}, corresponding to \cnmfracpercttot{} of the total \hi\ mass). Although CNM is detected ubiquitously, the CNM fraction along most lines of sight is $\lesssim 50\%$. We find that \unmfracpercttot{} of the total \hi\ mass is thermally unstable (\unmfracperct{} of \hi\ detected in absorption), with no significant variation with Galactic environment. Finally, although the WNM comprises \wnmfracpercttot{} of the total \hi\ mass, we detect little evidence for WNM absorption with $1000<T_s<4000\rm\,K$. Following spectral modeling, we detect a stacked residual absorption feature corresponding to WNM with $T_s\sim10^4\rm\,K$. We conclude that excitation in excess of collisions likely produces significantly higher WNM $T_s$ than predicted by steady-state models. 

\end{abstract}

\section{Introduction}
\label{sec:intro}

The formation of stars and evolution of galaxies relies on the cycle of interstellar matter between supernova-expelled plasma and molecule-rich gas. The center of this cycle is neutral hydrogen (\hi): fundamental fuel for star-forming clouds whose physical conditions comprise key constraints for theoretical models. 

Following the first astronomical observations of absorption and emission by the $21\rm\,cm$ transition of \hi\ \citep{muller1951, ewan1951,hagen1955}, clear differences in the observed velocity structure of $21\rm\,cm$ emission and absorption were attributed to significant variations in temperature and density of the gas along the line of sight \citep[e.g.,][]{clark1965, dickey1978}. Theoretical models of steady-state ISM heating and cooling quantified the nature of this thermal phase structure, predicting two thermally-stable phases: the cold neutral medium (CNM) and warm neutral medium (WNM), with density and kinetic temperatures of 
($n,T_k) = (7$--$70\rm\,cm^{-3}$, $60$--$260\rm\,K$), and $(n,T_k)=(0.2$--$0.9\rm\,cm^{-3}$, $5000$--$8300\rm\,K$) respectively \citep{mckee1977, wolfire2003}. However, subsequent analytical models and numerical simulations determined that time-dependent or dynamical processes such as turbulence and supernova shocks are likely very important, and will generate a significant amount of thermally unstable gas (UNM) in the intervening parameter space between CNM and WNM, thus throwing into question the validity of the steady-state paradigm of the ISM \citep[e.g.,][]{dalgarno1972, vazquez2000, audit2005}. 

However, the physical properties, mass fractions and ionization state of the diffuse neutral gas phases (i.e., WNM and UNM) are particularly sensitive to macro- and microphysical heating and cooling processes \citep{heiles2003b}. These include magnetic wave dissipation \citep[e.g.,][]{ferriere1988}, magnetic reconnection \citep[e.g.,][]{vishniac1999}, turbulence \citep[e.g.,][]{audit2005}, supernovae \citep[e.g.,][]{maclow2005}, diffusion of photons from H\textsc{ii} regions, and diffusion of low-energy cosmic rays and X-rays from time-dependent stellar phenomena. As these processes originate from sources on a huge range of physical and temporal scales, understanding them as part of a self-consistent model of the ISM has proven challenging. 

Furthermore, observational constraints for the properties of the UNM and WNM have been historically limited by insufficient observational capabilities. To constrain the optical depth and excitation (or spin) temperature of \hi\ --- crucial for determining the thermodynamic state of the gas --- measurements of both emission and absorption at $21\rm\,cm$ are necessary. Due to their low densities, detecting the absorbing properties of the WNM and UNM requires extremely high sensitivity to \hi\ optical depth. For example, past observations of 21cm absorption with single-dish and interferometric telescopes were primarily sensitive to detecting absorption by the CNM with $T_s=60-80\rm\,K$ \citep{hughes1971, radhakrishnan1972, crovisier1978}. Improved sensitivity to optical depth revealed absorption by \hi\ with warmer temperatures, up to $\sim 600\rm\,K$ \citep{lazareff1975, dickey1977}. Only a handful of detections of WNM with $T_s\gtrsim1000\rm\,K$ exist \citep{carilli1998, dwarakanath2002, murray2015}.

Considering the expense of high-sensitivity absorption measurements, and the fact that they are limited by the availability of sources of background continuum radiation, warm \hi\ (i.e., UNM and WNM) properties are typically indirectly estimated from $21\rm\,cm$ emission alone. Kinetic temperatures inferred from the Gaussian line widths of decomposed $21\rm\,cm$ emission profiles indicate that a significant fraction of the \hi\ mass is thermally unstable, with $T_s\sim3000\rm\,K$ \citep[e.g.,][]{verschuur1994, haud2007}. The Millennium Arecibo $21\rm\,cm$ Absorption-Line Survey detected absorption from gas with excitation temperatures of $\sim10-600\rm\,K$, and inferred from the emission that $\sim48\%$ of the remaining material detected in emission (i.e., $\sim30\%$ of the \emph{total} column density) is thermally unstable \citep[][hereafter HT03]{heiles2003b}. From a high-sensitivity survey of $21\rm\,cm$ absorption towards 35 sources, \citet{roy2013b} estimated that at most $28\%$ of \hi\ is unstable. However this result is based on \hi\ emission data from the Leiden Argentine Bonn (LAB) survey, whose $36'$ resolution probes much larger scales and thus different \hi\ populations than their sub-arcminute interferometric absorption measurements. 

Improving on previous observational efforts to tackle the properties of the diffuse \hi\ and constrain the UNM mass fraction requires expanded samples of $21\rm\,cm$ absorption lines at high sensitivity and careful attention to systematic uncertainties in analysis techniques. 

\subsection{The 21-SPONGE Survey}

In this paper, we present the final data products from the largest survey for Galactic \hi\ absorption to date at the Karl G. Jansky Very Large Array (VLA), titled 21-cm Observations of Neutral Gas with the EVLA (21-SPONGE). With superb sensitivity to $21\rm\,cm$ absorption, in combination with $21\rm\,cm$ emission from the Arecibo Observatory, 21-SPONGE is sensitive to CNM, UNM and WNM temperatures and densities in the Galactic ISM. In \citet[][; hereafter M15]{murray2015} we presented the observation and data analysis strategies for 21-SPONGE, as well as preliminary analysis of $21\rm\,cm$ spectral line pairs. We demonstrated that the exceptional optical depth sensitivity of 21-SPONGE ($\sigma_{\tau}<0.001$ per $0.42\rm\,km\,s^{-1}$ channels) enables direct detections of higher \hi\ spin temperatures than previous observational studies by more than a factor of two (e.g., HT03).

\begin{figure}[t!]
\begin{center}
\includegraphics[width=0.4\textwidth]{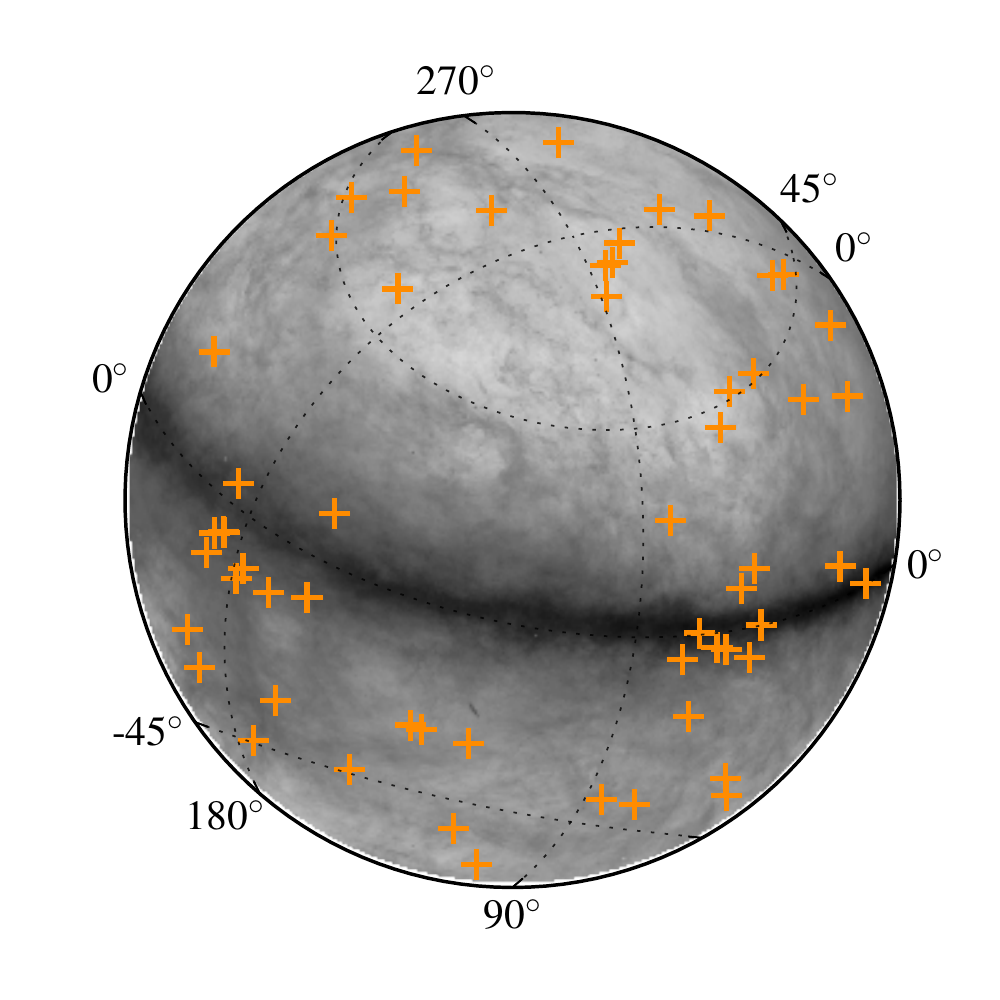}
\caption{All-Northern-sky $N({\rm HI})$ map from EBHIS in zenith-equal-area projection \citep{winkel2016}, with coordinates for the 57 21-SPONGE LOS overlaid (orange plusses). Grid lines denote Galactic coordinates. }
\label{f:map}
\end{center}
\end{figure}

Following the first half of the 21-SPONGE survey, we detected little evidence for WNM with $T_s > 1000\rm\, K$ ($< 10\%$ by number; M15). To improve sensitivity to shallow, broad absorption features further, we adapted a spectral stacking method to Galactic \hi\ spectral line pairs and detected a pervasive population of WNM gas with $T_s = 7200^{+1800}_{-1200}\rm\, K$ \citep{murray2014}. This excitation temperature is significantly higher than predictions from standard ISM models based on collisional \hi\ excitation \citep[e.g.,][]{liszt2001}, and suggests that additional excitation mechanisms such as resonant Ly$\alpha$ scattering \citep[i.e., the Wouthuysen-Field (WF) effect;][]{wouthuysen1952,field1958} are important for determining the thermodynamic properties of diffuse, neutral gas. Furthermore, enhanced $21\rm\,cm$ excitation has important implications for understanding \hi\ signals from early epochs of cosmic time when \hi\ dominated the baryonic content of the Universe \citep{pritchard2012}. 

With the full 21-SPONGE survey now complete, our goal is to measure the mass distribution of \hi\ as a function of temperature in the Galactic ISM. To prepare for this effort, in \citet{murray2017} we considered the biases imposed by our analysis techniques in detail. Specifically, we analyzed synthetic \hi\ absorption and emission spectral lines from a high-resolution, three-dimensional hydrodynamical simulation \citep[][hereafter KOK13 and KOK14]{kim2013, kim2014} to assess the power of our observational methods for revealing the inherent state of the ISM. For the first time, we computed the completeness of \hi\ structure recovery by Gaussian spectral line features, and quantified the decline in completeness with latitude due to velocity crowding. Furthermore, we showed that the physical gas properties inferred from $21\rm\,cm$ spectral lines agree with the ``true" simulated values within a factor of $<2$ for the majority of gas structures. We also identified a population of synthetic spectral features in KOK14 which are inconsistent with properties of $21\rm\,cm$ spectra from 21-SPONGE, motivating improved treatment of \hi\ excitation and feedback from star formation \citep{murray2017}. 

In this work we will compare gas properties inferred from the KOK14 synthetic spectra to underlying physical properties of the KOK13 simulation to estimate the bias imposed by our spectral analysis methods on the overall mass distribution of \hi\ as a function of temperature. Whereas in \citet{murray2017} we focused on a subset of the KOK13 simulation --- considering individual gas structures selected along simulated LOS and how their inherent properties were recovered by individual spectral line features --- in this work we consider \emph{all} gas along each simulated LOS. We will then use these comparisons to estimate uncertainties in the observed 21-SPONGE distribution, to produce new constraints on the mass distribution of \hi\ as a function of temperature.

This paper is organized as follows. In Section~\ref{sec:data}, we discuss the 21-SPONGE observations and synthetic data products used in our analysis. In Section~\ref{sec:gauss} we discuss our revised Gaussian decomposition and radiative transfer approach, derived from HT03 and \citet{murray2017}. In Section~\ref{sec:results}, we present the results of our analysis of 21-SPONGE, including comparison with synthetic spectra from KOK14. In Section~\ref{sec:discussion} we discuss these results and we summarize our conclusions in Section~\ref{sec:summary}. 

\section{Data} 
\label{sec:data}

\subsection{VLA 21cm Absorption Spectra}

The observing strategy for the 21-SPONGE \hi\ absorption is described in M15 and summarized here for clarity. Our targets are bright ($S_{\rm 1.4\,GHz}>3\rm\,Jy$), radio continuum sources from the NRAO/VLA Sky Survey \citep[NVSS; ][]{condon1998} at high Galactic latitude (generally, $|b|>10^{\circ}$) with compact angular sizes ($<1'$) to minimize the complexity of observed \hi\ profiles, and to avoid resolving substantial continuum flux. 

For all VLA observations, we use three separate $500\rm\,kHz$ bands with $1.95\rm\,kHz$ channel spacing centered on the \hi\ line ($1.42040575\rm\,GHz$) and $\pm1.5\rm\,MHz$ respectively. We use the off-line bands to avoid strong \hi\ lines at Galactic velocities in the directions of our calibrator sources and to perform bandpass calibration via frequency switching. Bandpass calibration is of particular importance for 21-SPONGE, as our primary interest is to detect broad, shallow absorption lines associated with high-temperature \hi. For details on our bandpass calibration strategies, we refer the reader to Section 2.2 of M15.

\begin{deluxetable*}{lllccccc|cc}
\setlength{\tabcolsep}{0.05in} 
\tablecolumns{10}  
\tabletypesize{\scriptsize}
\tablewidth{0pt}
\tablecaption{VLA Observation Information\label{tab:obs} }
\tablehead{
\colhead{Source}  & \colhead{RA (J2000)}  & \colhead{Dec (J2000)}  & \colhead{l}      & \colhead{b }     & \colhead{$S_{1.4\rm\,GHz}$}    & \colhead{Synthesized beam} &  \colhead{$\sigma_{\tau}$ }    & \colhead{$\tau_{\rm peak}$}  &  \colhead{$\int \tau$ dv}  \\ 
\colhead{(name) }      &  \colhead{(hh:mm:ss) }      &  \colhead{(dd:mm:ss)  }   & \colhead{($^{\circ}$)}  & \colhead{($^{\circ}$)}  &  \colhead{(Jy)}       & \colhead{($'' \times ''$)} &    \colhead{(x10$^{-3}$) }  &   & \colhead{($\rm\,km\,s^{-1}$) }     } 
\startdata
J0022  &  00:22:25.4  &  +00:14:56.2  &  107.462  &  -61.748  &  3.01  &  2.2 x 1.4  &  0.8  &  0.025$\pm$0.001  &  0.15$\pm$0.01    \\  
3C018A  &  00:40:50.7  &  +10:03:05.0  &  118.623  &  -52.732  &  4.60  &  15.5 x 13.7  &  1.5  &  0.623$\pm$0.004  &  2.40$\pm$0.04    \\  
3C018B  &  00:40:49.5  &  +10:03:50.0  &  118.616  &  -52.719  &  4.60  &  15.5 x 13.7  &  2.4  &  0.642$\pm$0.006  &  2.30$\pm$0.05    \\  
3C041A  &  01:26:44.8  &  +33:13:02.3  &  131.379  &  -29.075  &  3.71  &  1.7 x 1.4  &  2.4  &  0.039$\pm$0.004  &  0.34$\pm$0.04    \\  
3C041B  &  01:26:43.8  &  +33:13:21.8  &  131.374  &  -29.070  &  3.71  &  1.7 x 1.4  &  3.8  &  0.057$\pm$0.006  &  0.32$\pm$0.06    \\  
3C48  &  01:37:41.3  &  +33:09:35.1  &  133.963  &  -28.719  &  16.02  &  1.3 x 1.2  &  0.7  &  0.050$\pm$0.001  &  0.38$\pm$0.02    \\  
4C15.05  &  02:04:50.4  &  +15:14:11.0  &  147.930  &  -44.043  &  4.07  &  3.3 x 3.0  &  0.6  &  0.086$\pm$0.001  &  0.72$\pm$0.02    \\  
3C78  &  03:08:26.2  &  +04:06:39.0  &  174.858  &  -44.514  &  5.75  &  4.1 x 2.2  &  2.2  &  1.366$\pm$0.006  &  4.99$\pm$0.07    \\  
4C16.09  &  03:18:57.8  &  +16:28:32.7  &  166.636  &  -33.596  &  8.03  &  1.4 x 1.2  &  0.6  &  0.539$\pm$0.002  &  3.02$\pm$0.02    \\  
3C111A  &  04:18:21.3  &  +38:01:35.8  &  161.676  &  -8.820  &  7.73  &  13.0 x 5.3  &  1.7  &  0.888$\pm$0.005  &  10.17$\pm$0.08    \\  
3C111B  &  04:18:30.4  &  +38:02:30.4  &  161.686  &  -8.788  &  4.31  &  13.0 x 5.3  &  1.6  &  1.150$\pm$0.005  &  10.23$\pm$0.07    \\  
3C111C  &  04:18:15.5  &  +38:00:48.2  &  161.671  &  -8.843  &  2.92  &  13.0 x 5.3  &  3.0  &  1.125$\pm$0.009  &  11.37$\pm$0.13    \\  
3C120  &  04:33:11.1  &  +05:21:15.6  &  190.373  &  -27.397  &  3.44  &  4.7 x 4.3  &  0.9  &  2.033$\pm$0.003  &  10.62$\pm$0.03    \\  
3C123A  &  04:37:04.9  &  +29:40:10.2  &  170.584  &  -11.660  &  49.73  &  20.3 x 5.3  &  0.6  &  1.750$\pm$0.002  &  9.03$\pm$0.02    \\  
3C123B  &  04:37:04.0  &  +29:40:28.2  &  170.578  &  -11.659  &  49.73  &  20.3 x 5.3  &  0.7  &  1.783$\pm$0.003  &  8.79$\pm$0.03    \\  
3C132  &  04:56:43.5  &  +22:49:16.3  &  178.862  &  -12.522  &  3.43  &  1.8 x 1.5  &  1.1  &  1.614$\pm$0.005  &  7.37$\pm$0.06    \\  
3C133  &  05:02:58.1  &  +25:16:26.6  &  177.725  &  -9.913  &  5.77  &  1.2 x 1.1  &  2.7  &  1.645$\pm$0.010  &  8.99$\pm$0.13    \\  
3C138  &  05:21:010.0  &  +16:38:22.1  &  187.405  &  -11.343  &  8.60  &  14.5 x 5.1  &  1.0  &  1.115$\pm$0.004  &  5.95$\pm$0.04    \\  
PKS0531  &  05:34:44.5  &  +19:27:21.4  &  186.762  &  -7.108  &  7.02  &  1.3 x 1.1  &  0.5  &  0.535$\pm$0.002  &  3.31$\pm$0.03    \\  
3C147  &  05:42:36.1  &  +49:51:07.2  &  161.686  &  10.298  &  22.88  &  4.4 x 3.8  &  0.5  &  0.796$\pm$0.001  &  5.00$\pm$0.02    \\  
3C154  &  06:13:49.0  &  +26:04:36.7  &  185.592  &  4.003  &  5.00  &  13.7 x 12.8  &  0.7  &  1.704$\pm$0.004  &  14.76$\pm$0.06    \\  
PKS0742  &  07:45:33.1  &  +10:11:12.7  &  209.797  &  16.592  &  3.51  &  1.9 x 1.4  &  0.6  &  0.011$\pm$0.001  &  0.03$\pm$0.01    \\  
3C225A  &  09:42:15.3  &  +13:45:51.3  &  220.010  &  44.008  &  3.34  &  4.5 x 1.6  &  1.2  &  0.828$\pm$0.002  &  1.49$\pm$0.03    \\  
3C225B  &  09:42:15.6  &  +13:45:49.3  &  220.011  &  44.009  &  3.34  &  4.5 x 1.6  &  2.3  &  0.791$\pm$0.004  &  1.43$\pm$0.04    \\  
3C236  &  10:06:01.8  &  +34:54:10.4  &  190.065  &  53.980  &  3.24  &  4.8 x 1.8  &  0.6  &  0.003$\pm$0.001  &  0.00$\pm$0.00    \\  
3C237  &  10:08:00.0  &  +07:30:16.6  &  232.117  &  46.627  &  6.52  &  6.5 x 4.4  &  1.0  &  0.410$\pm$0.002  &  0.62$\pm$0.02    \\  
3C245A  &  10:42:44.6  &  +12:03:31.3  &  233.124  &  56.300  &  3.31  &  1.6 x 1.4  &  1.3  &  0.016$\pm$0.002  &  0.05$\pm$0.01    \\  
3C245B  &  10:42:44.3  &  +12:03:31.6  &  233.123  &  56.299  &  3.31  &  1.6 x 1.4  &  4.2  &  0.024$\pm$0.006  &  0.03$\pm$0.02    \\  
1055+018  &  10:58:29.6  &  +01:33:58.8  &  251.511  &  52.774  &  3.22  &  0.1 x 0.0  &  0.9  &  0.008$\pm$0.001  &  0.03$\pm$0.01    \\  
3C263.1  &  11:43:25.1  &  +22:06:56.1  &  227.201  &  73.766  &  3.13  &  7.5 x 4.4  &  0.7  &  0.020$\pm$0.001  &  0.06$\pm$0.01    \\  
3C273  &  12:29:06.1  &  +02:03:08.6  &  289.945  &  64.359  &  54.99  &  7.8 x 4.2  &  0.4  &  0.026$\pm$0.001  &  0.09$\pm$0.01    \\  
4C32.44  &  13:26:16.5  &  +31:54:09.5  &  67.234  &  81.048  &  4.86  &  2.8 x 1.2  &  0.7  &  0.020$\pm$0.001  &  0.06$\pm$0.01    \\  
4C25.43  &  13:30:37.7  &  +25:09:11.0  &  22.468  &  80.988  &  7.05  &  2.8 x 1.2  &  1.1  &  0.004$\pm$0.001  &  0.00$\pm$0.00    \\  
3C286  &  13:31:08.3  &  +30:30:33.0  &  56.524  &  80.675  &  14.90  &  5.2 x 3.3  &  0.4  &  0.007$\pm$0.001  &  0.06$\pm$0.01    \\  
4C12.50  &  13:47:33.4  &  +12:17:24.2  &  347.223  &  70.172  &  5.40  &  4.3 x 1.5  &  0.9  &  0.091$\pm$0.002  &  0.29$\pm$0.01    \\  
3C298  &  14:19:08.2  &  +06:28:34.8  &  352.160  &  60.666  &  6.10  &  2.4 x 1.4  &  0.6  &  0.020$\pm$0.001  &  0.08$\pm$0.01    \\  
UGC09799  &  15:16:44.5  &  +07:01:17.8  &  9.417  &  50.120  &  5.50  &  1.7 x 1.3  &  6.9  &  0.066$\pm$0.011  &  0.13$\pm$0.05    \\  
4C04.51  &  15:21:14.4  &  +04:30:22.0  &  7.292  &  47.747  &  3.93  &  9.4 x 3.8  &  0.8  &  0.068$\pm$0.001  &  0.32$\pm$0.01    \\  
3C327.1A  &  16:04:44.9  &  +01:17:52.8  &  12.181  &  37.006  &  4.08  &  3.5 x 2.4  &  3.2  &  0.505$\pm$0.008  &  2.26$\pm$0.07    \\  
3C327.1B  &  16:04:45.6  &  +01:17:47.6  &  12.182  &  37.003  &  4.08  &  3.5 x 2.4  &  3.0  &  0.448$\pm$0.008  &  2.13$\pm$0.07    \\  
PKS1607  &  16:09:13.3  &  +26:41:29.0  &  44.171  &  46.203  &  4.91  &  0.9 x 0.4  &  0.6  &  0.177$\pm$0.001  &  0.92$\pm$0.01    \\  
J1613  &  16:13:41.1  &  +34:12:47.9  &  55.151  &  46.379  &  4.02  &  4.1 x 3.5  &  1.0  &  0.005$\pm$0.001  &  0.00$\pm$0.00    \\  
3C345  &  16:42:58.8  &  +39:48:37.0  &  63.455  &  40.949  &  7.10  &  3.4 x 1.5  &  0.9  &  0.008$\pm$0.001  &  0.01$\pm$0.00    \\  
3C346  &  16:43:48.6  &  +17:15:49.3  &  35.332  &  35.769  &  3.66  &  1.0 x 0.7  &  2.0  &  0.288$\pm$0.004  &  1.16$\pm$0.04    \\  
3C390  &  18:45:37.6  &  +09:53:45.0  &  41.112  &  5.773  &  4.51  &  3.4 x 2.8  &  1.2  &  0.157$\pm$0.004  &  2.65$\pm$0.06    \\  
4C33.48  &  19:24:17.5  &  +33:29:29.7  &  66.389  &  8.371  &  3.77  &  4.8 x 1.7  &  2.9  &  0.408$\pm$0.007  &  2.42$\pm$0.09    \\  
3C409A  &  20:14:27.5  &  +23:34:55.4  &  63.398  &  -6.121  &  13.68  &  1.6 x 1.4  &  1.4  &  1.190$\pm$0.005  &  8.70$\pm$0.07    \\  
3C409B  &  20:14:27.7  &  +23:34:50.2  &  63.398  &  -6.122  &  13.68  &  1.6 x 1.4  &  1.3  &  1.303$\pm$0.005  &  8.50$\pm$0.06    \\  
3C410A  &  20:20:06.6  &  +29:42:14.8  &  69.212  &  -3.769  &  2.88  &  3.2 x 1.7  &  1.4  &  3.501$\pm$0.007  &  17.74$\pm$0.10    \\  
3C410B  &  20:20:06.7  &  +29:42:09.6  &  69.211  &  -3.770  &  6.39  &  3.2 x 1.7  &  2.0  &  3.146$\pm$0.010  &  16.99$\pm$0.13    \\  
B2050  &  20:52:52.1  &  +36:35:35.3  &  78.858  &  -5.124  &  5.14  &  4.3 x 2.0  &  0.8  &  0.331$\pm$0.002  &  2.65$\pm$0.04    \\  
3C433  &  21:23:44.6  &  +25:04:02.2  &  74.475  &  -17.697  &  10.33  &  9.7 x 5.7  &  3.0  &  0.467$\pm$0.008  &  1.88$\pm$0.07    \\  
PKS2127  &  21:30:32.9  &  +05:02:17.5  &  58.652  &  -31.815  &  4.10  &  4.0 x 1.7  &  0.7  &  0.128$\pm$0.001  &  0.56$\pm$0.01    \\  
J2136  &  21:36:38.6  &  +00:41:54.2  &  55.473  &  -35.578  &  3.47  &  8.3 x 5.0  &  1.2  &  0.143$\pm$0.002  &  0.88$\pm$0.02    \\  
J2232  &  22:32:36.4  &  +11:43:50.9  &  77.438  &  -38.582  &  7.20  &  5.2 x 4.3  &  1.0  &  0.156$\pm$0.002  &  1.05$\pm$0.03    \\  
3C454.3  &  22:53:58.0  &  +16:08:52.4  &  86.112  &  -38.185  &  12.66  &  2.2 x 1.5  &  1.1  &  0.317$\pm$0.002  &  1.81$\pm$0.03    \\  
3C459  &  23:16:35.2  &  +04:05:18.1  &  83.040  &  -51.285  &  4.68  &  5.7 x 4.5  &  0.9  &  0.142$\pm$0.002  &  1.15$\pm$0.02    \\  [5pt]
\hline
\hline

\multicolumn{5}{l}{Sources rejected as overly resolved (res) or saturated (sat)} \\ [3pt]

J0407 (res)   &   04:07:25.5   &	+03:40:47.3	&   187.651   &  -33.604   &    3.27  & \nodata  &  \nodata & \nodata  &  \nodata            \\   
J0534  (res)   &  05:34:34.9   &  +22:02:07.2     &   184.591   &  -5.759     &   13.81  &  \nodata &\nodata & \nodata  &    \nodata             \\   
J1651  (res) &   16:51:03.9    & +04:59:41.9      &    23.039    &   28.967   &   11.2   & \nodata  &\nodata & \nodata  & \nodata               \\   
PKS1944 (sat)  &  19:46:47.9   &  +25:12:45.0   &   61.472    & 0.096     & 4.9  &\nodata    & \nodata & \nodata  &  \nodata            \\   
J2021 (sat)       &  20:21:38.7   &  +37:31:10.1   &  75.833    &  0.402    & 6.6  &  \nodata   & \nodata & \nodata  &   \nodata           
\enddata

\tablecomments{ Col. (1):  Source name. Cols. (2) through (5): R.A. and Dec, $l$ and $b$ coordinates. Col. (6): Flux density at $1.4\rm\,GHz$ \citep{condon1998}. Col (7): Synthesized beam size. Col. (8): RMS uncertainty in optical depth, measured in off-line channels ($0.42\rm\,km\,s^{-1}$ channel spacing). Col. (9): Peak optical depth. Col. (10): Integrated optical depth.} 
\end{deluxetable*}

We reduce all 21-SPONGE data using the Astronomical Image Processing System \citep[AIPS\footnote{http://www.aips.nrao.edu/; }][]{greisen2003}. For a full description of the data reduction strategy, see Section 2 of M15. For each source, we produce a cleaned, calibrated data cube and continuum image. We then extract the absorption spectrum from the pixel of maximum flux density, and divide by the continuum flux density at the pixel location to compute $\exp({-\tau(v)})$. Our channel spacing of $1.95\rm\,kHz$ at the \hi\ frequency ($0.42\rm\,km\,s^{-1}$ channel spacing in velocity) corresponds to a velocity resolution of $0.5\rm\,km\,s^{-1}$ \citep{rohlfs2004}. 

Of the original 58 target sources from the 21-SPONGE observing program\footnote{VLA project codes: 10C-196, 12A-256, 13A-205}, 10 were removed upon inspection of preliminary data products for being overly resolved (3 sources), displaying saturated absorption (2 sources), or for not receiving integration time following the conclusion of the observing program (5 sources). An additional 9 sources were resolved into multiple continuum peaks, thereby providing additional sources for extracting \hi\ absorption, albeit at degraded sensitivity due to the loss of continuum flux density. A gallery of continuum images for the 48 final targets, demonstrating the range of complexity in source structure, is included in the Appendix (Figure~\ref{apB:f:cont}). Overall, we extracted 57 \hi\ spectra from the 48 targets, including the spectra extracted from multiple continuum peaks (denoted by $A,\,B,\,C$ lettering). Figure~\ref{f:map} displays an all-Northern-sky map of \hi\ column density from the Effelsberg-Bonn \hi\ Survey \citep[EBHIS;][]{winkel2016}, with the 57 final 21-SPONGE lines of sight (LOS) coordinates overlaid. Table~\ref{tab:obs} displays detailed source information, including coordinates, NVSS flux density \citep{condon1998}, and RMS noise in optical depth ($\sigma_{\tau}$) computed from offline channels ($0.42\rm\,km\,s^{-1}$ channel spacing). We include source information for the overly resolved and saturated sources at the bottom of Table~\ref{tab:obs}. 

In comparison with Table~2 of M15, we have improved $\sigma_{\tau}$ for many sources by including additional integration time and/or re-processing the original files. Generally, $\sigma_{\tau}<1\times10^{-3}$ (median value = $9\times10^{-4}$) which makes 21-SPONGE among the highest-sensitivity surveys for \hi\ absorption ever undertaken, and covers more sources than previous high-sensitivity surveys by almost a factor of two \citep[e.g.,][]{roy2013}. The outliers are due to degraded sensitivity from lack of flux density in the cases of our 9 resolved sources. Overall,  exceptional sensitivity to optical depth makes 21-SPONGE sensitive to absorption by \hi\ in all stable and thermally unstable ISM phases according to predictions from standard steady-state ISM models.

In M15, we demonstrated excellent agreement between 21-SPONGE and other \hi\ absorption studies by comparing the integrated \hi\ optical depths for sources which overlap with the Millennium Arecibo \hi\ Absorption Line Survey \citep[][hereafter HT03]{heiles2003}, \citet{stanimirovic2005} and \citet{roy2013}. Of our 48 targets, we overlap with 22/78 from HT03, 9/35 from \citet{roy2013} and 9/104 from \citet{mohan2004}. In Table 1 of M15 we summarized these and other external surveys. We find consistent agreement with these studies at the level of our uncertainties.

All 21-SPONGE VLA spectra will be made publicly available, and are accessible via their permanent Digital Object Identifier (DOI) at this link: \url{http://dx.doi:10.7910/DVN/BWFKL6}. [note: link not public yet]

\subsection{Matching \hi\ Emission Spectra}

To estimate the temperatures and column densities of \hi\ structures using radiative transfer calculations, we need information about the brightness temperature of \hi\ probed by our VLA absorption spectra. Observing \hi\ emission on the same angular scale as the \hi\ absorption measurement is ideal, however these measurements are prohibitively expensive to conduct at an interferometric facility such as the VLA. Therefore, we obtain the expected \hi\ brightness temperature spectra ($T_{B,\rm exp}(v)$) along the same LOS as the VLA targets by interpolating emission spectra from neighboring LOS across the target position, following the strategy outlined by HT03. We obtain $21\rm\,cm$ emission data from the $305\rm\,m$ Arecibo Observatory, whose $\sim4'$ beam at $1.4\rm\,GHz$ allows us to minimize the effects of mismatched beam sizes on interpreting \hi\ spectra. 31 sources were observed as part of project A2770 at Arecibo, and 11 sources were obtained from publicly-available data from HT03. Emission spectra for the remaining five sources, which lie outside of the Arecibo field of view, were obtained from the next-highest resolution survey available: EBHIS \citep[$10.'8$ resolution at $21\rm\,cm$;][]{winkel2016}. 

In Section 2.4 of M15, we describe our treatment of $T_{B,\rm exp}(v)$ spectra from Arecibo. For this work, to estimate the beam efficiency factor for converting antenna temperature to brightness temperature, we compare the integrated antenna temperatures to those derived from averaging brightness temperature spectra from Galactic Arecibo L-band Feed Array \citep[GALFA-\hi;][]{peek2011,peek2018} survey, which is flux-calibrated based on LAB, in annuli of radius 2 pixels (16 pixels) around each target pixel. From this comparison, we derived a new beam efficiency correction factor which we applied to the full dataset, equal to 0.94, which ensures that our $T_{B,\rm exp}(v)$ spectra are consistent with previous surveys (i.e., EBHIS). 

Furthermore, we note that the Arecibo \hi\ emission spectra in this work have not been corrected for radiation entering the main telescope beam from higher-order side lobes, an effect known as ``stray radiation." Unlike single-dish radio telescopes whose beam shapes can be accurately modeled for removing this effect \citep[e.g., Effelsberg, Parkes;][]{kalberla2005, mcg2009}, Arecibo has a very complex beam structure that varies with azimuth and elevation, and therefore stray radiation is extremely difficult to remove. Comparing $21\rm\,cm$ emission from the GALFA-\hi\ survey with the stray-corrected LAB survey, stray radiation likely contributes $\sim500\rm\,mK$ over $\sim50\rm\,km\,s^{-1}$ to the observed \hi\ brightness temperature \citep{peek2011}. Considering this effect, we are explicitly careful to not over-fit our $21\rm\,cm$ profiles from Arecibo, and emphasize that stray radiation does not affect $21\rm\,cm$ absorption from the VLA.

\subsection{Uncertainty Arrays}
\label{sec:data:uncert}

The uncertainty in each spectral channel depends on the system temperature, which can be significantly increased by strong brightness temperature at Galactic velocities. To determine the frequency-dependent uncertainty arrays for each LOS, we follow the methods described in Section 3.2 of M15, which were derived following \citet{roy2013}. In summary, for each LOS, the uncertainty array in absorption is a combination of on-source noise ($\sigma_{\rm on}(v)$) and off-source noise from the frequency-switched bandpass solution ($\sigma_{\rm BP}$). The on-source noise is computed by scaling the RMS uncertainty in $\exp{-\tau(v)}$ (i.e., our measured absorption quantity) by $(T_{B,\rm LAB}(v) + T_{\rm sys, VLA})/T_{\rm sys, VLA}$, where $T_{B,\rm LAB}(v)$ is the brightness temperature computed from adjacent pixels to each target source from the LAB survey, whose telescope at Dwingeloo is of similar size to a VLA antenna, with an assumed system temperature at the VLA of $T_{\rm sys, VLA}=25\rm\,K$. The uncertainty in $\exp{-\tau(v)}$ is then computed by solving $\sigma_{\exp{-\tau}}(v)^2 = \sigma_{\rm on}(v)^2 + \sigma_{\rm BP}^2$. From this, we solve for the uncertainty in $\tau(v)$ (i.e., $\sigma_{\tau}(v)$) for subsequent spectral analysis. For each emission spectrum, the uncertainty array ($\sigma_{T_B}(v)$) is estimated by scaling the RMS noise in $T_{B,\rm exp}(v)$ computed from offline channels by ($T_{B,\rm exp}(v)+T_{\rm sys,em})/T_{\rm sys,em}$ for an assumed system temperature of $T_{\rm sys,em}=30\rm\,K$.

\subsection{Line of Sight Properties}

In Table~\ref{tab:obs}, we list parameters of the 21-SPONGE VLA spectra. First, we include the peak optical depth ($\tau_{\rm peak}$) along the LOS, with uncertainty equal to the value of the $\sigma_{\tau}(v)$ at the velocity of $\tau_{\rm peak}$. We find a median $\tau_{\rm peak}=0.32$ and mean $\tau_{\rm peak}=0.61$. We observe $\tau_{\rm peak}\geq 3$ in only two cases (3C410A and 3C410B). Our sources lie generally at high Galactic latitude by design to avoid strong velocity crowding associated with the Galactic plane, and therefore the generally small $\tau_{\rm peak}$ is consistent with expectations. We also list the integrated optical depth ($\int \tau \, {\rm d}v$), with uncertainties computed by adding the uncertainty in each spectral channel in quadrature.

\subsection{Synthetic \hi\ Spectra}

To consider the performance of our analysis methods, we will compare the 21-SPONGE spectral line pairs with a sample of synthetic $21\rm\,cm$ spectral line pairs from KOK14. These synthetic spectra were constructed from the 3D hydrodynamical simulation of KOK13, which includes time-varying heating and cooling of interstellar gas, momentum feedback from supernovae, self-gravity, differential rotation and external gravity from dark matter and stars. 
From this simulation, KOK14 selected $10^4$ randomly distributed mock sight lines at $|b|>4^{\circ}.9$ within the simulated volume and extracted the number density ($n$), temperature (kinetic, $T_k$ and spin, $T_s$), and velocity ($v$) as a function of distance along the LOS. Using analytic radiative transfer and simple line excitation considerations, KOK14 constructed synthetic $21\rm\,cm$ brightness temperature ($T_B$) and optical depth ($\tau$) spectra from each LOS. We refer the reader to Section 2.3 of KOK14 for details of synthetic spectra construction.

In \citet{murray2017}, we found that the implementation of the WF effect has a significant effect on the line widths of KOK14 synthetic spectral lines and the resulting WNM spin temperature distribution. KOK14 constructed three sets of synthetic $21\rm\,cm$ data for different WF prescriptions: no WF, constant WF and maximum WF (i.e., $T_s=T_k$). We found the constant WF case, wherein the Ly$\alpha$ radiation field density was fixed at $10^{-6}\rm\,photoons/cm^{-3}$, resulted in a narrow spin temperature distribution ($T_s\sim4000K$ for $T_k>4000K$) for the WNM (c.f., Figure 2 of KOK14). Via comparison with 21-SPONGE spectra, we found that KOK14 spectra --- whether with no WF or with constant WF --- feature a significant population of large-amplitude, wide absorption components not observed in 21-SPONGE yet well above our sensitivity limits (c.f., Figure 11 of Murray et al.\,2017). Furthermore, these components correspond to WNM properties not observed by 21-SPONGE (i.e., $T_s\sim3000-4000\rm\,K$). From this comparison, we concluded that a more sophisticated treatment of the WF effect is likely necessary to produce realistic synthetic spectral lines from future simulations. 

For this study, we select the maximum WF dataset, which we determined by similar analysis as done in \citet{murray2017} features spectral components whose line widths agree best with those detected in 21-SPONGE, and therefore maximizes consistency between observed and synthetic datasets. To build the synthetic dataset, we select spectra without saturated ($\tau \geq 3$) absorption, for a final catalog of \numberkok{} \hi\ spectral pairs. To simulate the same observational properties of 21-SPONGE spectral line pairs, we add Gaussian noise to each synthetic $21\rm\,cm$ spectrum (RMS $\sigma_{\tau}=10^{-3}$ for absorption, RMS $\sigma_{T_{\rm B}}=0.2\,\rm K$ for emission) as done in \citet{murray2017}.

\section{Analysis} 
\label{sec:gauss}

To derive physical properties of individual \hi\ structures along each LOS, we decompose all \hi\ emission and absorption spectral line pairs into Gaussian functions. In the following section, we describe our method for autonomously decomposing $21\rm\,cm$ spectra.  

\subsection{Gaussian Decomposition}
\label{sec:gauss_fit}

We begin by decomposing the VLA \hi\ absorption spectra uniformly using the Autonomous Gaussian Decomposition algorithm \citep[AGD;][]{lindner2015} and its Python implementation, GaussPy\footnote{GaussPy; https://github.com/gausspy/gausspy}. AGD implements derivative spectroscopy and supervised machine learning to produce efficient, reproducible guesses for the basic parameters of Gaussian functions, including the number of components, and their amplitudes, positions and widths. Following the method described in \citet{lindner2015} and employed in \citet{murray2017}, we train the algorithm using a synthetic absorption line dataset constructed from spectral line parameters from HT03. From the training process, we determine optimal values of the two-phase smoothing parameters, $\alpha_1=1.12$ and $\alpha_2=2.75$, required by AGD to compute spectral line parameter guesses. We then decompose the 21-SPONGE absorption lines using these values with an imposed signal to noise ratio of $S/N =3$. To avoid aliasing narrow components, we first resample the spectra to a velocity resolution of $0.1\rm\,km\,s^{-1}$ \citep{lindner2015}. As shown in \citet{lindner2015} and \citet{murray2017} the resulting parameters of the decomposition are statistically indistinguishable from those found in the by-hand analysis of 21-SPONGE sources. Although no Gaussian decomposition represents a unique solution, we emphasize the benefits of the AGD: to eliminate subjective biases of human-derived guesses, and to ensure that the decomposition results are completely reproducible.

After decomposing each \hi\ absorption spectrum using AGD into $N$ components, we produce a model for the optical depth along the LOS:

\begin{equation}
\tau (v)_{\rm AGD}= \sum_{n=0}^{N-1} \tau_{0,n} \cdot e^{- 4 \ln{2} (v-v_{0,n})^2/ \Delta v_n^2},
\label{e:tau}
\end{equation}

\noindent where ($\tau_{0,n}$, $v_{0,n}$, $\Delta v_n$) are the amplitude, mean velocity and full width at half maximum (FWHM) of the $n^{\rm th}$ component.

\begin{figure*}
\begin{center}
\includegraphics[width=\textwidth]{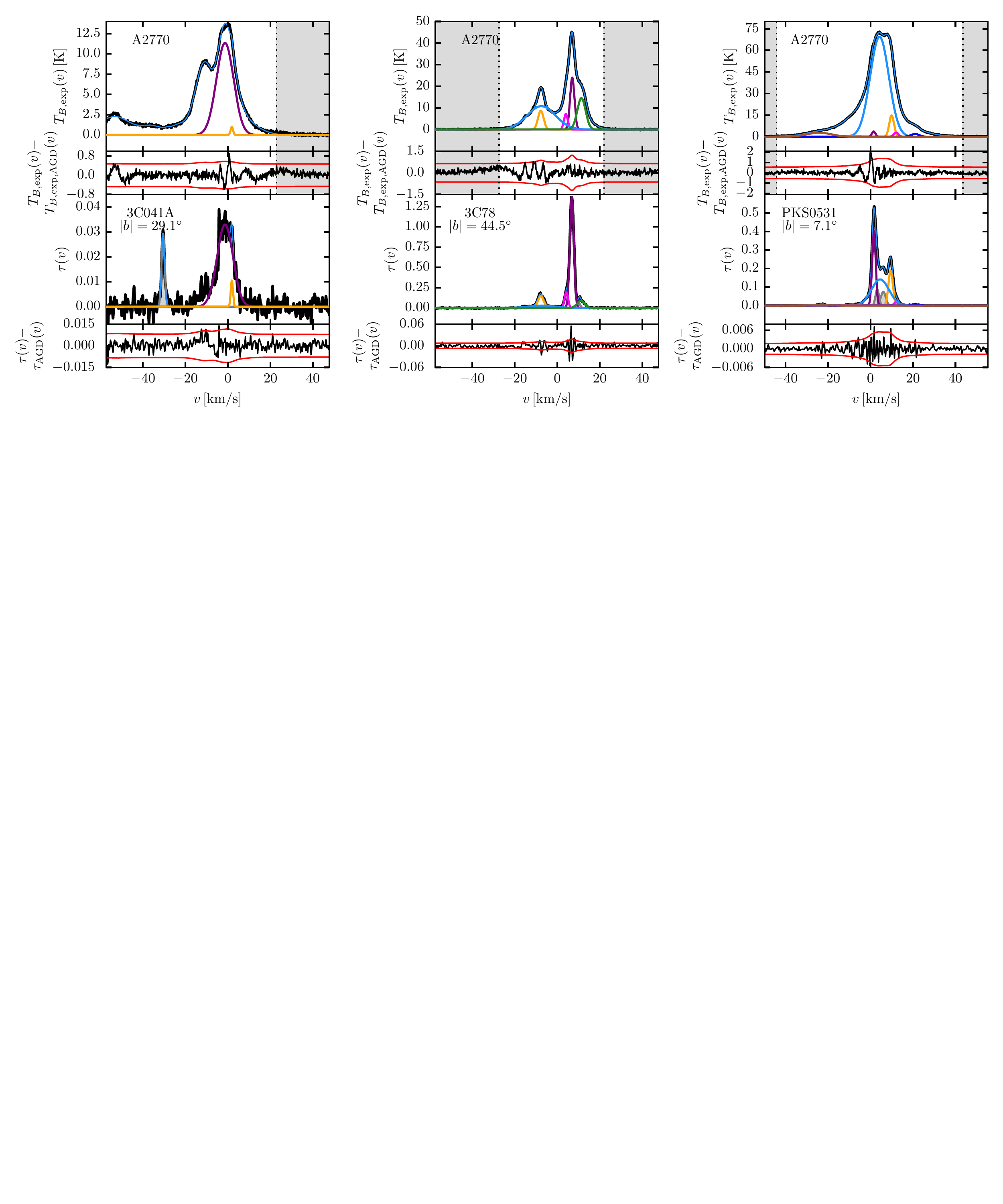}
\vspace{-400pt}
\caption{ Examples displaying the Gaussian fits to 21-SPONGE \hi\ emission and absorption spectral line pairs described in Section~\ref{sec:gauss}. In each panel, we plot the emission ($T_{B,\rm exp}(v)$; top) and absorption ($\tau(v)$; bottom) spectra. The residual spectra following the Gaussian fits are included below each panel, with $\pm 3\times $ the noise spectra for $T_{B,\rm exp}$ and $\tau$ respectively (red). We plot all fitted absorption components in the bottom panel. Components whose derived spin temperatures are unphysical (i.e., $\leq 10\rm\,K$) are plotted in shaded grey, and components with $T_s > 10\rm\,K$ are plotted in matching colors in the middle and top panels. The total fits, $T_{B, \rm exp, \,AGD}(v)$ and $\tau_{\rm AGD}(v)$ are displayed in thin, blue lines. The source of $T_{B,\rm exp}(v)$, whether from Arecibo (A2770 or HT03) or EBHIS is printed in the top panels. In the bottom panels, we print the source name and the absolute Galactic latitude ($|b|$). Finally, in the top panels we shade in grey the velocities where $T_{B,\rm exp}(v) \leq 3\cdot \sigma_{T_B}(v)$ to illustrate the range over which LOS column densities are computed (i.e., the unshaded region is used). If no vertical shading is present, the full displayed velocity range is used. Plots for all sources are included in the Appendix.}
\label{f:sources_panel}
\end{center}
\end{figure*}

To fit these components to the expected brightness temperature along the LOS ($T_{B, \rm exp}(v)$), we assume a two-component \hi\ medium, wherein some clouds contribute both opacity and brightness temperature (i.e., detected in emission and absorption), and some clouds are dominated by the WNM and contribute only brightness temperature to the LOS \citep[e.g., ][; HT03, M15]{mebold1997, dickey2000}. For each LOS, we solve, 

\begin{equation}
T_{B, \rm exp, \, AGD}(v) = T_{B, \rm abs, \, AGD} (v) + T_{B, \rm em, \,AGD}(v).
\label{e:tb}
\end{equation}

To determine the contributions of absorption-detected and emission-only components to $T_{B, \rm exp}(v)$, we implement a new method to fit all components to $T_{B, \rm exp}(v)$ using AGD, based on the strategy described in HT03. We note that whereas the original HT03 method involved fitting components by eye, our new method is autonomous.
The method involves the following steps:

\begin{enumerate}
\item Fit all $N$ components from $\tau(v)$ to $T_{B, \rm exp}(v)$ via least-squares fit. The mean velocities and widths are allowed to vary by $\pm10\%$ to simulate small random fluctuations and their amplitudes are constrained so that $0< T_{B,n} = T_{s,n}\cdot (1-e^{-\tau_{0,n}})$ and $T_{s,n} \leq T_{k,{\rm max}, n} = 21.866\cdot \Delta v_n^2$, to produce realistic spin temperatures.
\item Subtract the best fit model in step (1) from $T_{B, \rm exp}(v)$ to produce a residual emission spectrum, which contains only emission not detected in absorption.
\item Apply GaussPy to fit $K$ new components to the residual emission spectrum from (2), using the trained one-phase value of $\alpha=3.75$ and $S/N=5$ from previous analysis of HT03 emission spectra \citep{murray2017}. The S/N requirement is more strict in the fit to emission than the initial absorption fit so that we avoid over-fitting the emission residuals in the presence of stray radiation. We also remove any component guesses whose mean velocities agree with previously-detected absorption components within 1 spectral channel (i.e., $0.42\rm\,km\,s^{-1}$) so as not to spuriously overfit $T_{B, \rm exp}(v)$ (this occurs only in the presence of strong residuals following the subtraction of absorption components in complex LOS). 
\item Combine the $N+K$ Gaussian components from steps (1) and (3) and execute a final least squares fit to $T_{B, \rm exp}(v)$. In this final fit, we allow all mean velocities and widths to vary by $10\%$, and constrain all amplitudes such that $T_{B, \rm exp} > 0$. In this step, initial estimates of $T_s$ for the $N$ absorption components and the Gaussian parameters of the $K$ emission-only components are computed. 
\end{enumerate}

Given a final list of $N+K$ Gaussian components fitted to $T_{B, \rm exp}(v)$ from the procedure described above, we solve Equation~\ref{e:tb} for all possible orderings of the $N$ absorption components along the LOS, and for varying absorption properties of the $K$ emission-only components, following HT03. In detail, we solve:

\begin{equation}
T_{B,\rm abs, \, AGD}(v) = \sum_{n=0}^{N-1} T_{s,n} (1-e^{-\tau_n(v)}) e^{-\sum_{m=0}^M \tau_m(v)},
\label{e:order}
\end{equation}

\noindent where the subscript ``$m$" refers to all components which lie in front of the $n^{\rm th}$ component, and,

\begin{multline}
T_{B, \rm em, \, AGD}(v) = \sum_{k=0}^{K-1} [ \mathscr{F}_k + (1- \mathscr{F}_k) \,e^{-\tau(v)} ] \, \cdot \\ \cdot \,T_{0,k} \,e^{\frac{- 4 \ln{2} (v-v_{0,k})^2}{ \Delta v_k^2}},
\label{e:fval}
\end{multline}

\noindent where ($T_{0,k}$, $v_{0,k}$, $\Delta v_k$) are the amplitude, mean velocity and FWHM of the $k^{\rm th}$ component fitted only in emission, and $\mathscr{F}_k$ is the fraction of each component lying in front of all absorption components. Previous analysis has shown that $\mathscr{F}$ is highly uncertain  \citep[e.g., HT03;][; M15]{stanimirovic2014}, and yet it has a significant effect on the derived spin temperatures. Therefore, following HT03, we allow $\mathscr{F}_k$ to have one of three values, $\mathscr{F}_k = 0.0, \,0.5$ or $1.0$ for all $K$ emission-only components.

\begin{figure*}
\begin{center}
\vspace{-200pt}
\includegraphics[width=\textwidth]{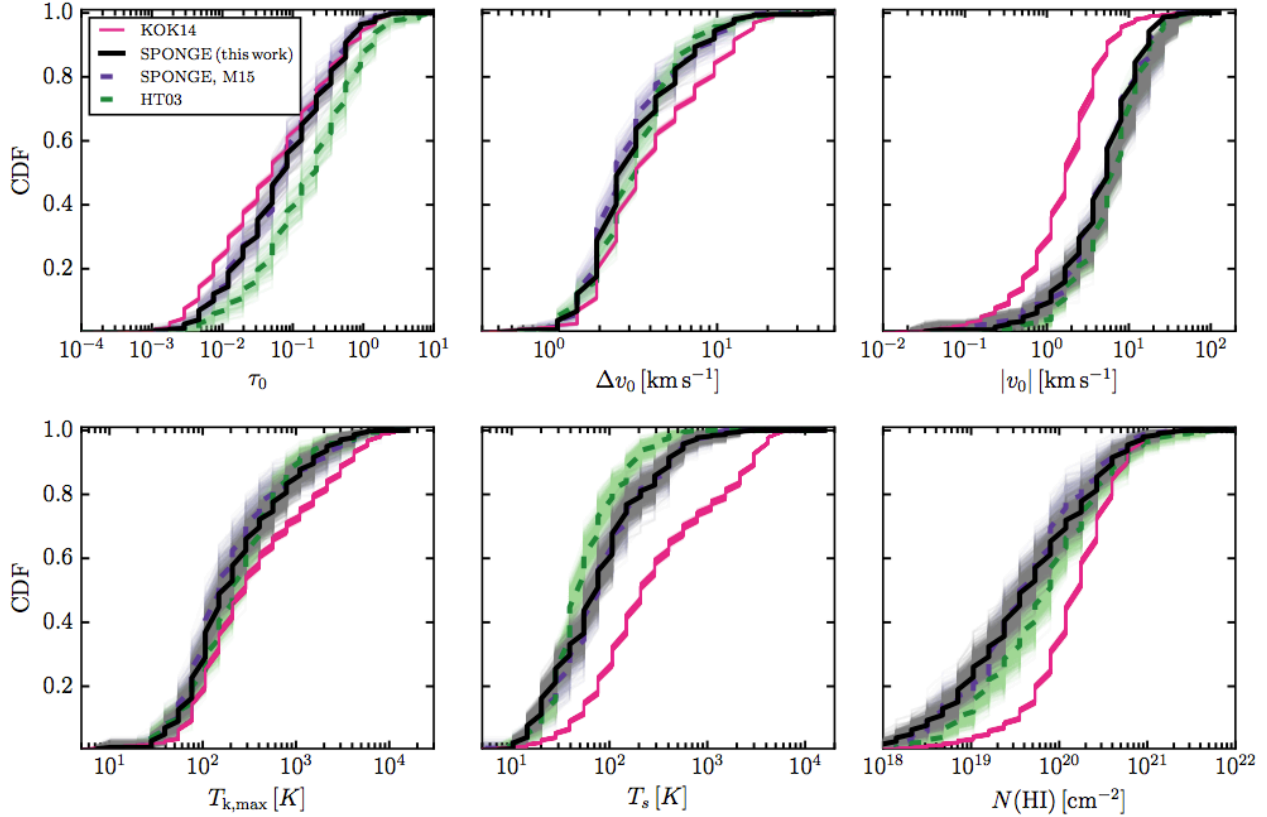}
\vspace{-180pt}
\caption{Parameters derived from the Gaussian fits to the 57 21-SPONGE absorption lines (\numberabs{} components) via AGD (Section~\ref{sec:gauss_fit}; black). These include: optical depth amplitude ($\tau_0$), FWHM ($\Delta v_0$) and mean velocity ($|v_0|$), maximum kinetic temperature ($T_{k, \rm max}$), spin temperature ($T_s$) and column density ($N(\rm H\textsc{i})$). For comparison, we include the results of HT03 (green dashed), the first half of the 21-SPONGE survey (M15; purple dashed) and a fit to KOK14 synthetic \hi\ spectral line pairs using the same methodology (pink). }
\label{f:params}
\end{center}
\end{figure*}

Furthermore, the order of the absorption components along the LOS will only affect $T_{B,\rm abs}(v)$ in the cases of components which overlap significantly in velocity (HT03). For each LOS, there are a maximum of $N!$ possible orderings, but we select only the unique orderings corresponding to components which overlap in area by more than $3\sigma_{\tau} \, dv$. Therefore, there are a total of $3\times N!$ possible iterations for the final fit to Equation~\ref{e:tb}, but in practice there are many fewer for each LOS.

We determine a final estimate for the spin temperature of each absorption-detected component by computing the weighted mean and standard deviation over all ordering trials (c.f, Equations 21a and 21b of HT03).  We fit a total of \numberabs{} absorption components and \numberem{} emission-only components to the 57 lines of sight. Figure~\ref{f:sources_panel} displays three examples of the fitting process described above. Similar plots for all 57 LOS are included in Appendix Figures~\ref{f:sources0} through~\ref{f:sources6}.

\subsection{Synthetic \hi\ Decomposition}
\label{s:kok}

We use the same methodology described above to fit 9355 synthetic \hi\ absorption and emission spectral line pairs from KOK14 with maximum WF. 
In contrast to the analysis of \citet{murray2017}, wherein we presented a method for matching Gaussian spectral lines to ``true" gas structures along the LOS, the new method presented here improves the statistics of components for which we can derive $T_s$ and $N(\rm H\textsc{i})$. The method described in \citet{murray2017} selected only those components with unambiguous signatures in $\tau(v)$, $T_{B, \rm exp}(v)$ and $n/T_s$, and therefore resulted in fewer detected components per LOS. Here we fit a total of \numberkokabs{} components to the \numberkok{} synthetic absorption lines. 

\section{Fitting Results}
\label{sec:results}

\subsection{Properties of Fitted Components}

For each absorption-detected component, we compute the maximum kinetic temperature ($T_{k, \rm max}$) via,

\begin{equation}
T_{k,\rm max} = \frac{m_H}{8 k_B \ln{2}} \Delta v_0^2 = 21.866 \cdot \Delta v_0^2,
\end{equation}

\noindent for hydrogen mass $m_H$, and Boltzmann's constant $k_B$ \citep{draine2011}. 

Next, we compute the \hi\ column density per absorption component, given by,

\begin{equation}
N({\rm H\textsc{i}})_{\rm abs} =  C_0 \int \tau \, T_s \, dv  = 1.064 \cdot C_0 \cdot \tau_0 \cdot \Delta v_0 \cdot T_s,
\label{e:nhi_comp}
\end{equation}

\noindent where $C_0 = 1.823\times 10^{18} \rm\,cm^{-2} / (km\,s^{-1}\,K)$, $\Delta v_0$ is measured in $\rm km\,s^{-1}$ and $1.064$ converts the product to the area under a Gaussian function with the given height and width.

Figure~\ref{f:params} displays cumulative distribution functions (CDFs) of the fitted parameters for all absorption-detected components with physically reasonable values of $T_s$, defined as $T_s > 10\rm\,K$. This limit for ``reasonable" temperatures is defined conservatively based on the estimated contribution of the cosmic microwave and Galactic synchrotron backgrounds at the locations of our sources, which we found to vary between $2.76\rm\,K$ and $2.85\rm\,K$ for sources from the first half of the 21-SPONGE survey (M15), plus an estimated minimum $T_k$ of the CNM \citep[$\sim 7\rm\,K$;][]{wolfire2003}. Out of \numberabs{} components, \numberabscut{} have $T_s>10\rm\,K$. 

\begin{deluxetable*}{l|ccc | cccc}
\setlength{\tabcolsep}{0.05in} 
\tablecolumns{10}  
\tabletypesize{\scriptsize}
\tablewidth{0pt}
\tablecaption{Column Densities\label{tab:coldens} }
\tablehead{
\colhead{Source}       & \colhead{$N({\rm HI})_{\rm thin}$}  & \colhead{$N({\rm HI})_{\rm iso}$}  & \colhead{$N({\rm HI})_{\rm total}$}  &  \colhead{$N({\rm HI})_{\rm CNM}$}   & \colhead{$N({\rm HI})_{\rm UNM}$}  & \colhead{$f_{\rm CNM}$}      & \colhead{$f_{\rm UNM}$ }   \\
\colhead{(name) }      & \colhead{($10^{20}\rm\,cm^{-2}$) } & \colhead{($10^{20}\rm\,cm^{-2}$) } &  \colhead{($10^{20}\rm\,cm^{-2}$) }  &    \colhead{($10^{20}\rm\,cm^{-2}$) }   &    \colhead{($10^{20}\rm\,cm^{-2}$) }    &   &      }
\startdata
J0022  &  2.63$\pm$0.14  &  2.64$\pm$0.14  &  2.63$\pm$0.08  &  0.11  &  0.89  &  0.04$\pm$0.07  &  0.34$\pm$0.37    \\  
3C018A  &  5.65$\pm$0.02  &  6.29$\pm$0.03  &  6.71$\pm$0.24  &  3.36  &  0.00  &  0.50$\pm$0.21  &  0.00$\pm$0.28    \\  
3C018B  &  5.65$\pm$0.02  &  6.27$\pm$0.03  &  6.67$\pm$0.25  &  3.36  &  0.00  &  0.50$\pm$0.22  &  0.00$\pm$0.42    \\  
3C041A  &  5.43$\pm$0.23  &  5.46$\pm$0.23  &  5.46$\pm$0.12  &  0.02  &  2.04  &  0.00$\pm$0.10  &  0.37$\pm$0.51    \\  
3C041B  &  5.43$\pm$0.23  &  5.47$\pm$0.23  &  5.47$\pm$0.08  &  0.90  &  0.00  &  0.16$\pm$0.15  &  0.00$\pm$0.76    \\  
3C48  &  4.24$\pm$0.26  &  4.27$\pm$0.26  &  4.26$\pm$0.05  &  0.16  &  1.56  &  0.04$\pm$0.14  &  0.37$\pm$0.21    \\  
4C15.05  &  4.33$\pm$0.15  &  4.42$\pm$0.16  &  4.42$\pm$0.24  &  0.40  &  3.18  &  0.09$\pm$0.04  &  0.72$\pm$0.18    \\  
3C78  &  9.76$\pm$0.22  &  11.44$\pm$0.26  &  11.37$\pm$0.87  &  4.52  &  3.24  &  0.40$\pm$0.20  &  0.28$\pm$0.24    \\  
4C16.09  &  9.96$\pm$0.16  &  11.01$\pm$0.17  &  10.93$\pm$0.28  &  4.45  &  2.42  &  0.41$\pm$0.24  &  0.22$\pm$0.08    \\  
3C111A  &  24.25$\pm$1.06  &  28.43$\pm$1.24  &  28.60$\pm$1.25  &  20.24  &  0.70  &  0.71$\pm$0.18  &  0.02$\pm$0.09    \\  
3C111B  &  24.25$\pm$1.06  &  28.60$\pm$1.24  &  28.04$\pm$1.71  &  10.19  &  9.05  &  0.36$\pm$0.22  &  0.32$\pm$0.09    \\  
3C111C  &  24.25$\pm$1.06  &  29.21$\pm$1.27  &  24.14$\pm$3.27  &  13.45  &  0.00  &  0.56$\pm$0.22  &  0.00$\pm$0.19    \\  
3C120  &  10.02$\pm$0.03  &  15.85$\pm$0.04  &  14.07$\pm$1.12  &  7.64  &  0.00  &  0.54$\pm$0.24  &  0.00$\pm$0.08    \\  
3C123A  &  15.04$\pm$0.71  &  19.62$\pm$0.93  &  19.75$\pm$3.54  &  5.69  &  5.47  &  0.29$\pm$0.24  &  0.28$\pm$0.10    \\  
3C123B  &  15.04$\pm$0.71  &  19.51$\pm$0.92  &  19.68$\pm$3.68  &  5.86  &  6.20  &  0.30$\pm$0.22  &  0.31$\pm$0.19    \\  
3C132  &  22.83$\pm$0.27  &  27.15$\pm$0.32  &  24.95$\pm$1.98  &  8.69  &  0.00  &  0.35$\pm$0.14  &  0.00$\pm$0.07    \\  
3C133  &  25.11$\pm$0.07  &  30.07$\pm$0.09  &  28.80$\pm$2.15  &  3.07  &  16.27  &  0.11$\pm$0.09  &  0.56$\pm$0.14    \\  
3C138  &  18.55$\pm$0.18  &  21.32$\pm$0.20  &  20.82$\pm$2.33  &  7.31  &  0.91  &  0.35$\pm$0.12  &  0.04$\pm$0.07    \\  
PKS0531  &  25.13$\pm$0.36  &  27.10$\pm$0.39  &  27.15$\pm$1.75  &  1.64  &  14.96  &  0.06$\pm$0.26  &  0.55$\pm$0.07    \\  
3C147  &  17.26$\pm$0.33  &  18.71$\pm$0.35  &  18.43$\pm$1.54  &  7.29  &  0.00  &  0.40$\pm$0.18  &  0.00$\pm$0.04    \\  
3C154  &  34.11$\pm$0.05  &  46.48$\pm$0.07  &  38.87$\pm$10.72  &  2.31  &  27.17  &  0.06$\pm$0.38  &  0.70$\pm$0.32    \\  
PKS0742  &  3.13$\pm$0.19  &  3.13$\pm$0.20  &  3.11$\pm$0.08  &  0.10  &  0.00  &  0.03$\pm$0.05  &  0.00$\pm$0.25    \\  
3C225A  &  3.53$\pm$0.03  &  3.65$\pm$0.03  &  3.63$\pm$0.10  &  0.43  &  0.66  &  0.12$\pm$0.08  &  0.18$\pm$0.40    \\  
3C225B  &  3.53$\pm$0.03  &  3.65$\pm$0.03  &  3.65$\pm$0.13  &  0.64  &  1.02  &  0.17$\pm$0.14  &  0.28$\pm$0.71    \\  
3C236  &  0.78$\pm$0.15  &  0.78$\pm$0.15  &  0.78$\pm$0.15  &  0.00  &  0.00  &  0.00$\pm$0.20  &  0.00$\pm$0.99    \\  
3C237  &  2.16$\pm$0.03  &  2.19$\pm$0.03  &  2.19$\pm$0.08  &  0.13  &  0.64  &  0.06$\pm$0.11  &  0.29$\pm$0.57    \\  
3C245A  &  2.19$\pm$0.13  &  2.20$\pm$0.13  &  2.19$\pm$0.08  &  0.00  &  0.41  &  0.00$\pm$0.15  &  0.19$\pm$0.73    \\  
3C245B  &  2.19$\pm$0.13  &  2.20$\pm$0.13  &  1.92$\pm$0.13  &  0.00  &  0.00  &  0.00$\pm$0.45  &  0.00$\pm$1.00    \\  
1055+018  &  2.94$\pm$0.16  &  2.94$\pm$0.16  &  2.94$\pm$0.08  &  0.00  &  0.85  &  0.00$\pm$0.07  &  0.29$\pm$0.37    \\  
3C263.1  &  1.81$\pm$0.18  &  1.81$\pm$0.18  &  1.81$\pm$0.00  &  0.03  &  0.00  &  0.02$\pm$0.16  &  0.00$\pm$0.82    \\  
3C273  &  2.11$\pm$0.05  &  2.11$\pm$0.05  &  2.10$\pm$0.08  &  0.02  &  0.30  &  0.01$\pm$0.05  &  0.14$\pm$0.24    \\  
4C32.44  &  1.17$\pm$0.14  &  1.17$\pm$0.14  &  1.16$\pm$0.02  &  0.12  &  0.07  &  0.10$\pm$0.14  &  0.06$\pm$0.69    \\  
4C25.43  &  1.07$\pm$0.12  &  1.07$\pm$0.12  &  1.04$\pm$0.12  &  0.00  &  0.00  &  0.00$\pm$0.25  &  0.00$\pm$1.00    \\  
3C286  &  1.10$\pm$0.15  &  1.10$\pm$0.15  &  1.10$\pm$0.01  &  0.09  &  0.00  &  0.08$\pm$0.10  &  0.00$\pm$0.48    \\  
4C12.50  &  2.07$\pm$0.10  &  2.10$\pm$0.11  &  2.09$\pm$0.06  &  0.33  &  0.67  &  0.16$\pm$0.11  &  0.32$\pm$0.54    \\  
3C298  &  2.05$\pm$0.11  &  2.06$\pm$0.12  &  2.06$\pm$0.25  &  0.15  &  0.00  &  0.07$\pm$0.08  &  0.00$\pm$0.38    \\  
UGC09799  &  2.90$\pm$0.14  &  2.91$\pm$0.14  &  2.91$\pm$0.03  &  0.18  &  0.00  &  0.06$\pm$0.49  &  0.00$\pm$1.00    \\  
4C04.51  &  4.02$\pm$0.17  &  4.06$\pm$0.17  &  3.99$\pm$0.04  &  1.16  &  0.00  &  0.29$\pm$0.05  &  0.00$\pm$0.24    \\  
3C327.1A  &  7.21$\pm$0.17  &  7.87$\pm$0.18  &  7.87$\pm$0.20  &  3.41  &  0.00  &  0.43$\pm$0.12  &  0.00$\pm$0.47    \\  
3C327.1B  &  7.21$\pm$0.17  &  7.83$\pm$0.18  &  7.87$\pm$0.21  &  3.33  &  0.00  &  0.42$\pm$0.11  &  0.00$\pm$0.44    \\  
PKS1607  &  3.78$\pm$0.14  &  3.88$\pm$0.14  &  3.86$\pm$0.11  &  0.91  &  0.00  &  0.24$\pm$0.09  &  0.00$\pm$0.20    \\  
J1613  &  1.50$\pm$0.14  &  1.50$\pm$0.14  &  1.50$\pm$0.14  &  0.00  &  0.00  &  0.00$\pm$0.16  &  0.00$\pm$0.80    \\  
3C345  &  0.71$\pm$0.06  &  0.71$\pm$0.06  &  0.71$\pm$0.06  &  0.00  &  0.00  &  0.00$\pm$0.29  &  0.00$\pm$1.00    \\  
3C346  &  5.00$\pm$0.15  &  5.19$\pm$0.16  &  5.21$\pm$0.10  &  1.48  &  0.00  &  0.29$\pm$0.09  &  0.00$\pm$0.44    \\  
3C390  &  25.05$\pm$1.13  &  25.91$\pm$1.17  &  26.00$\pm$0.79  &  3.12  &  10.74  &  0.12$\pm$0.02  &  0.41$\pm$0.08    \\  
4C33.48  &  13.48$\pm$0.39  &  14.04$\pm$0.40  &  13.80$\pm$0.44  &  1.62  &  0.00  &  0.12$\pm$0.06  &  0.00$\pm$0.27    \\  
3C409A  &  23.00$\pm$0.80  &  27.74$\pm$0.96  &  27.05$\pm$2.18  &  13.23  &  0.64  &  0.49$\pm$0.19  &  0.02$\pm$0.08    \\  
3C409B  &  23.00$\pm$0.80  &  27.66$\pm$0.96  &  25.63$\pm$1.94  &  7.26  &  7.83  &  0.28$\pm$0.21  &  0.31$\pm$0.08    \\  
3C410A  &  38.53$\pm$2.81  &  54.10$\pm$3.94  &  53.13$\pm$7.21  &  24.98  &  1.01  &  0.47$\pm$0.20  &  0.02$\pm$0.14    \\  
3C410B  &  38.53$\pm$2.81  &  53.31$\pm$3.88  &  42.87$\pm$17.87  &  16.51  &  0.00  &  0.39$\pm$0.32  &  0.00$\pm$0.08    \\  
B2050  &  23.59$\pm$1.20  &  24.56$\pm$1.25  &  24.54$\pm$0.85  &  3.57  &  3.87  &  0.15$\pm$0.11  &  0.16$\pm$0.05    \\  
3C433  &  8.38$\pm$0.04  &  8.88$\pm$0.04  &  8.86$\pm$0.28  &  2.16  &  0.00  &  0.24$\pm$0.10  &  0.00$\pm$0.39    \\  
PKS2127  &  4.62$\pm$0.28  &  4.71$\pm$0.29  &  4.70$\pm$0.24  &  0.68  &  1.62  &  0.15$\pm$0.04  &  0.35$\pm$0.19    \\  
J2136  &  4.36$\pm$0.05  &  4.49$\pm$0.06  &  4.40$\pm$0.21  &  0.70  &  0.00  &  0.16$\pm$0.07  &  0.00$\pm$0.34    \\  
J2232  &  4.88$\pm$0.16  &  5.00$\pm$0.16  &  4.98$\pm$0.08  &  1.51  &  0.00  &  0.30$\pm$0.10  &  0.00$\pm$0.26    \\  
3C454.3  &  6.84$\pm$0.08  &  7.05$\pm$0.08  &  6.91$\pm$0.54  &  2.01  &  0.00  &  0.29$\pm$0.10  &  0.00$\pm$0.21    \\  
3C459  &  5.43$\pm$0.16  &  5.60$\pm$0.16  &  5.56$\pm$0.68  &  0.33  &  4.36  &  0.06$\pm$0.33  &  0.78$\pm$0.29    \\  

\enddata

\tablecomments{ Col. (1):  Source name. Col. (2): Optically thin \hi\ column density (Equation~\ref{e:nhi_thin}). Col. (3): Isothermal \hi\ column density (Equation~\ref{e:nhi_iso}). Col. (4): Total \hi\ column density following autonomous computation of $T_s$ and $N({\rm HI})_{\rm abs}$ for individual spectral components (Equation~\ref{e:nhi_tot}). Col. (5): Sum of $N({\rm HI})_{\rm abs}$ in the CNM ($T_s \leq250\rm\,K$). If no CNM components were detected within uncertainties, equal to 0.0. Col. (6): Sum of $N({\rm HI})_{\rm abs}$ in the UNM ($250<T_s\leq1000\rm\,K$). If no UNM components were detected within uncertainties, equal to 0.0. Col. (7): CNM fraction per LOS ($f_{\rm CNM} = N({\rm HI})_{\rm CNM}/ N({\rm HI})_{\rm total}$). Col. (8): UNM fraction per LOS ($f_{\rm UNM} = N({\rm HI})_{\rm UNM}/ N({\rm HI})_{\rm total}$).} 
\end{deluxetable*}

In the top row of Figure~\ref{f:params}, we display parameters from the best fit to $\tau(v)$, including: optical depth amplitude ($\tau_0$), FHWM ($\Delta v_0$), absolute mean velocity ($|v_0|$). In the bottom row, we include derived physical properties. In all panels of Figure~\ref{f:params}, we bootstrap each sample over 100 trials and include the resampled CDFs to illustrate the effect of outliers on the distributions. For comparison, we include the results of HT03, M15, and a re-processing of the synthetic \hi\ spectral line pair database from KOK14 with maximum WF (Section~\ref{s:kok}). The parameters for all 21-SPONGE sources are listed in Table~\ref{tab:params} in Appendix A. The uncertainties for all parameters are computed as part of the least squares AGD fit, except for the uncertainty in $T_s$ (and, subsequently $N({\rm HI})_{\rm abs}$) which is computed following the iterations over LOS component ordering. We set the minimum uncertainty in optical depth amplitudes equal to $\sigma_{\tau}(v)$ at the position of each component, and the minimum uncertainty in the mean velocity and FWHM is equal to $0.1\rm\,km\,s^{-1}$. We note that for one absorption component towards 3C111A, the uncertainties from the AGD fit are extremely large ($\gg 1000$)--- this component was not recovered in the fit to $T_{B,\rm exp}(v)$, and we identify it by setting the uncertainties in its component parameters equal to 99 in Table~\ref{tab:params}. 

In comparison with the previous observations shown in Figure~\ref{f:params}, we find that our decomposition results (black) are statistically indistinguishable from the by-hand analysis of the first half of the 21-SPONGE survey \citet[purple dashed; ][]{murray2017}, wherein we analyzed a subset of the 21-SPONGE sample presented in this work. We are also generally consistent with HT03, except in the case of $\tau_0$, wherein the superior sensitivity of 21-SPONGE allows us to probe smaller \hi\ optical depths. In addition, we detect higher $T_s$ in the present analysis than found by HT03, which is also attributable to our improved observational sensitivity. 

In Figure~\ref{f:params} we find that the observed (SPONGE, HT03) and synthetic (KOK14) $T_s$ and $N({\rm HI})$ distributions are significantly different. We will discuss this further in Section~\ref{sec:discussion}.

\subsection{Correspondence between $21\rm\,cm$ Absorption and Emission}

Overall, most ($78\%$) of the $N$ fitted components to $\tau(v)$ have corresponding components in the fit to $T_{B, \rm exp}(v)$. Even when all $\tau(v)$ components are forced to be included in the fit to $T_{B, \rm exp}(v)$  (e.g., as in the method of HT03, M15), $\sim10\%$ of components end up with unphysical spin temperatures. Components with $T_s\leq10\rm\,K$ are displayed in shaded grey in Figures~\ref{f:sources0}-\ref{f:sources6}. 

However, the overall fraction of absorption components with corresponding detected emission is generally high: for $22\%$ of the 57 LOS, 100\% of the absorption-fitted components correspond to components in the fit to $T_{B, \rm exp}(v)$, and for $98\%$ of LOS, $\geq 50\%$ of absorption components correspond to $T_{B, \rm exp}(v)$ components. The LOS featuring the lowest fraction of corresponding components between absorption and emission tend to lie at low Galactic latitude, where velocity blending of spectral lines is strongest. We find consistent statistics in the decomposition of KOK14 spectra. We will discuss the implications of the observed correspondence between absorption and emission further in Section~\ref{sec:discussion_cnm}.

An important effect in producing absorption components with no corresponding emission components (i.e.,  components with $T_s<10\rm\,K$) is beam mismatch between absorption and emission. In addition, $21\rm\,cm$ emission profiles are necessarily measured using adjacent LOS from the absorption profile in order to avoid the background continuum source, which means that the two profiles are not sampling identical populations of \hi\ structures. For emission spectra from Arecibo (e.g., A2770, HT03), $T_{B, \rm exp}(v)$ is computed on $\sim4'$ scales, and for those from EBHIS, $T_{B, \rm exp}(v)$ is computed on $10.'8$ scales, in contrast with sub-arcminute scales for absorption from the VLA. An example of this effect is shown in the case of 3C041A (Figure~\ref{f:sources0}). The absorption line clearly detected at $v\sim-30\rm\,km\,s^{-1}$ is not recovered in $T_{B, \rm exp}(v)$, likely due to a beam mismatch or LOS effect. 

\begin{figure*}[t!]
\begin{center}
\includegraphics[width=0.9\textwidth]{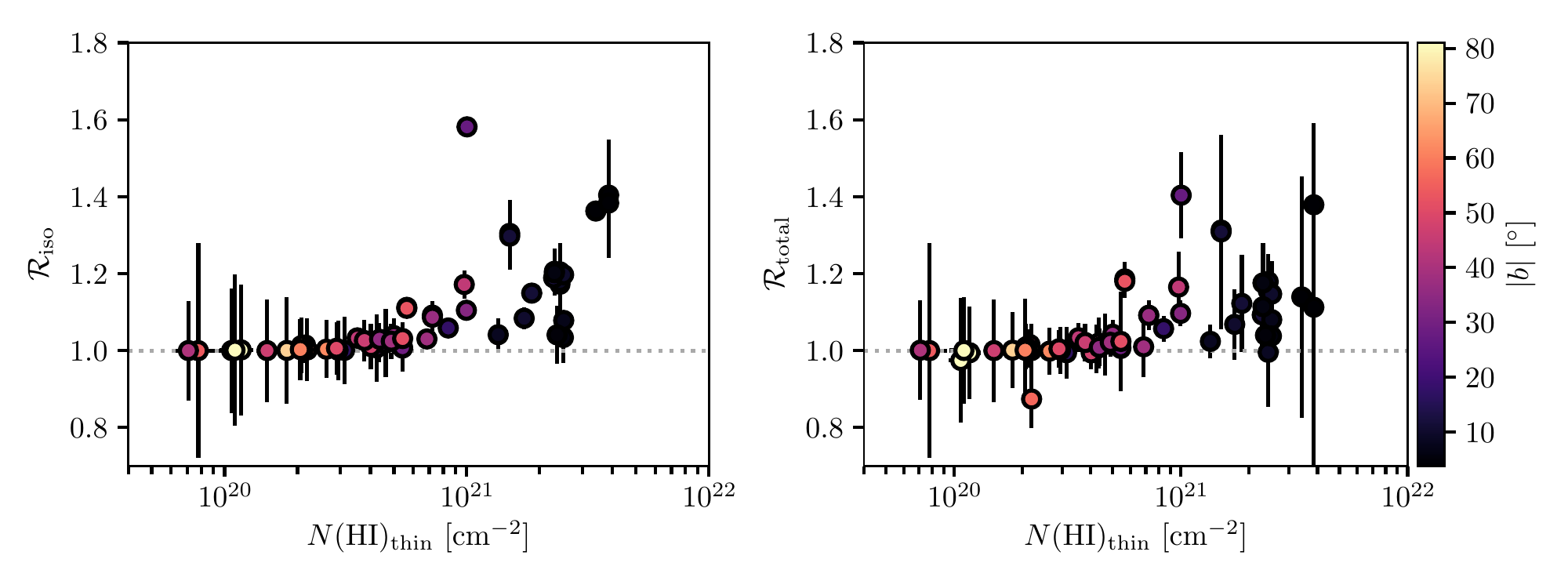} 
\caption{The effect of optical depth on the LOS column density. Left:  Isothermal column density correction factor ($\mathcal{R}_{\rm iso} = N({\rm HI})_{\rm iso}/N({\rm HI})_{\rm thin}$) versus optically thin column density ($N({\rm HI})_{\rm thin}$). Right: ``Total" column density correction factor ($\mathcal{R}_{\rm total} = N({\rm HI})_{\rm total}/N({\rm HI})_{\rm thin}$) versus $N({\rm HI})_{\rm thin}$. Data points are colored by the absolute Galactic latitude of the LOS ($|b|$). }
\label{f:coldens_comp}
\end{center}
\end{figure*}

We find that varying the decomposition scheme described in Section~\ref{sec:gauss_fit} did not have a significant effect on the results or derived physical parameters. Allowing the component parameters to vary by between 1\% and 20\% in step (1) has the largest effect on the number and properties of the fitted emission-only components. For example, the fraction of absorption-detected components recovered in the fit to $T_{B,\rm exp}(v)$ and their derived $T_s$ values do not vary significantly for different allowed variances. 

\section{Column Densities}

Following decomposition, we compute and compare different estimates of the total column density along 21-SPONGE LOS. Given that the majority of our $21\rm\,cm$ brightness temperatures have not been corrected for stray radiation, we conservatively restrict our analysis of $T_{B,\rm exp}(v)$ to channels above the uncertainty array for each LOS, specifically: $T_{B,\rm exp}(v)\geq 3\, \sigma_{T_B}(v)$.  In the top panels of each source plot in Figures~\ref{f:sources_panel} and~\ref{f:sources0}-\ref{f:sources6}), we shade the velocity channels which are not used in the column density analysis in light grey. The minimum and maximum velocities satisfying $T_{B,\rm exp}(v)\geq 3\, \sigma_{T_B}(v)$ are referred to in subsequent discussion as $v_{\rm em, \, min}$ and $v_{\rm em,\,max}$ respectively. 

In the absence of optical depth information, it is common in the literature to assume that neutral gas is optically thin so that the LOS column density can be computed via,

\begin{equation}
N({\rm HI})_{\rm thin} = C_0 \int_{v_{\rm em,\,min}}^{v_{\rm em,\,max}}T_{B,\rm exp}(v) \, {\rm d}v ,
\label{e:nhi_thin}
\end{equation}

\noindent where ${\rm d} v$ is measured in $\rm km\,s^{-1}$. Incorporating the optical depth information from 21-SPONGE, assuming that all \hi\ within each velocity channel is isothermal \citep[e.g.,][]{dickey1982, chengalur2013}, the column density is given by,

\begin{equation}
N({\rm HI})_{\rm iso} = C_0 \int_{v_{\rm em,\,min}}^{v_{\rm em,\,max}} \frac{\tau(v) \, T_{B,\rm exp}(v)}{1-e^{-\tau(v)}} \, {\rm d}v .
\label{e:nhi_iso}
\end{equation}

However, considering significant overlap in velocity of individual spectral features in Figures~\ref{f:sources0}-\ref{f:sources6}, there may be multiple \hi\ structures with different physical properties within the same radial velocity channels, potentially invalidating the isothermal approximation in Equation~\ref{e:nhi_iso}. Using the results of our spectral decomposition, we compute the ``total" \hi\ column density,

\begin{multline}
N({\rm HI})_{\rm total} = \sum_{n=0}^{N-1} N({\rm HI})_{{\rm abs},n} \,+ \\ + C_0 \int_{v_{\rm em,\,min}}^{v_{\rm em,\,max}} \left ( \sum_{k=0}^{K-1} T_{0,k} e^{\frac{- 4 \ln{2} (v-v_{0,k})^2}{ \Delta v_k^2}}\right ) \, {\rm d} v,
\label{e:nhi_tot}
\end{multline}

\noindent where we first sum over the column densities of all $N$ components detected in both absorption and emission ($N({\rm HI})_{\rm abs}$; Equation~\ref{e:nhi_comp}) and then add the total brightness temperature of the $K$ components detected in emission only for channels above $3\sigma_{T_B}(v)$ (i.e., between $v_{\rm em,\,min}$ and $v_{\rm em,\,max}$) under the optically thin approximation. We restrict the integration of the emission-only components so that $N({\rm HI})_{\rm total}$ can be compared  with $N({\rm HI})_{\rm thin}$ and $N({\rm HI})_{\rm iso}$. The uncertainty in $N({\rm HI})_{\rm total}$ is computed as the standard deviation over 1000 Monte Carlo trials for each LOS wherein all Gaussian component parameters are drawn from a normal distribution around each parameter \citep[e.g.,][]{heiles2003b, stanimirovic2014}. 

In Table~\ref{tab:coldens} we list $N({\rm HI})_{\rm thin}$, $N({\rm HI})_{\rm iso}$ and $N({\rm HI})_{\rm total}$ for all 57 21-SPONGE LOS, with uncertainties propagated in quadrature from $\sigma_{T_B}(v)$, $\sigma_{\tau}(v)$ for $N({\rm HI})_{\rm thin}$ and $N({\rm HI})_{\rm iso}$. We observe that $N({\rm HI})_{\rm total}$ and $N({\rm HI})_{\rm iso}$ are consistent within uncertainties, indicating that our autonomous decomposition method is performing well and recovering the majority of $\tau(v)$ for all LOS. 

To compare these column densities, we explore the effect of optical depth on the LOS column density by estimating the ``correction factor" to be applied to $N({\rm HI})_{\rm thin}$ to account for optical depth effects. Specifically, we compute two versions, one for the isothermal column density, $\mathcal{R}_{\rm iso} = N({\rm HI})_{\rm iso}/N({\rm HI})_{\rm thin}$, and one for the ``total" column density, $\mathcal{R}_{\rm total} = N({\rm HI})_{\rm total}/N({\rm HI})_{\rm thin}$. In Figure~\ref{f:coldens_comp}, we plot $\mathcal{R}_{\rm iso}$ and $\mathcal{R}_{\rm total}$ as a function of $N({\rm HI})_{\rm thin}$. 
For low \hi\ column densities, $N({\rm HI})_{\rm thin}\lesssim 5\times10^{20}\rm\,cm^{-2}$,  $\mathcal{R}_{\rm iso}$ and $\mathcal{R}_{\rm total}$ are equal to or consistent with unity within uncertainties. The scatter is larger (and uncertainties are higher) in the case of $\mathcal{R}_{\rm total}$ due to the uncertainty in fitting components between absorption and emission. In particular, the uncertainty in $\mathcal{R}_{\rm total}$ is only significant at low Galactic latitudes where significant line-blending yields large uncertainties in fitted parameters. 

\begin{figure}[t!]
\begin{center}
\includegraphics[width=0.45\textwidth]{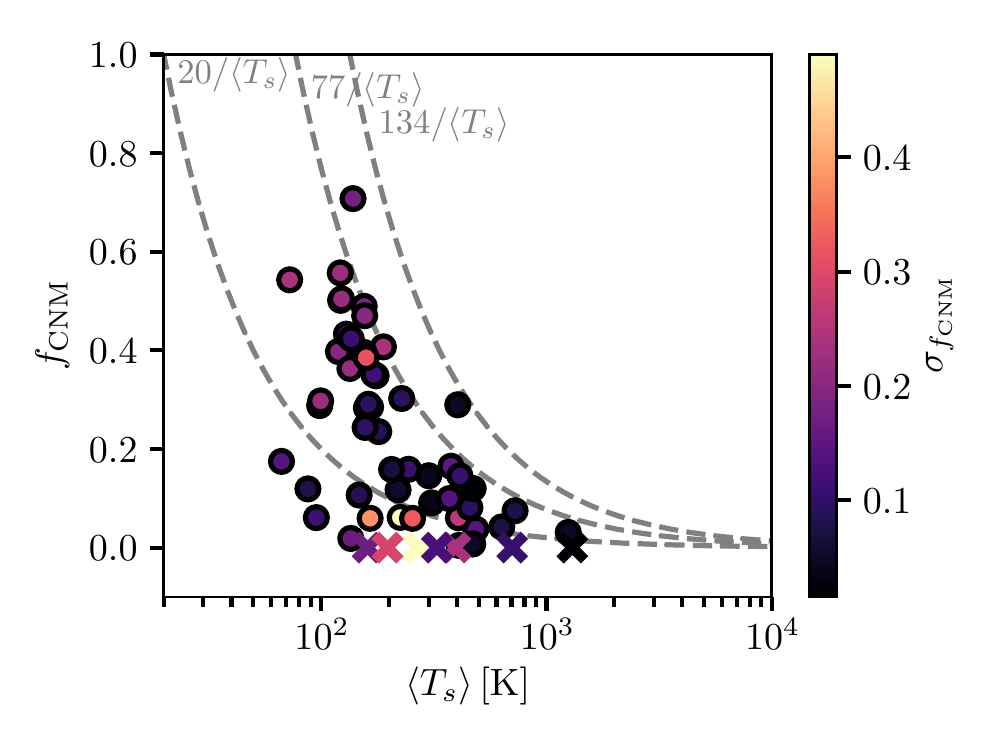}
\caption{Optical depth-weighted average spin temperature along the LOS ($\langle T_s \rangle$) vs. CNM fraction ($f_{\rm CNM}$). Dashed lines denote $T_s/\langle T_s \rangle$ for $T_s = 77\pm 57\rm\,K$ (the mean $\pm$ the standard deviation $T_s$ for the CNM). Crosses indicate LOS with $f_{\rm CNM}=0$, and points are colored by the uncertainty in $f_{\rm CNM}$ ($\sigma_{f_{\rm CNM}}$). Despite large scatter, as the majority of observed data points are consistent with the dashed lines, this suggests that $\langle T_s \rangle$ may be used as a tracer of $f_{\rm CNM}$. }
\label{f:fcnm_intts}
\end{center}
\end{figure}

In Figure~\ref{f:coldens_comp}, $\mathcal{R}_{\rm iso}$ and $\mathcal{R}_{\rm total}$ exhibit the same overall behavior as a function of $N({\rm HI})_{\rm thin}$, which has also been pointed out and discussed as part of a recent study of \hi\ in the Perseus molecular cloud \citep{lee2015} and recent analysis of data from HT03 and the first half of the 21-SPONGE survey (Nguyen et al. \,2018, submitted). This agreement indicates that our autonomous spectral decomposition method recovers the velocity structure of $21\rm\,cm$ absorption and emission well for the majority of sources. We will discuss this further in Section~\ref{sec:discussion_cnm}.

In Table~\ref{tab:coldens}, we also include the total column density along each LOS in the CNM and UNM phases, $N({\rm HI})_{\rm CNM}$ and $N({\rm HI})_{\rm UNM}$, as the sum of all components with $T_{s,\rm CNM}\leq250\rm\,K$ and $250<T_{s,\rm UNM}\leq1000\rm\,K$ (defined following predictions from \citet{wolfire2003} in the Solar circle) respectively. We then estimate the fraction of $N({\rm HI})_{\rm total}$ in each phase, with the assumption that the remaining mass is in the WNM. These mass fraction estimates are uncertain, as both uncertainties in the observations and the AGD-based fit contribute. To estimate the uncertainties in $f_{\rm CNM}$ and $f_{\rm UNM}$, we compute the standard deviation of $N({\rm HI})_{\rm CNM}$ and $N({\rm HI})_{\rm UNM}$ over the Monte Carlo trials used to estimate the uncertainty in $N({\rm HI})_{\rm total}$. Next, we compute upper limits to the column density of CNM and UNM below our sensitivity limit by integrating the uncertainty spectrum ($\sigma_{\tau}(v)$) and assuming $T_s = 100\rm\,K$ and $T_s = 500\rm\,K$ for the CNM and UNM respectively. The final uncertainty is the maximum value between these two estimates. %, with a minimum uncertainty of $10\%$. 

\begin{figure*}[t!]
\begin{center}
\includegraphics[width=0.95\textwidth]{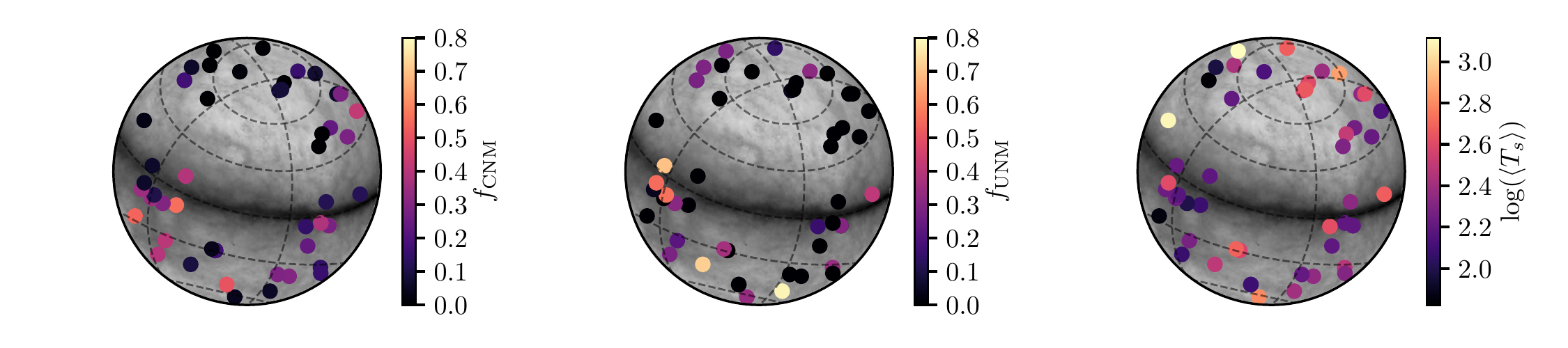}
\caption{All-sky plots of CNM fraction ($f_{\rm CNM}$; left), UNM fraction ($f_{\rm UNM}$; right) and optical depth-weighted harmonic mean spin temperature ($\langle T_s \rangle$; right), overlaid on \hi\ column density maps in ZEA projection from EBHIS \citep{winkel2016}.  }
\label{f:fracs_allsky}
\end{center}
\end{figure*}

Keeping in mind the large uncertainties in the LOS mass fractions, we observe most LOS are roughly less than or roughly half made up of CNM by mass. This is in agreement with a similar analysis of $21\rm\,cm$ absorption line pairs within and around the Perseus molecular cloud \citep{stanimirovic2014}, and also with HT03. All LOS feature a significant fraction of WNM by mass, detected only in emission. We will return to discuss the overall mass fractions in the CNM, UNM and WNM in Section~\ref{sec:discussion}. 

In Figure~\ref{f:fcnm_intts} we plot $f_{\rm CNM}$ versus the optical depth-weighted harmonic mean spin temperature, $\langle T_s \rangle$, given by,

\begin{equation}
\langle T_s \rangle = \frac{\int\, \tau(v) \cdot T_s(v) \, {\rm d} v}{\int \, \tau(v) \, {\rm d}v} = \frac{\int\, \tau(v) \cdot \frac{T_B(v)}{(1- \exp(-\tau(v))} \, {\rm d} v}{\int \, \tau(v) \, {\rm d}v} .
\end{equation}

\noindent Although the uncertainty in $f_{\rm CNM}$ is large, the observed data points cluster around a relation given by $f_{\rm CNM} = T_s/\langle T_s \rangle$ for $T_s = 77 \pm 57 \rm \,K$ (corresponding to the mean and standard deviation of $T_s$ for the CNM). This behavior is consistent with previous observational results \citep{stanimirovic2014, murray2015}, as well as KOK14, and suggests that $\langle T_s \rangle$ may be used as an alternative tracer for $f_{\rm CNM}$. Future observations at high latitude, where minimal line blending reduces the uncertainty in parameter decomposition, will be important for testing this hypothesis further. 

\begin{figure}[t!]
\begin{center}
\includegraphics[width=0.43\textwidth]{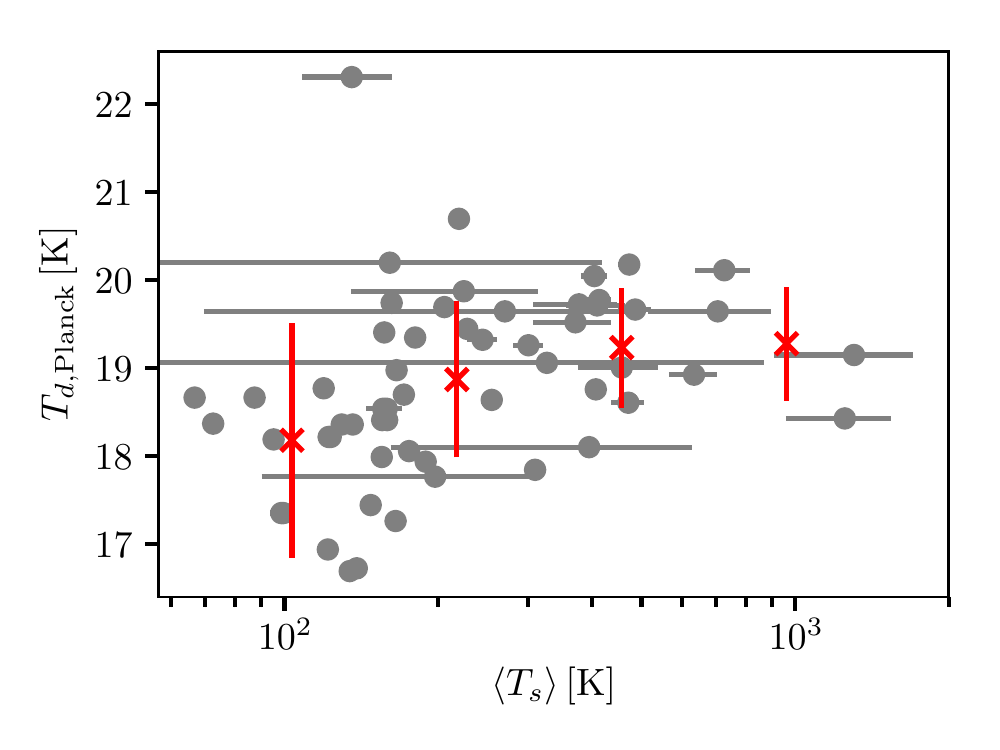} 
\vspace{-10pt}
\caption{Optical depth weighted average spin temperature ($\langle T_s\rangle$) versus dust temperature ($T_{d,\rm Planck}$). The mean and standard deviation over $T_{d,\rm Planck}$ in bins of increasing $\langle T_s \rangle$ are plotted as red crosses.}
\label{f:td_intts}
\end{center}
\end{figure}

In Figure~\ref{f:fracs_allsky} we plot the all-sky distributions of $f_{\rm CNM}$ (left), $f_{\rm UNM}$ (middle) and $\langle T_s \rangle$ (right). The highest Galactic latitudes are dominated by WNM (generally low $f_{\rm CNM}$ and $f_{\rm UNM}$), whereas low Galactic latitudes feature the largest $f_{\rm CNM}$. Sight lines with the largest $\langle T_s \rangle$, as shown in Figure~\ref{f:fcnm_intts}, feature small $f_{\rm CNM}$ and intermediate $f_{\rm UNM}$. 

We explore the effect of interstellar environment further in Figure~\ref{f:td_intts}, where we compare $\langle T_s \rangle$ with dust temperature $T_{d, \rm Planck}$, derived from modified black-body fits to \emph{Planck} observations at $353$, $545$ and $857\rm\,GHz$, as well as \emph{IRAS} $100\rm\,\mu m$ emission \citep{planck2014}. The dust temperature is a first-order approximation of the interstellar radiation field, which will affect the neutral gas temperature distribution. There is considerable scatter within a narrow range of observed $T_{d,\rm Planck}$, however there is a very weak discernible linear trend ($p=0.005$) of increasing $\langle T_s \rangle$ with increasing $T_{d,\rm Planck}$, illustrated by the mean and standard deviation $T_{d,\rm Planck}$ in bins of increasing $\langle T_s \rangle$ (red crosses). Taken together with the trends observed in Figure~\ref{f:fracs_allsky}, this provides very tentative evidence that stronger interstellar radiation fields result in smaller CNM fractions.

\subsection{Mass distribution of \hi\ as a function of $T_s$}

In Figure~\ref{f:tsnh}, we display a histogram (top panel) and CDF (bottom panel) of the fraction of the total column density detected in absorption and emission over all LOS per $T_s$ bin (${\rm d }\,N({\rm HI})/{\rm d }\, T_s$), a quantity that we denote the ``gas fraction" as a function of spin temperature. In the bottom panel, we include CDFs of the results of M15 (purple dashed) for comparison. We bootstrap the observed distributions over 1000 trials and include each resampled CDF to illustrate the effect of outliers on the distribution shapes. 

To test the effect of interstellar environment on the observed gas fractions, we isolate the \numberabshigh{} components found at high latitudes ($|b|>10^{\circ}$), and include the results for this subsample (black dotted). Sight lines at low Galactic latitude feature increased spectral line complexity due to blending of overlapping structures in radial velocity, and also probe different Galactic conditions. We find no significant difference in the observed gas fraction as a function of spin temperature between the full and high-latitude samples. We will discuss this further in Section~\ref{sec:discussion_unm}.

In Table~\ref{tab:frac} we list the total mass fractions of gas detected in absorption  from each ISM phase in 21-SPONGE. These fractions and their uncertainties were computed as the mean and standard deviation over bootstrapped trials (Figure~\ref{f:tsnh}). In each trial, using the range of predicted $T_s$ from \citet{wolfire2003} (c.f., their Table 3) we allowed the maximum spin temperature definition of the CNM to vary from $T_s< 150-350\rm\,K$ and the definition of the WNM to vary from $T_s>1000-4000\rm\,K$ (with the UNM defined as the intervening temperatures) to incorporate the uncertainty in these definitions. We also include the fractions of the total \hi\ column density in each phase, with the assumption that the $50\%$ of the total \hi\ column density detected by 21-SPONGE in emission alone is from the WNM.  

\subsection{Estimating observational bias}

\begin{figure}[t!]
\begin{center}
\vspace{-30pt}
\includegraphics[width=0.5\textwidth]{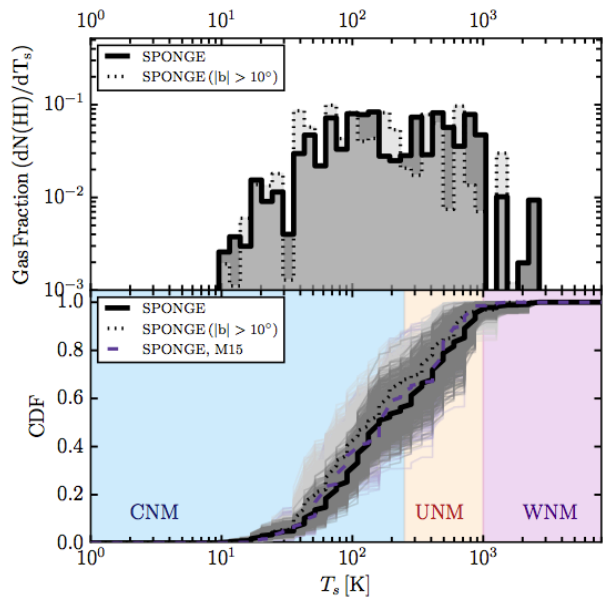} 
\vspace{-60pt}
\caption{Fraction of total column density detected by 21-SPONGE in absorption as a function of spin temperature for the full sample of \numberabscut{} absorption components (solid) and the \numberabshigh{} components detected in high-latitude ($|b|>10^{\circ}$) LOS (dotted). The results from the first half of the 21-SPONGE survey are shown for comparison (purple dashed; M15). Top: histogram; Bottom: cumulative distribution function (CDF). Approximate spin temperature ranges corresponding to the CNM ($T_s\leq250\rm\,K$), UNM ($250<T_s<1000\rm\,K$) and WNM ($T_s>1000\rm\,K$) are shaded in blue, orange and purple respectively.}

\label{f:tsnh}
\end{center}
\end{figure}
\begin{figure}[t!]
\begin{center}
\includegraphics[width=0.5\textwidth]{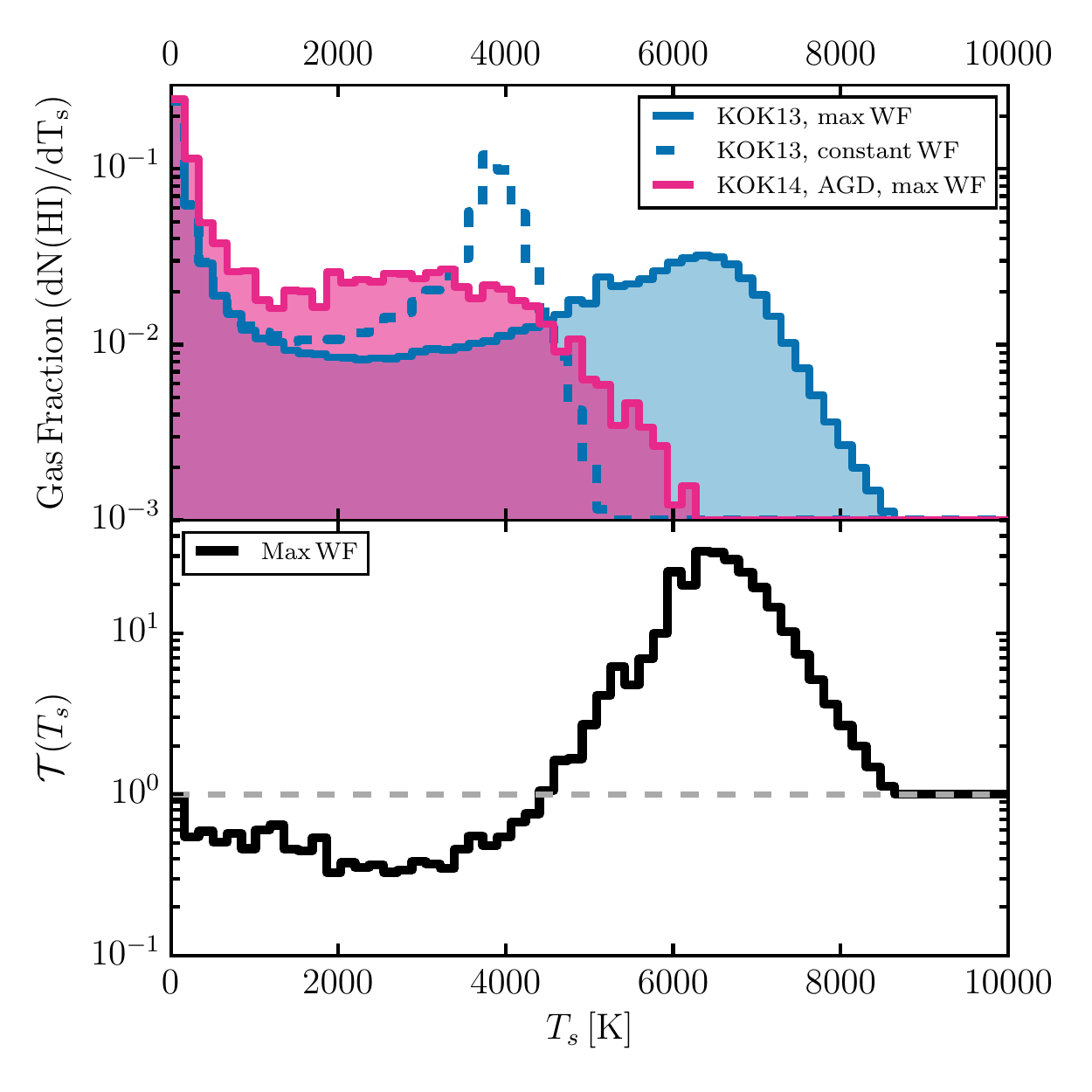} 
\caption{Top: Fraction of total column density as a function of spin temperature computed from the KOK13 simulation with maximum WF (i.e., $T_s = T_k$; solid blue), with constant WF (dashed blue), and from the AGD analysis of KOK14 (pink). Bottom: Transfer function, $\Tau(T_s)$, derived by comparing the temperatures and densities of ISM simulated by KOK13 with those inferred by the synthetic observations of KOK14 (Equation~\ref{e:tauv}). } 
\label{f:transfer}
\end{center}
\end{figure}

To address the bias imposed by our analysis method on the observed 21-SPONGE mass distribution (i.e., bias towards certain spectral line shapes imposed by AGD, or observational sensitivity limits), we compare \hi\ mass distributions from simulated data. First, in an identical manner to the 21-SPONGE distributions shown in Figure~\ref{f:tsnh}, we compute the fraction of total column density in each $T_s$ bin from the Gaussian decomposition of the \numberkok{} synthetic spectral line pairs from KOK14 with maximum WF. For each LOS, we then extract the density ($n$) and temperature ($T$) as a function of distance ($s$) from the KOK13 simulation used to construct the synthetic spectra in KOK14. To compare with the ``observed" mass distribution of \hi\ as a function of $T_s$, we estimate the ``true" underlying mass distribution by computing the fraction of total $n$ per $T_s$ bin for all gas along the KOK13 simulated LOS from which the KOK14 spectra were computed. We display both distributions in the top panel of Figure~\ref{f:transfer}.

To emphasize the influence of the WF prescription on these results, in the top panel of Figure~\ref{f:transfer} we also include the mass distribution as a function of $T_s$ from the KOK13 simulation under the assumption of a constant Ly$\alpha$ radiation field density ($10^{-6}\rm\,cm^{-3}$; dashed blue) for comparison. Although the mass fractions in the CNM are unchanged between ``constant" and ``maximum" WF effect implementation, the peak of the WNM spin temperature distribution changes from $\sim4000\rm\,K$ to $\sim6000\rm\,K$. Clearly, the implementation of the WF effect has the potential to change the WNM spin temperature distribution dramatically.

By comparing the KOK13 and KOK14 distributions in Figure~\ref{f:transfer} (with max WF) in detail, we can quantify the differences between observed and simulated \hi\ mass distributions. Specifically, we compute a ``transfer function", $\Tau(T_s)$, where,

\begin{equation}
\left (\frac{{\rm d }\,N({\rm HI})}{{\rm d }\, T_s}\right )_{\rm true} =  \Tau(T_s) \cdot \left( \frac{{\rm d }\, N({\rm HI})}{{\rm d }\, T_s } \right )_{\rm obs} .
\end{equation}

\noindent Assuming that KOK13 and KOK14 trace the ``true"  and observed gas fractions respectively, we have: 

\begin{equation}
\Tau(T_s) =  \left (\frac{{\rm d }\,N({\rm HI})}{{\rm d }\, T_s}\right )_{\rm KOK13} / \left (\frac{{\rm d }\,N({\rm HI})}{{\rm d }\, T_s}\right )_{\rm KOK14, AGD} .
\label{e:tauv}
\end{equation}

We display $\Tau(T_s)$ for all bins with ${\rm d }\,N({\rm HI})/{\rm d }\, T_s > 0.001$ in the bottom panel of Figure~\ref{f:transfer}. The shape of the transfer function encodes biases in our AGD-based analysis method's recovery of a realistic \hi\ mass distribution. In addition, $\Tau(T_s)$ encodes the radiative transfer and WF prescription required to go from $(n,T_k)$ in KOK13 to synthetic spectra in KOK14, as well as observational sensitivity limitations imposed by adding synthetic noise to the KOK14 spectra. We emphasize that $\Tau(T_s)$ will be used in this work to qualitatively assess these limitations. In the future, a full library of simulations and synthetic observations is required to find the best fit with observations, and correct observed gas fractions quantitatively.

\begin{deluxetable}{l | c | c } 
\tablecolumns{4}  
\tablewidth{3in}
\tablecaption{ \hi\ Mass Fractions  \label{tab:frac} } 
\tablehead{
\colhead{Phase}  & \colhead{Absorption}& \colhead{Total} }
\startdata
CNM & \cnmfrac{} & 0.28 \\
UNM & \unmfrac{} & 0.20 \\
WNM & \wnmfrac{} & 0.52 
\tablecomments{To compute the ``Absorption" mass fractions, we adopt definitions of each phase as follows: over all bootstrapped trials displayed in Figure~\ref{f:tsnh}, the maximum $T_s$ varies between $150$ and $350\rm\,K$ \citep[][ c.f., Table 3]{wolfire2003}, the minimum WNM $T_s$ varies between $1000$ and $4000\rm\,K$ \citep{liszt2001} and the UNM occupies intervening $T_s$. The ``Total" mass fractions are computed by incorporating the $50\%$ of the total \hi\ mass detected in emission only, which we assume is WNM. }
\enddata
\end{deluxetable}

\section{Discussion}
\label{sec:discussion}

\begin{figure*}
\begin{center}
\includegraphics[width=0.7\textwidth]{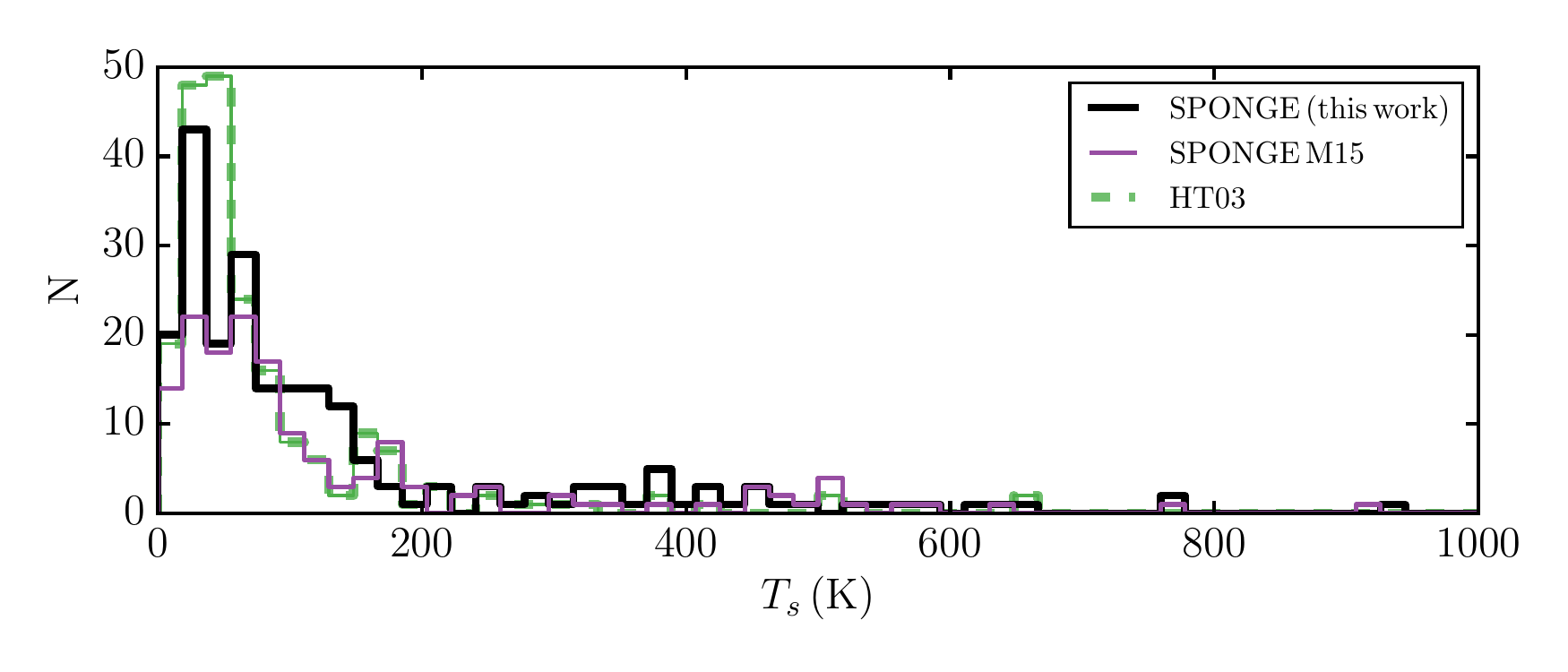}
\caption{Histogram of spin temperatures ($T_s$) for components with $T_s\leq1000\rm\,K$ from 21-SPONGE (57 LOS; thick black), the first half of the 21-SPONGE survey (M15; thin purple) and HT03 (79 LOS, dashed green). The higher sensitivity of 21-SPONGE has the effect of broadening the $T_s$ distribution for the CNM ($T_s\lesssim 200\rm\,K$) in both by-hand analysis following the HT03 method (M15) and the decomposition method presented here. }
\label{f:ts}
\end{center}
\end{figure*}

With the preceding analysis in hand, we will discuss the observed properties of Galactic \hi\ from 21-SPONGE, aided by the concurrent analysis of synthetic $21\rm\,cm$ spectral lines from numerical simulations. As numerical simulations produce models of the ISM with increasing resolution, precision and accuracy, and as wide field surveys at next-generation observatories roll in, this is likely to become an increasingly common practice. We emphasize that the AGD-based decomposition and radiative transfer approach presented here was designed to be applied to real and synthetic data volumes which are too large to analyze by hand, so that this study may serve as a pilot for future surveys. 

So, what have we learned from 21-SPONGE? In this section we discuss salient properties of the three canonical \hi\ phases, the CNM, UNM and WNM. 

\subsection{The CNM is ubiquitous}
\label{sec:discussion_cnm}

Of all the \hi\ phases, we are most successful in recovering the physical properties of the CNM. From a detailed comparison between individual clouds along simulated KOK13 LOS and decomposed spectral features from KOK14 synthetic spectra, we demonstrated that our spectral analysis method not only recovers the majority of ``real" CNM clouds, but successfully infers their column densities and spin temperatures \citep{murray2017}. Furthermore, we observe that the transfer function in Figure~\ref{f:transfer} is $\sim 1$ in the CNM regime at low $T_s$, indicating that our the AGD-plus-radiative transfer method is relatively complete in its recovery of the overall fraction of \hi\ mass in the CNM. 

21-SPONGE has established that with improved sensitivity in optical depth, signatures of the CNM in the form of weak, narrow absorption lines are detected ubiquitously. This agrees with \citet{stanimirovic2005}, who found that increased integration time on several non-detection sight lines from HT03 revealed CNM absorption features. Out of 57 total LOS, we have only 7 non-detections of the CNM ($\sim88\%$ detection rate). For five of these sources (3C236, 3C245B, 4C25.43, J1613, 3C345), no absorption was detected above $3\sigma_{\tau}(v)$, and for two (3C245A, 1055+018), no detected components featured $T_s\leq 250\rm\, K$ within uncertainties from the AGD-based fit. Furthermore, we detect the majority of \hi\ mass in absorption from the CNM \cnmfracperct{} ($28\%$ of the total \hi\ mass including WNM detected only in emission). 

However, despite the ubiquity of the CNM, the integrated optical depths are low enough that the correction to the total column density for the presence of cold \hi\ is relatively small. As shown in Figure~\ref{f:coldens_comp}, it is only sources at low Galactic latitudes ($|b|\lesssim10^{\circ}$) which exhibit correction factors significantly greater than unity. Our results agree with previous studies of $21\rm\,cm$ absorption to infer the optical depth correction to the \hi\ column density budget \citep[][Nguyen et al.\,2018, submitted]{heiles2003, lee2015, reach2017}. For example, \citet{lee2015} analyzed 26 sight lines within and around the Perseus molecular cloud to estimate the contribution of optically-thick \hi\ to the ``dark gas" budget (i.e. gas undetected in \hi\ or CO emission). Applying a similar trend as observed in Figure~\ref{f:coldens_comp}, they found that the total \hi\ mass of Perseus increases by only $10\%$ due to optically-thick \hi. Together our results contrast recent studies inferring significant corrections ($\gtrsim200\%$) to the total gas column density from optically-thick \hi\ based on observed far-infrared properties of dust \citep[e.g.,][]{fukui2015}, and indicate that optically thick \hi\ is not likely a dominant component of CO-dark gas in the local ISM. 

\begin{deluxetable*}{c  c | c | c c | c c || c | c c | c c}[b!]
\tablecolumns{12}  
\tablecaption{CNM and UNM $T_s$ Statistics \label{tab:ts_cnm} } 
\tablehead{\colhead{} & \colhead{} &\colhead{} &  \multicolumn{2}{c}{\emph{By $N_{\rm CNM}$} [K]} & \multicolumn{2}{c}{\emph{By $N({\rm HI})$ }[K]} & \colhead{} &  \multicolumn{2}{c}{\emph{By $N_{\rm UNM} $} [K]} & \multicolumn{2}{c}{\emph{By $N({\rm HI})$} [K]}\\
\colhead{Latitude} 	& \colhead{Survey} 			&  \colhead{$N_{\rm CNM}$}  & \colhead{Mean} & \colhead{Median} & \colhead{Mean} & \colhead{Median}  & \colhead{$N_{\rm UNM}$}  & \colhead{Mean} & \colhead{Median} & \colhead{Mean} & \colhead{Median} }

\startdata
High 				& 21SPONGE 	&  109 	 &	73 	& 	61  	& 	95 	& 	79 		& 	23  	&  	450 		&  390 		&  500	& 	460 	\\
($|b|>10^{\circ}$)		& HT03   		&  135	 &     63 	& 	45	& 	74 	& 	56 		&      11 	& 	420		& 380		& 400  	&	380  \\

\hline
Low				 & 21SPONGE 	&  71 	 &	 68 	& 	49 	& 	108 	& 	106 		& 	14  	&  	460 		& 430 		&  510	 & 	490 	\\
($|b|<10^{\circ}$)		& HT03 		&  51 	 &      67	& 	47 	& 	 86 	& 	60 		& 	2 	&   	300		& 300		&  260	 & 	260 

\enddata
\tablecomments{Means and medians by ``by $N$" are computed for all $N_{\rm CNM}$ CNM components ($T_s \leq 250\rm\,K$) and  all $N_{\rm UNM}$ UNM components ($250\leq T_s \leq 1000\rm\,K$) with no weighting; following \citet{heiles2003b}, the median ``by $N({\rm HI})$" is the $T_s$ for which half the total CNM or UNM column density lies above and half below, and the mean is weighted by $N({\rm HI})$. }

\end{deluxetable*}

To compare with previous studies of CNM properties, in Figure~\ref{f:ts} we display histograms of $T_s$ below $1000\rm\,K$ from 21-SPONGE (thick black), M15 (thin purple) and HT03 (dashed green). The HT03 distribution is strongly peaked at $\sim50\rm\,K$, indicative of a characteristic $T_s$ for the CNM. Although we observe a similar feature in 21-SPONGE, we also find evidence for a broader CNM $T_s$ distribution. To compare the statistics in detail, in Table~\ref{tab:ts_cnm} we compute mean and median values for the CNM $T_s$ from 21-SPONGE and HT03 (i.e., all components with $T_s\leq250\rm\,K$) with and without weighting by $N({\rm HI})_{\rm abs}$ for two latitude bins, above and below $|b|=10^{\circ}$, following Table 2 of \citet{heiles2003b}. Considering the considerable scatter (standard deviations $\gtrsim 50\rm\,K$), the values are generally consistent. We appear to detect higher mean and median CNM $T_s$ weighted by $N({\rm HI})_{\rm abs}$ in 21-SPONGE than HT03, consistent with the observed broadening of the $T_s$ histogram in Figure~\ref{f:ts}. 

Ultimately, improving the statistics of the CNM $T_s$ distribution will be crucial for understanding the sources of observed differences and scatter between distributions, as environmental effects are likely important. For example, photoelectric heating by dust grains may be enhanced in particularly dust-rich Galactic environments, or strong variations in turbulence may broaden the distribution. 
Upcoming $21\rm\,cm$ absorption surveys at the VLA and Australian Square Kilometer Array Pathfinder \citep[GASKAP;][]{dickey2013} will dramatically expand the known sample of Galactic \hi\ absorption properties. Although these studies will likely not reach the optical depth sensitivity in individual sight lines achieved by 21-SPONGE, they will be crucial for resolving detailed CNM properties as a function of Galactic environment.

\subsection{Thermally unstable gas fraction}
\label{sec:discussion_unm}

A key motivation for the 21-SPONGE survey was to determine the effect of improved optical depth sensitivity on the inferred fraction of \hi\ mass in the thermally unstable regime (e.g., $250\lesssim T_s\lesssim 1000\rm\,K$). There is substantial evidence in the literature for a significant population of thermally unstable gas in the ISM, however the majority of constraints on unstable spin temperatures are inferred as upper limits from line width-based kinetic temperatures \citep[e.g.,][]{verschuur1994, heiles2003, haud2007}. With detections in $\tau$ and $T_B$, we can constrain $T_s$ and more accurately assess the thermodynamic state of the neutral ISM.

However, as spectral features corresponding to warmer gas are characterized by broader line widths, spectral complexity due to velocity blending increases the difficulty in recovering accurate gas density and temperature from both $21\rm\,cm$ absorption and emission. We quantified this effect in \citet{murray2017} by showing that our decomposition and radiative transfer approach tends to overestimate the temperatures of non-CNM structures ($T_s\gtrsim400\rm\,K$). In that study, we were primarily sensitive to gas with $T_s<1000\rm\,K$ (i.e., CNM and UNM) for which we could unambiguously identify ``true" simulated counterpart structures along the simulated LOS. Considering this bias, the mass fraction of thermally-unstable \hi\ in absorption presented here is possibly an upper limit, as CNM with overestimated $T_s$ may contribute. We also note that in the expected range of thermally unstable temperatures ($250\lesssim T_s \lesssim1000\rm\,K$), we observe that $\Tau(T_s)$ is relatively flat but $<1$ (Figure~\ref{f:transfer}), indicating that our analysis method is not only sensitive to these conditions but that we tend to overestimate the true mass fraction. 

To illustrate the typical UNM properties detected by 21-SPONGE (in comparison with HT03), in Table~\ref{tab:ts_cnm} we display mean and median values for the UNM $T_s$ from 21-SPONGE and HT03 (i.e., all absorption components with $250<T_s<1000\rm\,K$) with and without weighting by $N({\rm HI})_{\rm abs}$ for high and low Galactic latitudes. The statistics are much poorer for comparing UNM properties (i.e., HT03 only find two UNM components at low latitudes, due to limited observational sensitivity), however, the observed UNM properties appear to be largely consistent with each other at both high and low latitudes. 

Our estimate of the thermally unstable gas fraction (\unmfracperct{} of \hi\ detected in absorption by mass, and $\sim20\%$ of the total observed \hi\ mass) is generally consistent with previous observational results. For example, a similar high-sensitivity \hi\ absorption line study from \citet{roy2013b} argued that at most $28\%$ of \hi\ is thermally unstable. HT03 asserted that $\sim48\%$ of the total WNM column density beyond the Galactic plane, or $\sim30\%$ of the \emph{total} out-of-plane column density, has thermally unstable temperatures. We note that this estimate was derived from components detected in emission for which HT03 estimated upper-limits to $T_s$ based on the spectral line width. For comparison with our UNM fraction, of the gas detected in absorption by HT03, only $9\%$ by mass has $250\leq T_s<1000\rm\,K$. Finally, KOK13 also find a substantial fraction of gas out of thermal equilibrium, $\sim 18\%$ by mass, due to strong turbulence, expanding shocks from supernovae, and a time-dependent heating rate. Below $T_s\sim1000\rm\,K$, our observed distribution qualitatively agrees with their results (c.f., Figure 8d of KOK13). 

In contrast with HT03, we do not observe a significant difference between the mass distribution of \hi\ (including the thermally unstable gas fraction) between the full sample and subsample of sight lines at high Galactic latitude ($|b|>10^{\circ}$). Considering that velocity blending and overall fitting uncertainties are higher at lower latitudes, we might expect some kind of bias towards certain gas populations in different regimes. To resolve variations with Galactic environments, we need larger samples of high-sensitivity absorption lines to improve statistics, which may distinguish between regimes where thermal instability, turbulence and dynamical processes dominate the regulation of the ISM \citep{wolfire2015}. 

\subsection{How ``warm" is the WNM?}
\label{sec:discussion_wnm}

From standard, steady-state ISM models, we expect the WNM to have kinetic temperature of $T_k=5000-10000\rm\,K$ and spin temperature of $T_s\sim1000-4000\rm\,K$ \citep{liszt2001, wolfire2003}. In the WNM, it is typically assumed that $T_s<T_k$ because collisions are insufficient at low densities for thermalizing the $21\rm\,cm$ transition. Our AGD analysis of KOK14 spectra demonstrates that we should easily detect WNM in the expected range of spin temperature, despite the limitations of our observational sensitivity. The prominent peak at $T_s\sim2500\rm\,K$ in the KOK14 histogram of Figure~\ref{f:tsnh} (pink) illustrates the sensitivity of the AGD method to this range of temperature. However, we detect very little \hi\ mass in absorption from the WNM (\wnmfracperct{}).

Consequently, that we do not detect a significant mass fraction of WNM from 21-SPONGE with $T_s\gtrsim1000-4000\rm\,K$ indicates that the WNM spin temperature may be higher than standard analytical predictions, which are based on \hi\ excitation via particle collisions alone \citep[e.g.,][]{liszt2001}. In agreement, \citet{murray2014} detected an unexpectedly ``warm" WNM population with $T_s=7200^{+1800}_{-1200}\rm\,K$, which was attributed to supplemental excitation beyond collisions from resonant scattering of Ly$\alpha$ photons \citep[the WF effect;][]{wouthuysen1952, field1958}. In \citet{murray2017}, we found that simple WF treatment in KOK14 spectra produces spectral features corresponding to expected WNM properties (i.e., $T_s\sim3000-4000\rm\,K$), but which are \emph{not} detected by 21-SPONGE, and this comparison is what led us to analyze synthetic spectra from KOK14 with ``maximum WF" in this study. Sophisticated theoretical treatment of the WF effect in future simulations is of utmost importance, as myriad environmental conditions, including metallicity and cosmic ray ionization rate, will affect Ly$\alpha$ pumping of the $21\rm\,cm$ transition \citep[e.g.,][]{shaw2017}.

\begin{figure}[t!]
\begin{center}
\includegraphics[width=0.45\textwidth]{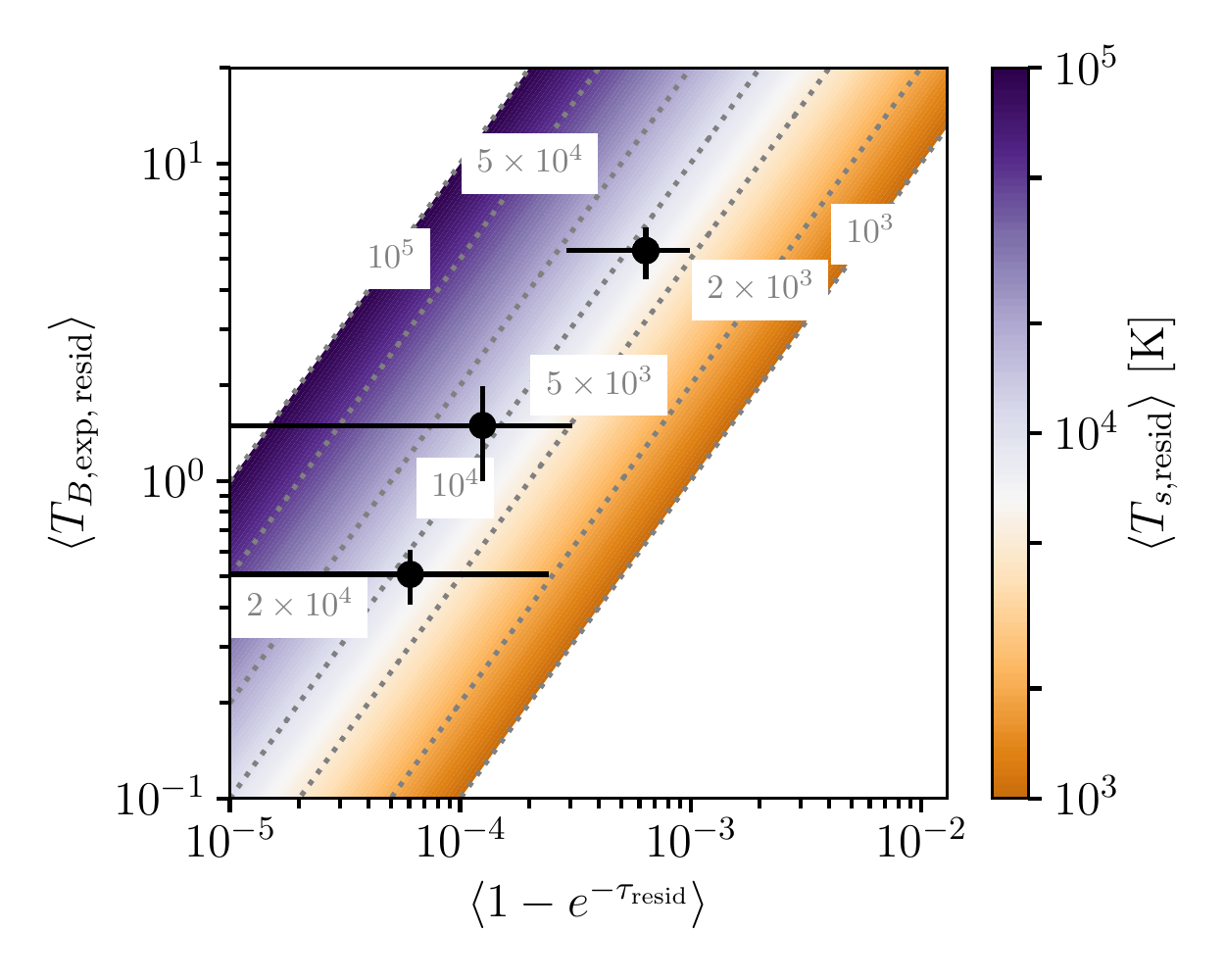} 
\caption{Stacked residual absorption $\langle 1-e^{-\tau_{\rm resid}} \rangle$ vs. stacked residual emission $\langle T_{B,\rm resid}\rangle$ for all spectral channels binned by residual emission: $3\sigma_{T_B} \leq T_{B, \rm resid}<0.9\rm\,K$, $0.9\leq T_{B,\rm resid}<  2.4\rm\,K$, and $T_{B, \rm resid}> 2.4\rm\, K$ (limits chosen to select significantly-detected residual emission in three bins of roughly equal sizes). Dotted grey lines, shading and inset labels denote constant harmonic mean spin temperature for the stacked residual channels ($\langle T_{s, \rm resid}\rangle $). For the bin of largest residual emission, we detect a significant residual absorption signal, corresponding to $\langle T_{s, \rm resid} \rangle \approx 10^4\rm\,K$.  }
\label{f:resids_wnm}
\end{center}
\end{figure}

Ultimately, considering that we analyze the ``maximum" WF case of the KOK14 synthetic spectra ($T_s=T_k$ for all \hi), we are likely limited by observational sensitivity to detect gas at higher spin temperatures (i.e., $T_s\gtrsim4000\rm\,K$: prominent in the KOK13 mass distribution but missing from the inferred distribution from KOK14). We added synthetic Gaussian noise with $\rm RMS = 1\times10^{-3}$ to the synthetic absorption spectra from KOK14 to mimic the 21-SPONGE sensitivity, and AGD is unable to recover lines with $T_s \gtrsim 4000\rm\,K$ at $\rm S/N \geq 3$ with this level of spectral uncertainty. As a result, the majority of \hi\ mass not detected in absorption ($50\%$ of the total LOS column density) is likely in the form of WNM and the mass fraction detected in absorption (\wnmfracperct{}) is an obvious lower limit. Future studies aiming to detect the absorbing properties of the WNM will need to target even better optical depth sensitivity than we have achieved with 21-SPONGE, or improve sensitivity by applying stacking techniques similar to \citet{murray2014}.

To test for the presence of WNM below our sensitivity limit, we performed a simple stacking experiment. First, we subtracted all components detected in both absorption and emission from $\tau(v)$ and $T_{B,\rm exp}(v)$ to produce residual absorption and emission spectra, $\tau_{\rm resid}(v)$ and $T_{B,\rm exp,\, resid}(v)$ for all LOS. The residual spectra contain components with unphysical $T_s$ (i.e., not successfully fitted to $T_{B,\rm exp}(v)$) and \hi\ below our sensitivity limit to detect in absorption (i.e., WNM at high temperature). We mask all velocity channels with components detected in $\tau(v)$ but not $T_{B,\rm exp}(v)$, as these features are likely CNM or UNM which we did not recover due primarily to beam mismatch effects. We further mask by-eye channels with strong residuals due to over-subtraction of Gaussian components (typically $\sim10$ channels for roughly half of the LOS). Next, we compute the average of all unmasked velocity channels whose $T_{B,\rm resid}$ falls into one of four bins (chosen to select significantly-detected residual emission in bins of roughly equal size), $3\sigma_{T_B}\leq T_{B, \rm exp,\,resid}< 0.9\rm\,K$, $0.9\leq T_{B,\rm exp, \,resid}<  2.4\rm\,K$, and $T_{B,\rm exp, \,resid} > 2.4\rm\,K$,
 weighted by $1/\sigma_{\tau}$. The number of channels in each bin is $2383$, $2604$ and $2423$ respectively from low to high $T_{B, \rm exp, \,resid}$. In Figure~\ref{f:resids_wnm} we plot the resulting weighted averages of the absorption channels, $\langle 1-e^{-\tau_{\rm resid}(v)} \rangle$, and the emission channels $\langle T_{B,\rm exp,\, resid}\rangle$. The weighted average absorption in the bins of smallest $T_{B, \rm exp,\,resid}$ are consistent with zero, however, we detect a significant residual absorption signal in the bin of highest $T_{B, \rm exp,\,resid}$. The uncertainties are computed as the standard deviation over $10^4$ trials wherein we bootstrap-resampled the LOS used in the stack with replacement. We denote constant harmonic mean $T_s$ ($\langle T_{s,\rm resid} \rangle$) with dotted lines and shading in Figure~\ref{f:resids_wnm}, and observe that the significant residual absorption feature is consistent with  $\langle T_{s,\rm\, resid} \rangle \sim10^4 \rm\,K$. 
 
The inferred spin temperature of the detected residual absorption signal in Figure~\ref{f:resids_wnm} is consistent with our previous findings \citep[$T_s=7200^{+1800}_{-1200}\rm\,K$][]{murray2014}. Although we do not detect a significant signal in the bins of smallest $T_{B,\rm resid}$, possibly due to over-subtraction of Gaussian components in complex LOS, we note that $\langle T_{s,\rm resid}\rangle$ is similar in all bins. This is expected if we are sampling broad spectral-line features from a diffuse, warm parent \hi\ population, rather than randomly sampling occasional peaks in $T_B$ or $\tau$ due to fitting imperfections. If the signal detected here \citep[and in ][]{murray2014} originates from the diffuse WNM with a constant temperature, bins of higher $T_{B,\rm resid}$ should correspond to higher $(1-e^{\tau_{\rm resid}})$. 

Overall, the detection of a residual absorption signal consistent with high WNM spin temperature further emphasizes the importance of supplementary diffuse neutral gas excitation (e.g., the WF effect) in producing such high $T_s$. In addition, future studies aiming to detect the absorbing properties of the WNM will need to target even better optical depth sensitivity than we have achieved with 21-SPONGE, or improve sensitivity by applying stacking techniques similar to those demonstrated in Figure~\ref{f:resids_wnm} and by \citet{murray2014}.

\section{Summary} 
\label{sec:summary}

In this work, we present the data release of 21-SPONGE, a large Karl G. Jansky VLA survey for high-sensitivity absorption by Galactic \hi\ to probe the temperature distribution of the neutral ISM. 21-SPONGE is distinguished among previous $21\,\rm cm$ studies of the Galactic \hi\ as a result of: (1) exceptional optical depth sensitivity ($\sigma_{\tau} < 10^{-3}$ per $0.42\rm\,km\,s^{-1}$ channels) thanks to careful calibration considerations and the upgraded capabilities of the VLA WIDAR correlator for producing wide spectral bandwidths and narrow velocity channels for resolving the cold neutral medium (CNM), unstable neutral medium (UNM) and warm neutral medium (WNM) simultaneously; (2) matching single-dish $21\rm\,cm$ emission spectra with the highest-possible angular resolution ($\sim 4'$) from the Arecibo Observatory, minimizing the mismatch with sub-arcminute interferometric VLA absorption measurements; (3) detailed comparisons with 3D numerical simulations of the ISM for assessing observational biases. In this work, we have presented a novel method for autonomously decomposing $21\rm\,cm$ absorption and emission spectra and deriving the physical properties (column density, temperature) for individual spectral features via detailed radiative transfer. The efficient, objective nature of the analysis method enables us to compare our results with thousands of synthetic observations from numerical simulations. Our main results are summarized here:

\begin{enumerate}

\item We demonstrate that with improved optical depth sensitivity, narrow absorption lines arising from the CNM are detected ubiquitously. The detection rate of $21\rm\,cm$ absorption from the CNM is $\sim 88\%$. However, the optical depth of these features is typically small so that contribution of cold, optically-thick gas to the \hi\ mass budget is typically small ($<20\%$). We find that the CNM fraction along typical 21-SPONGE LOS is $\lesssim 50\%$. 
\item To assess the biases of our observational methods, we apply the same analysis techniques to a sample of \numberkok{} synthetic \hi\ spectral line pairs from \citet{kim2014}, constructed from the 3D hydrodynamical simulation by \citet{kim2013}. We add spectral noise to the synthetic dataset to mimic the 21-SPONGE sensitivity limits. By comparing the underlying simulated gas properties with those inferred from the synthetic spectral lines, we construct a ``transfer function", $\Tau(T_s)$ between the true and observed mass distribution of \hi\ as a function of temperature (i.e., by dividing the two distributions). We find that for $T_s<4000\rm\,K$, $\Tau(T_s)\sim 1$, indicating that we are sensitive to \hi\ properties within this regime. At higher $T_s$ (i.e., $\gtrsim4000\rm\,K$), $\Tau(T_s)>1$, indicating that our synthetic spectral line analysis is likely missing a significant fraction of warm neutral medium (WNM) mass present in the KOK13 simulation. 
\item We compute the fractions of \hi\ mass detected in emission \textit{and} absorption (i.e., for which we have constraints on $T_s$ for measuring $N({\rm HI})$, corresponding to $50\%$ of the total \hi\ mass) in the cold neutral medium (CNM; \cnmfracperct{}), WNM (\wnmfracperct{}) and thermally unstable medium (UNM; \unmfracperct{}). Our UNM mass fraction, among the first observational constraints from direct \hi\ absorption detections, is generally consistent with previous indirect observational estimates \citep[e.g.][]{heiles2003b}. Incorporating the remaining $50\%$ of \hi\ mass detected in emission alone (i.e., for which we do not have constraints on $T_s$, which we assume is due to WNM), the mass fractions are $28\%$, $20\%$ and $52\%$ for the CNM, UNM and WNM respectively. 
\item Although the WNM comprises the majority of the total \hi\ mass ($52\%$), the lack of WNM absorption detected by 21-SPONGE in the spin temperature range expected from steady-state collisional excitation models \citep[$T_s=1000-4000\rm\,K$; ][]{liszt2001,wolfire2003} implies that the WNM spin temperature is higher, likely due to supplemental excitation from the Wouthuysen-Field (WF) effect. This is in agreement with previous analysis of 21-SPONGE and the KOK14 synthetic spectra \citep{murray2017}, as well as stacking analysis of 21-SPONGE spectra which revealed a high-spin temperature WNM population with $T_s=7200^{+1800}_{-1200}\rm\,K$ \citep{murray2014}. As a test of this hypothesis, following spectral line modeling, we stack residual absorption in bins of residual emission and detect a significant absorption feature with harmonic mean spin temperature $\sim10^4\rm\,K$. 

\end{enumerate}

Overall, larger samples of $21\rm\,cm$ absorption lines, as well as next-generation simulations with sophisticated WF treatment, are required to improve the statistical uncertainties and probe the effect of Galactic environment on these results. 
We emphasize that the autonomous, efficient nature of the AGD method presented here will enable detailed analysis of future real and synthetic data volumes which will be orders of magnitude larger than 21-SPONGE. However, even in the era of next-generation interferometers (e.g., SKA), future surveys will be unlikely to target the superb optical depth sensitivity reached by 21-SPONGE \citep[e.g.,][]{mcg2015}, ensuring that this dataset will provide an important benchmark for future work.

\acknowledgements{
This work was supported by the NSF Early Career 
Development (CAREER) Award AST-1056780. C.\,E.\,M. acknowledges 
support by the National Science Foundation Graduate Research 
Fellowship and the Wisconsin Space Grant Institution. 
S.\,S. thanks the Research Corporation for Science 
Advancement for their support.  
The authors would like to thank Bob Lindner and 
Carlos Vera-Ciro for developing AGD, writing GaussPy, and inspiring the methods
presented here.
C.\,E.\,M. thanks Elijah Bernstein-Cooper, Helga D\'enes, Katie Jameson, 
James Dempsey and Naomi McClure-Griffiths for helpful 
conversations and collaboration towards developing GaussPy. 
We also thank Van Hiep Nguyen and Joanne Dawson for
valuable discussions that led to improved beam efficiency considerations
for our expected brightness temperature spectra.
This work makes use of data from the Karl G. Jansky Very Large Array, operated by the National Radio Astronomy Observatory (NRAO). NRAO is a facility of the NSF operated under cooperative agreement by Associated Universities, Inc. EBHIS is based on observations with the 100-m telescope of the MPIfR (Max-Planck-Institut fŸr Radioastronomie) at Effelsberg. The Parkes Radio Telescope is part of the Australia Telescope which is funded by the Commonwealth of Australia for operation as a National Facility managed by CSIRO.  We acknowledge the use of the Legacy Archive for Microwave Background Data Analysis (LAMBDA), part of the High Energy Astrophysics Science Archive Center (HEASARC). HEASARC/LAMBDA is a service of the Astrophysics Science Division at the NASA Goddard Space Flight Center. This research has made use of the NASA/IPAC Extragalactic Database (NED) which is operated by the Jet Propulsion Laboratory, California Institute of Technology, under contract with the National Aeronautics and Space Administration.  This research has made use of NASA's Astrophysics Data System. This research made use of Astropy, a community-developed core Python package for Astronomy \citep{2013A&A...558A..33A}, NumPy \citep{van2011numpy}, and matplotlib, a Python library for publication quality graphics \citep{Hunter:2007}.
}

\bibliographystyle{apj}
\bibliography{ms}

\appendix{

\begin{figure*}
\begin{center}
\includegraphics[width=\textwidth]{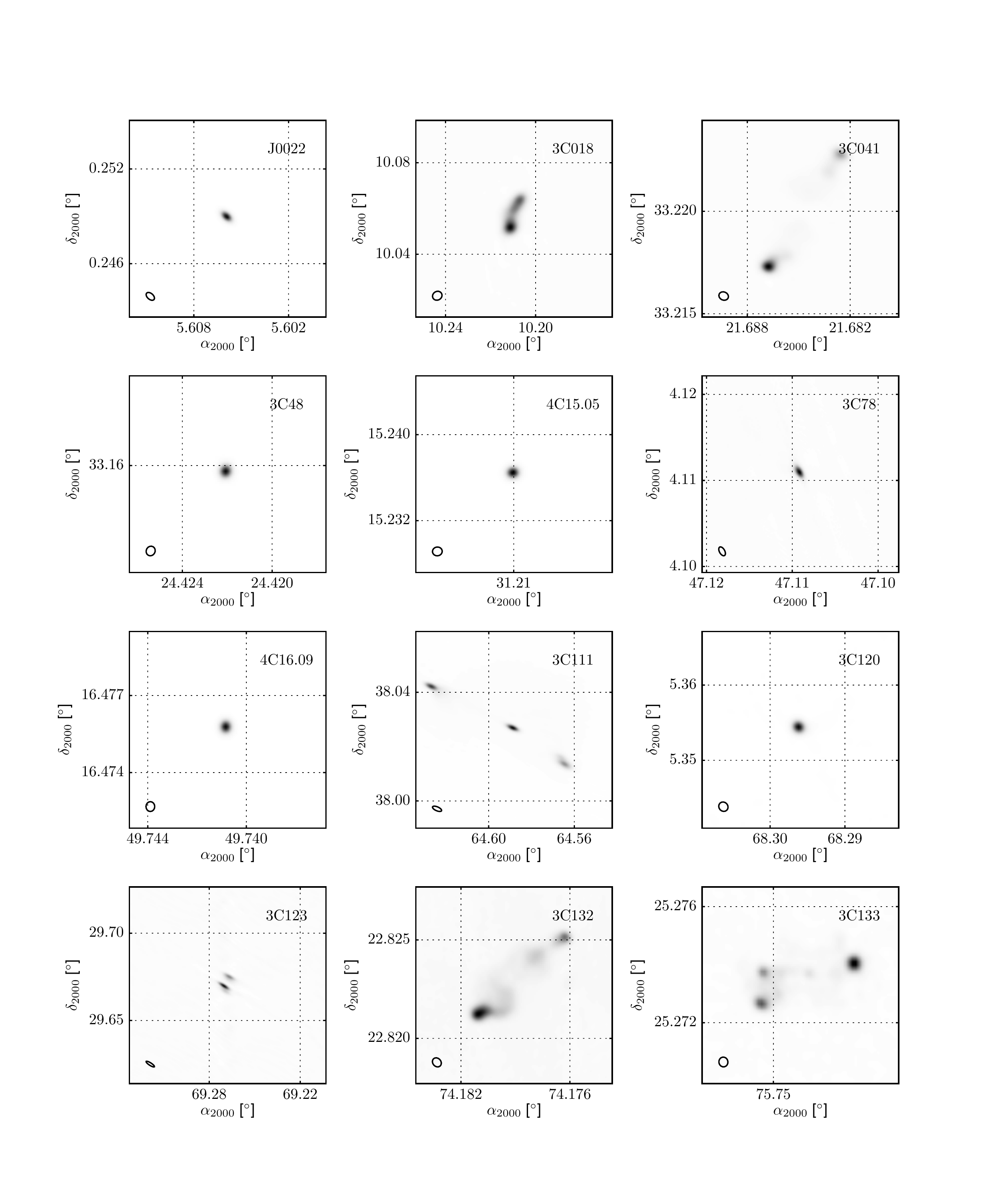}
\caption{21SPONGE $1.42040575\rm\,GHz$ continuum images. The source name is printed within each panel, and the synthesized beam used to construct each image is included in the bottom left corner. Each image is scaled so that the peak flux density is unity.}
\label{apB:f:cont}
\end{center}
\end{figure*}

\begin{figure*}
\begin{center}
\includegraphics[width=\textwidth]{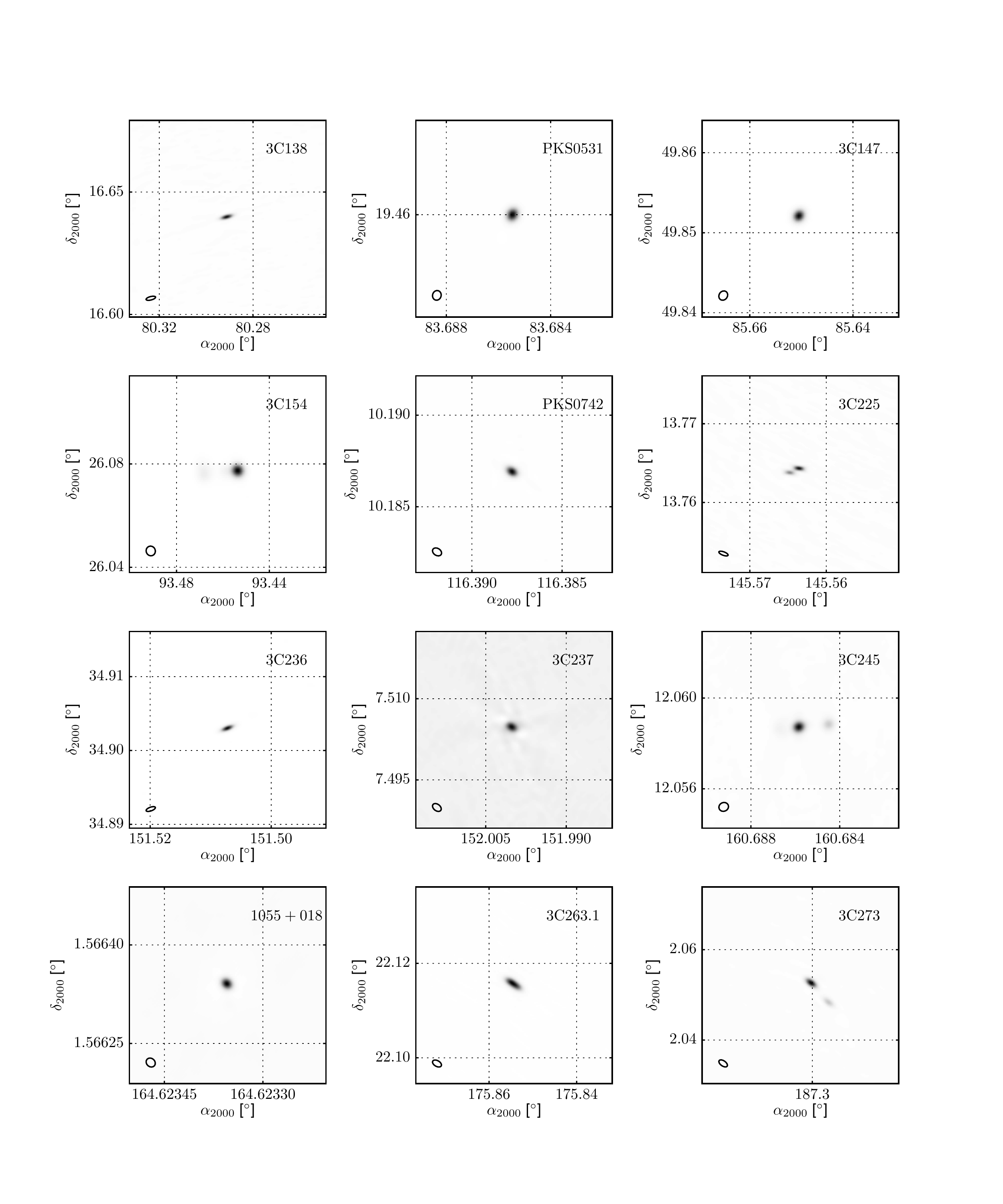}
%\caption[Appendix: 21-SPONGE continuum]{ 21SPONGE HI continuum images}
\caption{(contd.)}
%\label{chap5:f:sources2}
\end{center}
\end{figure*}

\begin{figure*}
\begin{center}
\includegraphics[width=\textwidth]{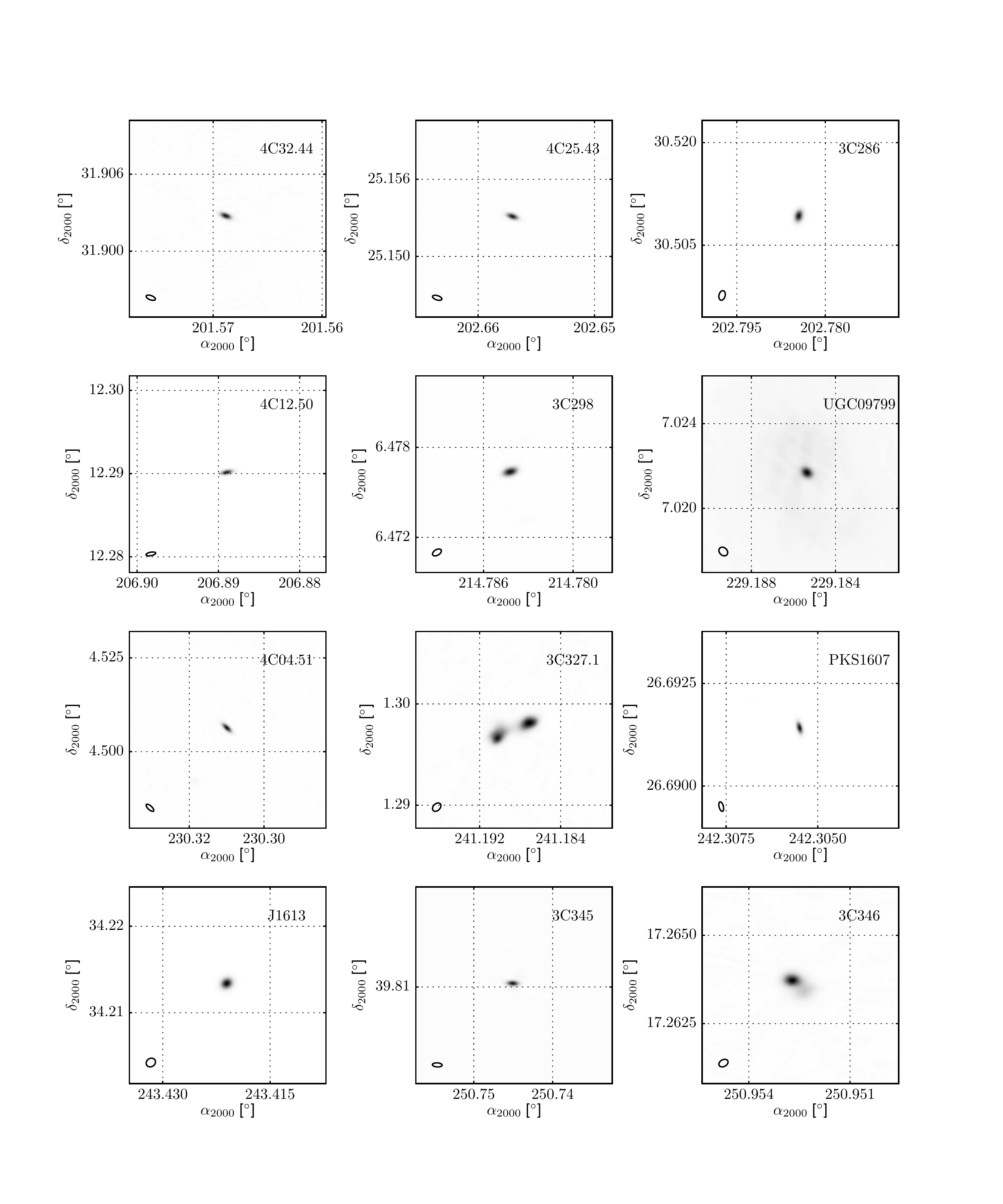}
%\caption[Appendix: 21-SPONGE continuum]{ 21SPONGE HI continuum images}
\caption{(contd.)}
%\label{chap5:f:sources2}
\end{center}
\end{figure*}

\begin{figure*}
\begin{center}
\includegraphics[width=\textwidth]{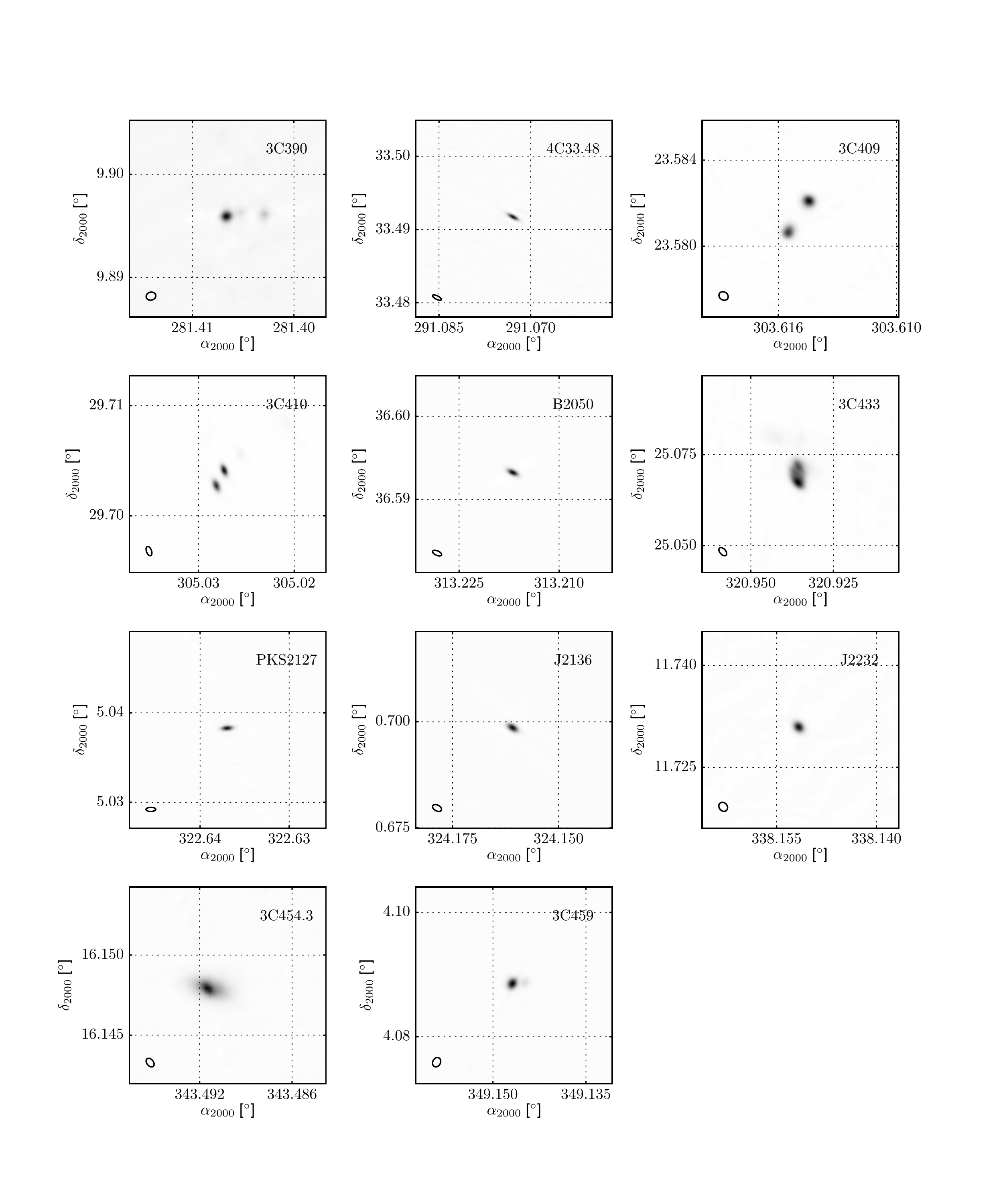}
%\caption[Appendix: 21-SPONGE continuum]{ 21SPONGE HI continuum images}
\caption{(contd.)}
%\label{chap5:f:sources2}
\end{center}
\end{figure*}

\begin{figure*}
\begin{center}
\includegraphics[width=\textwidth]{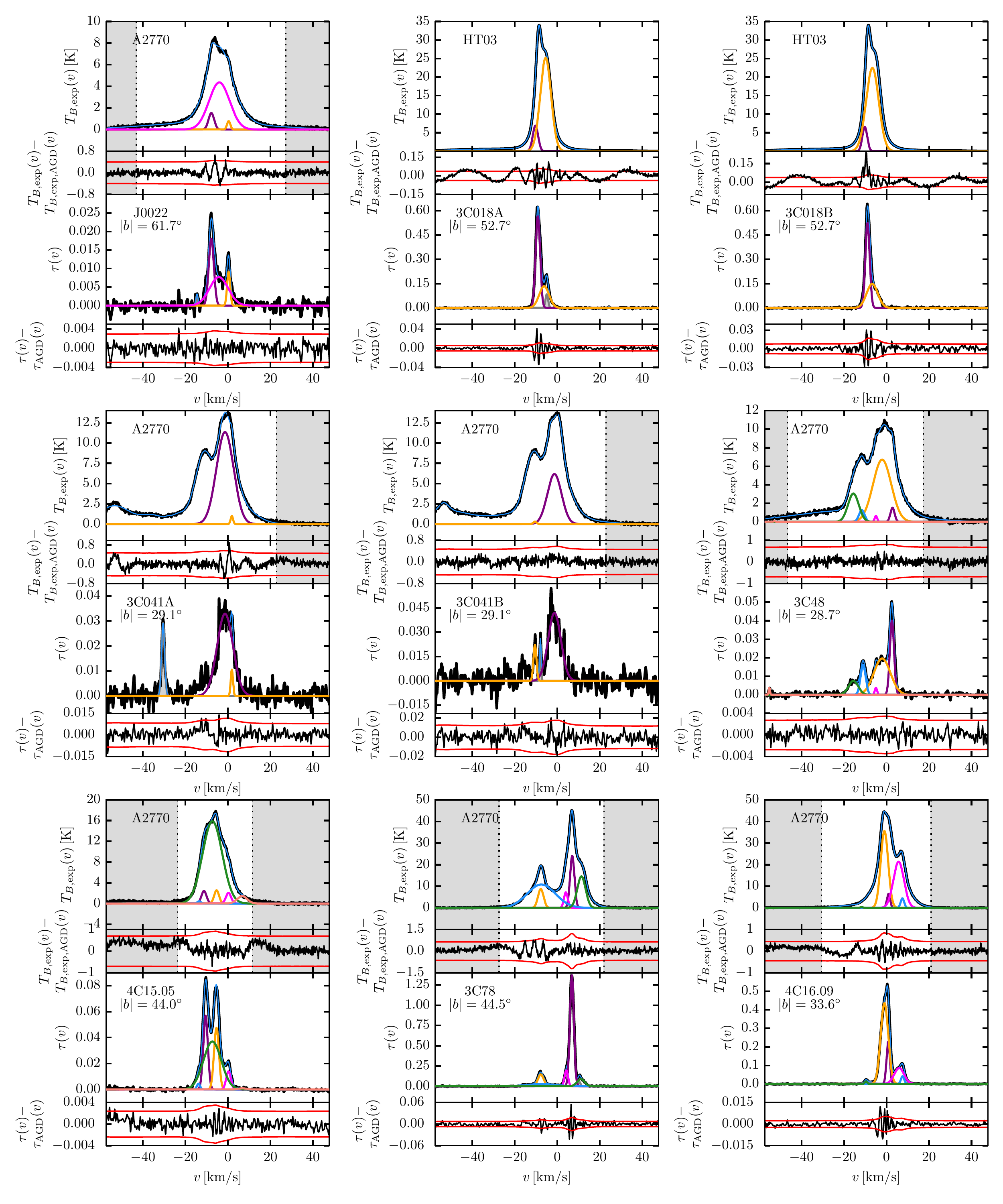}
%\vspace{-200pt}
\caption{ A summary of the Gaussian fits to 21-SPONGE \hi\ emission and absorption spectral line pairs described in Section~\ref{sec:gauss}. In each panel, we plot the emission ($T_{B,\rm exp}(v)$; top) and absorption ($\tau(v)$; bottom) profiles. The residual spectra following the Gaussian fits below are included below each panel, with$\pm 3\times $ the noise spectra for $T_B$ and $\tau$ respectively (red). We plot all fitted absorption components in the bottom panel. Components whose derived spin temperatures are unphysical (i.e., $\leq 10\rm\,K$) are plotted in shaded grey, and components with $T_s > 10\rm\,K$ are plotted in matching colors in the middle and top panels. The total fits, $T_{B, \rm exp, \,AGD}(v)$ and $\tau_{\rm AGD}(v)$ are displayed in thin, blue lines. The source of $T_{B,\rm exp}(v)$, whether from Arecibo (A2770 or HT03) or EBHIS is printed in the top panels. In the bottom panels, we print the source name and the absolute Galactic latitude ($|b|$). Finally, in the top panels we shade in grey the velocities where $T_{B,\rm exp}(v) \leq 3 \cdot \sigma_{T_B}(v)$ to illustrate the range over which LOS column densities are computed (if no vertical shading is present, the full displayed velocity range is used).}
\label{f:sources0}
\end{center}
\end{figure*}

\begin{figure*}
\begin{center}
\includegraphics[width=\textwidth]{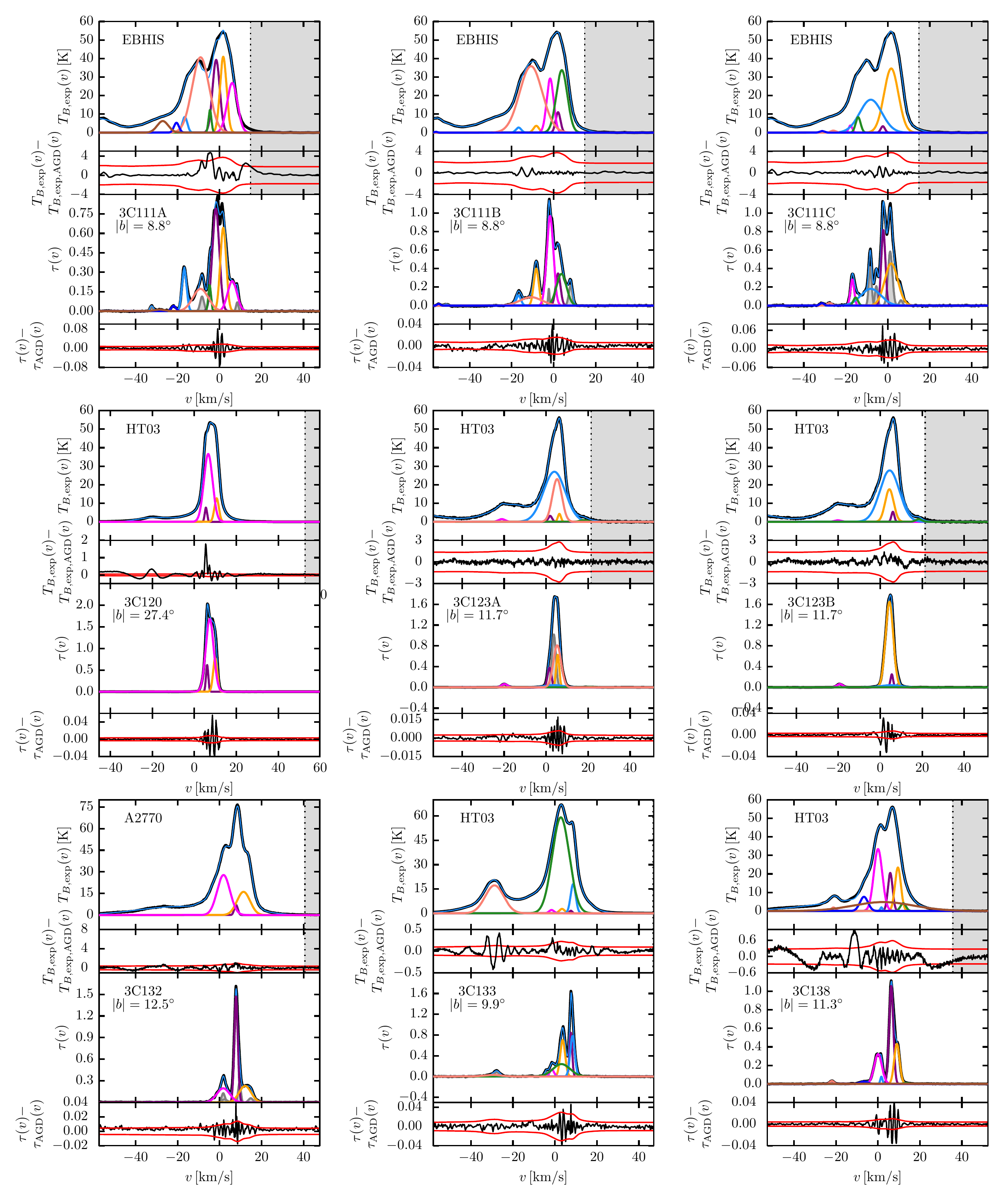}
\caption{See Figure~\ref{f:sources0} for details.}
\label{f:sources1}
\end{center}
\end{figure*}

\begin{figure*}
\begin{center}
\includegraphics[width=\textwidth]{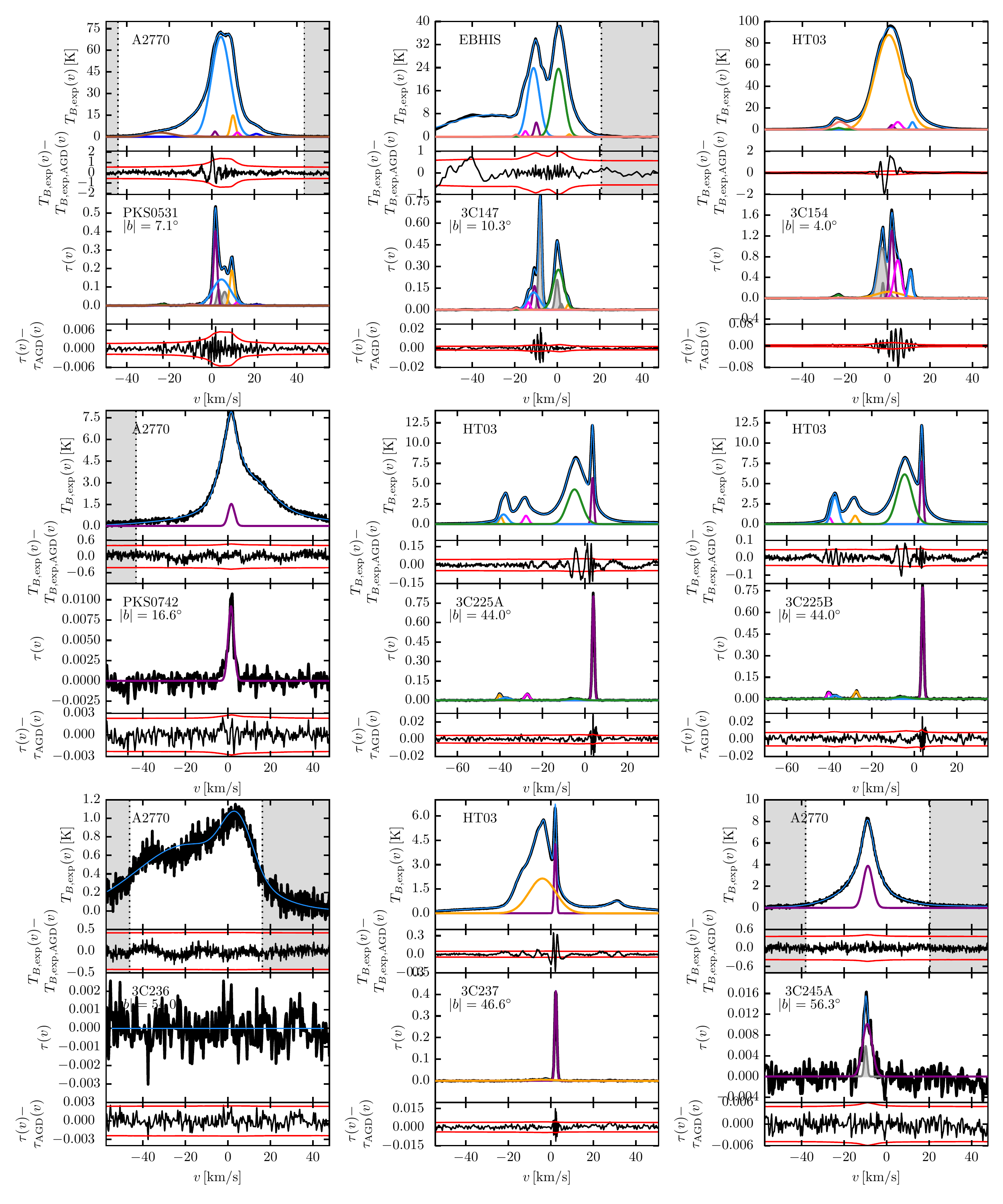}
%\caption[21-SPONGE spectral line pairs and fits (3)]{See caption to Figure~\ref{f:sources0}.}
\caption{See Figure~\ref{f:sources0} for details.}
\label{f:sources2}
\end{center}
\end{figure*}

\begin{figure*}
\begin{center}
\includegraphics[width=\textwidth]{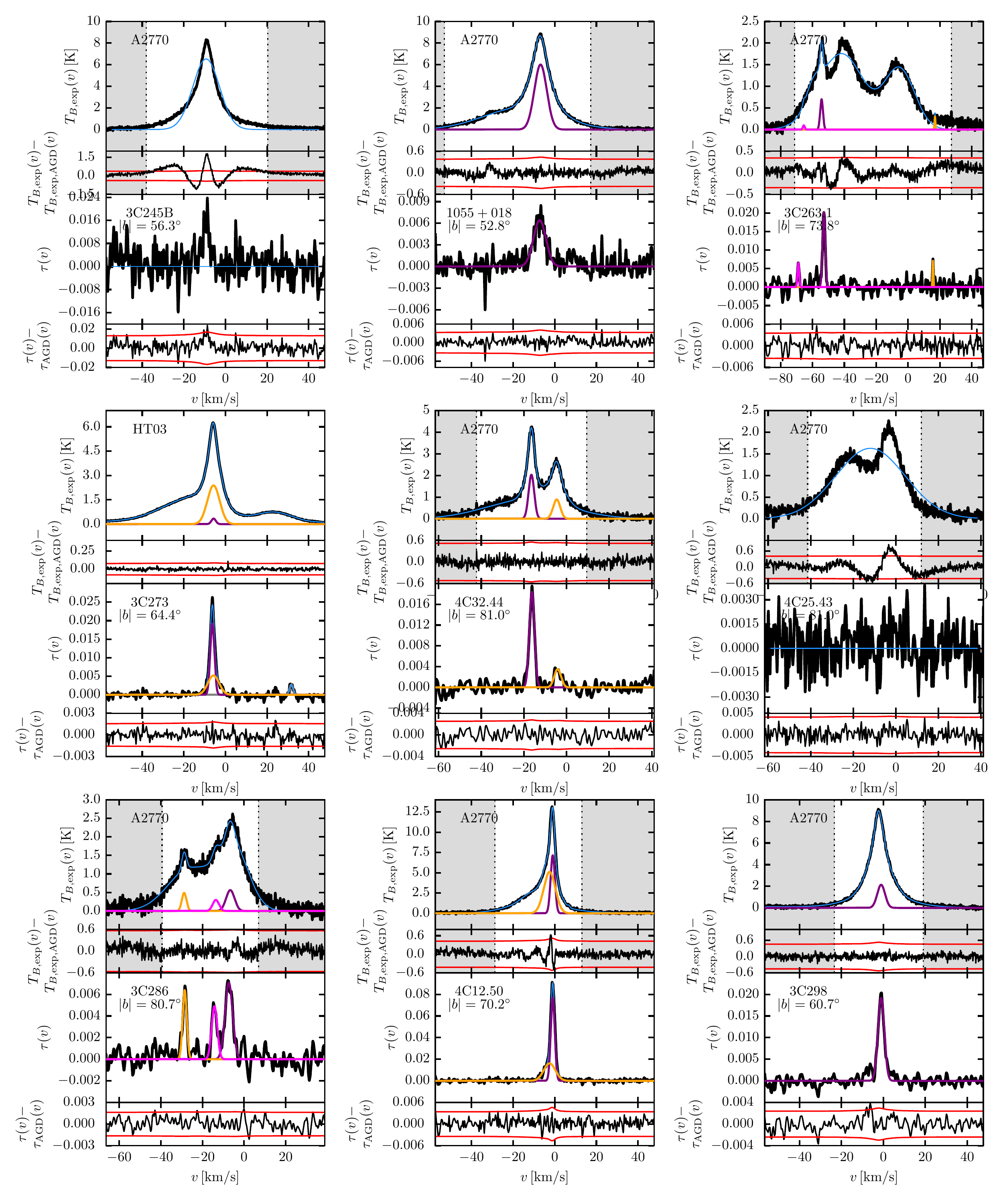}
%\caption[21-SPONGE spectral line pairs and fits (4)]{See caption to Figure~\ref{f:sources0}.}
\caption{See Figure~\ref{f:sources0} for details.}
\label{f:sources3}
\end{center}
\end{figure*}

\begin{figure*}
\begin{center}
\includegraphics[width=\textwidth]{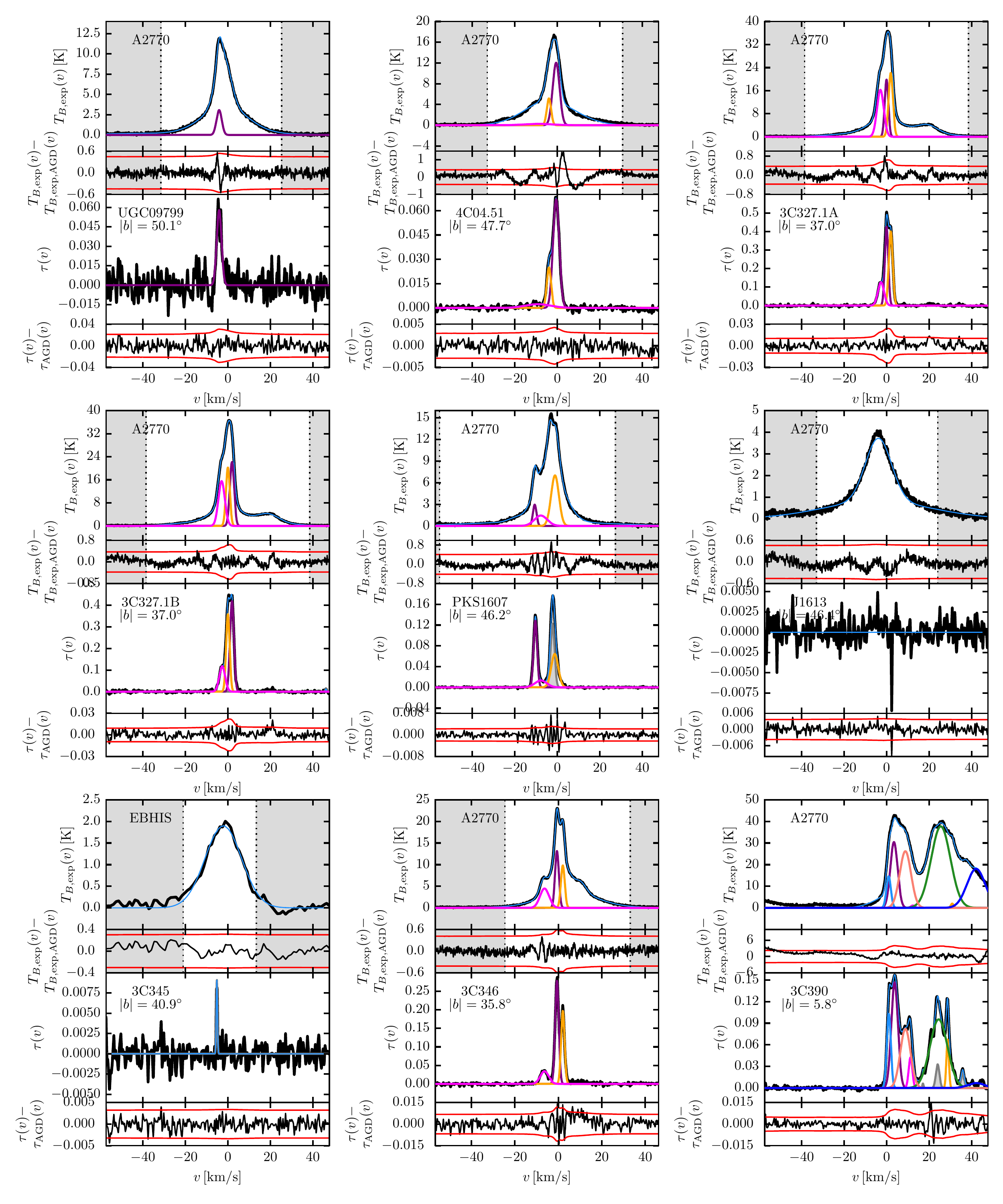}
%\caption[21-SPONGE spectral line pairs and fits (5)]{See caption to Figure~\ref{f:sources0}.}
\caption{See Figure~\ref{f:sources0} for details.}
\label{f:sources4}
\end{center}
\end{figure*}

\begin{figure*}
\begin{center}
\includegraphics[width=\textwidth]{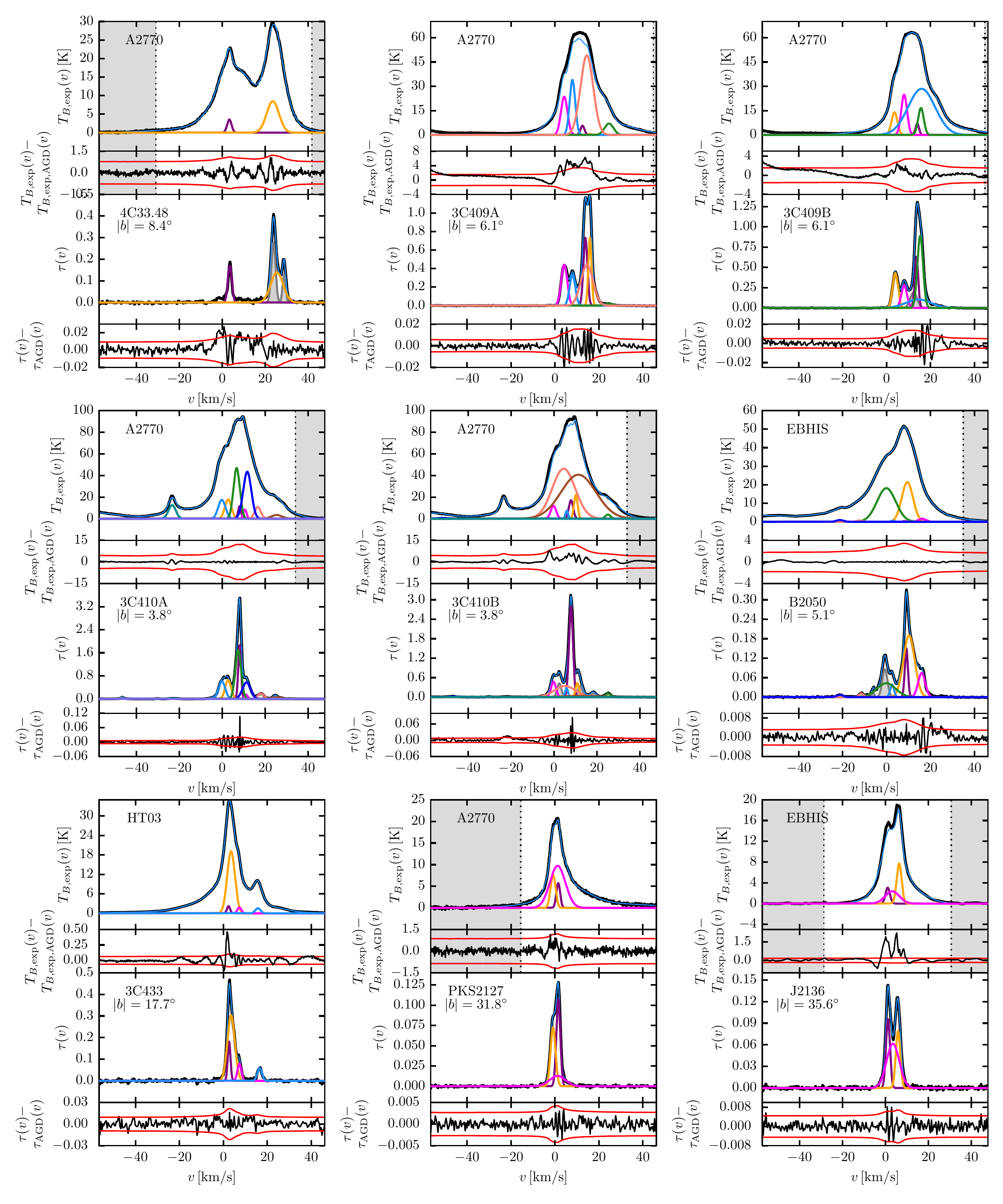}
%\caption[21-SPONGE spectral line pairs and fits (6)]{See caption to Figure~\ref{f:sources0}.}
\caption{See Figure~\ref{f:sources0} for details.}
\label{f:sources5}
\end{center}
\end{figure*}

\begin{figure*}
\begin{center}
\includegraphics[width=\textwidth]{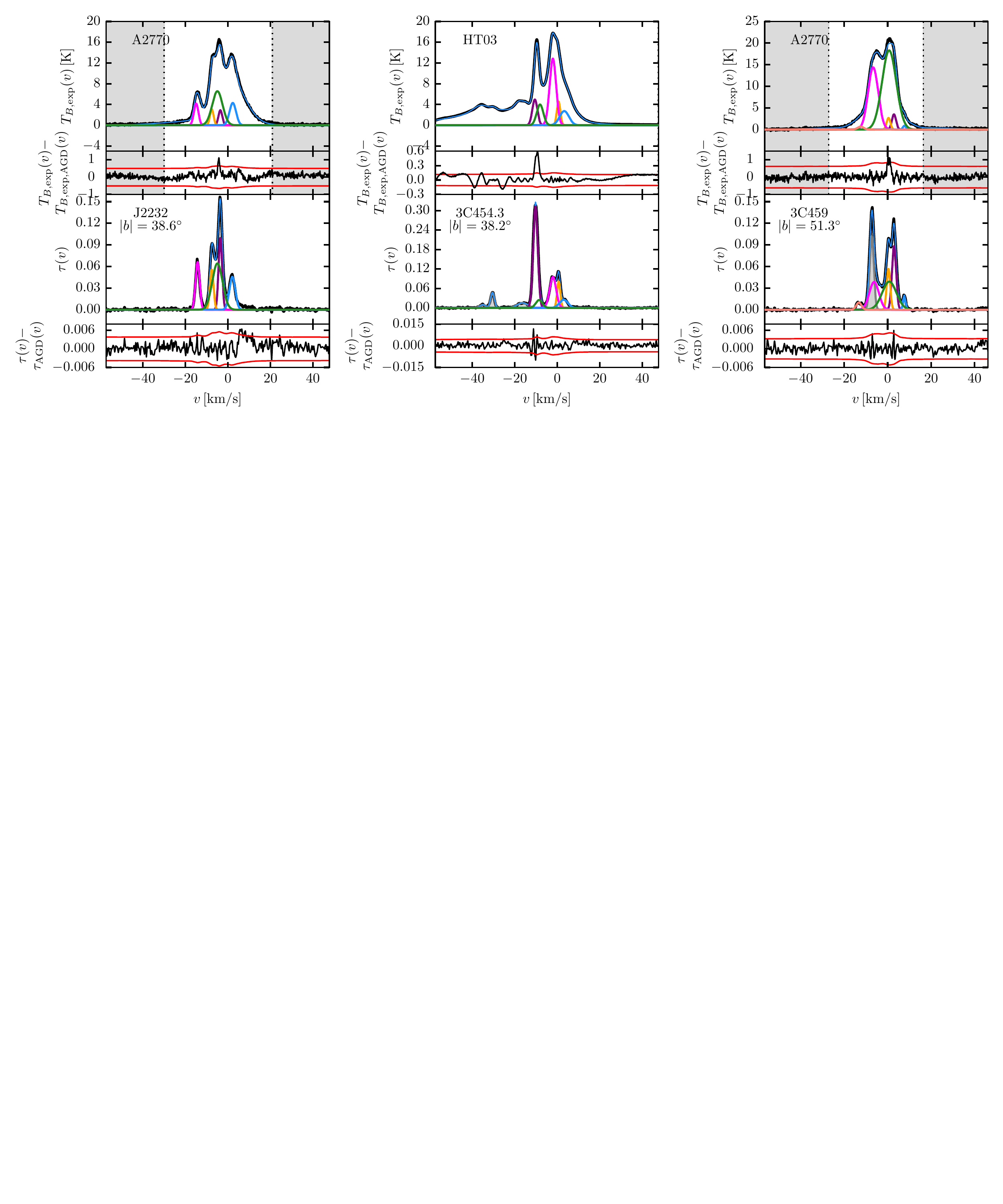}
\vspace{-400pt}
%\caption[21-SPONGE spectral line pairs and fits (7)]{See caption to Figure~\ref{f:sources0}.}
\caption{See Figure~\ref{f:sources0} for details.}
\label{f:sources6}
%\label{f:sources6}
\end{center}
\end{figure*}

\clearpage % optional
\LongTables

\begin{deluxetable*}{cccc|ccc|cccc}
\setlength{\tabcolsep}{0.05in} 
\tablecolumns{11} 
%\label{chap5:tab:comps} 
\tabletypesize{\scriptsize}
\tablewidth{0pt}
\tablecaption{Fitted Parameters\label{tab:params}}
\tablehead{
\colhead{Source}  &  \colhead{$\tau_0$} &  \colhead{$\Delta v_0$}         &  \colhead{$v_0$}        & \colhead{$T_{B,n}$}  &  \colhead{$\Delta v_{0,n}$}         &  \colhead{$v_{0,n}$}    & \colhead{$T_s$}       & \colhead{$N({\rm HI})_{\rm abs}$}    & \colhead{$\mathscr{O}$}  & \colhead{$\mathscr{F}$} \\ [1mm]
\colhead{(name)}   &          &  \colhead{($\rm km\,s^{-1}$)}  & \colhead{($\rm km\,s^{-1}$)}   & \colhead{($\rm K$)}  & \colhead{($\rm km\,s^{-1}$)}  & \colhead{($\rm km\,s^{-1}$)}  &  \colhead{($\rm K$)}  & \colhead{($10^{20}\rm\, cm^{-2}$)} }
\startdata
J0022  &  0.018$\pm$0.001  &  2.8$\pm$0.1  &  -7.8$\pm$0.1  &  1.5$\pm$0.0  &  3.1$\pm$0.1  &  -7.8$\pm$0.1  &  83$\pm$2  &  0.08$\pm$0.00  &  4  &  0.0    \\   
  &  0.008$\pm$0.001  &  10.3$\pm$0.3  &  -4.3$\pm$0.2  &  4.4$\pm$0.1  &  11.4$\pm$0.1  &  -4.1$\pm$0.1  &  567$\pm$19  &  0.89$\pm$0.05  &  4  &      \\   
  &  0.009$\pm$0.001  &  1.7$\pm$0.1  &  0.3$\pm$0.1  &  0.8$\pm$0.0  &  1.8$\pm$0.1  &  0.3$\pm$0.1  &  84$\pm$4  &  0.03$\pm$0.00  &  4  &      \\   
  &  0.003$\pm$0.001  &  1.0$\pm$0.2  &  -14.6$\pm$0.1  &  \nodata  &  \nodata  &  \nodata  &  \nodata  &  \nodata  &  \nodata  &    \\  
\hline   
3C018A  &  0.565$\pm$0.007  &  2.5$\pm$0.1  &  -9.1$\pm$0.1  &  6.9$\pm$0.1  &  3.0$\pm$0.1  &  -10.3$\pm$0.1  &  17$\pm$1  &  0.49$\pm$0.04  &  4  &  1.0    \\   
  &  0.134$\pm$0.003  &  5.5$\pm$0.2  &  -6.2$\pm$0.1  &  25.2$\pm$0.6  &  6.0$\pm$0.1  &  -5.5$\pm$0.1  &  196$\pm$5  &  2.87$\pm$0.13  &  2  &      \\   
  &  0.084$\pm$0.003  &  1.5$\pm$0.1  &  -5.0$\pm$0.1  &  \nodata  &  \nodata  &  \nodata  &  \nodata  &  \nodata  &  \nodata  &    \\  
  &  0.007$\pm$0.003  &  0.7$\pm$0.3  &  24.4$\pm$0.1  &  \nodata  &  \nodata  &  \nodata  &  \nodata  &  \nodata  &  \nodata  &    \\  
\hline   
3C018B  &  0.524$\pm$0.006  &  2.4$\pm$0.1  &  -9.0$\pm$0.1  &  6.6$\pm$0.9  &  2.7$\pm$0.3  &  -10.3$\pm$0.3  &  16$\pm$2  &  0.42$\pm$0.06  &  2  &  1.0    \\   
  &  0.149$\pm$0.005  &  6.2$\pm$0.1  &  -6.8$\pm$0.1  &  22.5$\pm$0.5  &  7.0$\pm$0.1  &  -6.7$\pm$0.1  &  162$\pm$4  &  2.95$\pm$0.10  &  1  &      \\   
\hline   
3C041A  &  0.033$\pm$0.004  &  8.9$\pm$0.2  &  -1.4$\pm$0.1  &  11.4$\pm$0.2  &  9.5$\pm$0.1  &  -1.4$\pm$0.1  &  351$\pm$7  &  2.04$\pm$0.07  &  3  &  1.0    \\   
  &  0.011$\pm$0.004  &  1.2$\pm$0.2  &  1.8$\pm$0.1  &  1.0$\pm$0.2  &  1.3$\pm$0.1  &  1.8$\pm$0.1  &  98$\pm$16  &  0.02$\pm$0.01  &  3  &      \\   
  &  0.029$\pm$0.003  &  1.7$\pm$0.1  &  -30.5$\pm$0.1  &  \nodata  &  \nodata  &  \nodata  &  \nodata  &  \nodata  &  \nodata  &    \\  
\hline   
3C041B  &  0.022$\pm$0.005  &  1.7$\pm$0.2  &  -10.5$\pm$0.1  &  0.3$\pm$0.0  &  1.8$\pm$0.1  &  -10.5$\pm$0.1  &  14$\pm$3  &  0.01$\pm$0.00  &  3  &  0.5    \\   
  &  0.042$\pm$0.006  &  7.1$\pm$0.2  &  -1.3$\pm$0.1  &  6.2$\pm$0.2  &  7.9$\pm$0.1  &  -1.3$\pm$0.1  &  150$\pm$5  &  0.89$\pm$0.05  &  3  &      \\   
  &  0.023$\pm$0.005  &  0.8$\pm$0.1  &  -8.0$\pm$0.1  &  \nodata  &  \nodata  &  \nodata  &  \nodata  &  \nodata  &  \nodata  &    \\  
\hline   
3C48  &  0.004$\pm$0.001  &  0.7$\pm$0.1  &  -55.1$\pm$0.1  &  0.1$\pm$0.0  &  0.7$\pm$0.1  &  -52.8$\pm$0.1  &  25$\pm$3  &  0.00$\pm$0.00  &  6  &  0.5    \\   
  &  0.008$\pm$0.001  &  5.4$\pm$0.3  &  -15.6$\pm$0.1  &  3.1$\pm$0.1  &  6.0$\pm$0.1  &  -15.6$\pm$0.1  &  406$\pm$11  &  0.33$\pm$0.02  &  6  &      \\   
  &  0.016$\pm$0.001  &  2.8$\pm$0.1  &  -11.0$\pm$0.1  &  1.3$\pm$0.0  &  3.1$\pm$0.1  &  -11.6$\pm$0.1  &  79$\pm$2  &  0.07$\pm$0.00  &  6  &      \\   
  &  0.004$\pm$0.001  &  1.3$\pm$0.2  &  -5.0$\pm$0.1  &  0.6$\pm$0.1  &  1.4$\pm$0.1  &  -5.0$\pm$0.1  &  169$\pm$21  &  0.02$\pm$0.00  &  6  &      \\   
  &  0.020$\pm$0.001  &  9.1$\pm$0.1  &  -2.1$\pm$0.1  &  6.7$\pm$0.1  &  10.1$\pm$0.1  &  -2.1$\pm$0.1  &  346$\pm$3  &  1.23$\pm$0.02  &  6  &      \\   
  &  0.040$\pm$0.001  &  2.3$\pm$0.1  &  2.6$\pm$0.1  &  1.5$\pm$0.0  &  2.3$\pm$0.1  &  2.8$\pm$0.1  &  38$\pm$2  &  0.07$\pm$0.00  &  6  &      \\   
\hline   
4C15.05  &  0.005$\pm$0.001  &  1.7$\pm$0.2  &  -13.9$\pm$0.1  &  0.4$\pm$0.0  &  1.9$\pm$0.1  &  -13.6$\pm$0.1  &  82$\pm$24  &  0.01$\pm$0.00  &  6  &  0.0    \\   
  &  0.057$\pm$0.001  &  2.6$\pm$0.1  &  -10.5$\pm$0.1  &  2.4$\pm$0.0  &  2.8$\pm$0.1  &  -11.3$\pm$0.1  &  46$\pm$3  &  0.13$\pm$0.01  &  2  &      \\   
  &  0.037$\pm$0.001  &  9.9$\pm$0.1  &  -7.3$\pm$0.1  &  15.9$\pm$0.4  &  10.9$\pm$0.1  &  -7.3$\pm$0.1  &  438$\pm$12  &  3.18$\pm$0.13  &  1  &      \\   
  &  0.048$\pm$0.001  &  2.7$\pm$0.1  &  -5.4$\pm$0.1  &  2.6$\pm$0.0  &  2.7$\pm$0.1  &  -5.3$\pm$0.1  &  59$\pm$6  &  0.15$\pm$0.02  &  6  &      \\   
  &  0.014$\pm$0.001  &  2.6$\pm$0.1  &  0.3$\pm$0.1  &  2.1$\pm$0.1  &  2.8$\pm$0.1  &  0.3$\pm$0.1  &  145$\pm$5  &  0.10$\pm$0.01  &  6  &      \\   
  &  0.001$\pm$0.001  &  5.0$\pm$1.1  &  6.7$\pm$0.5  &  1.5$\pm$0.3  &  5.5$\pm$0.1  &  6.7$\pm$0.1  &  1551$\pm$298  &  0.14$\pm$0.05  &  6  &      \\   
\hline   
3C78  &  0.147$\pm$0.004  &  3.1$\pm$0.1  &  -7.8$\pm$0.1  &  8.7$\pm$0.1  &  3.1$\pm$0.1  &  -7.8$\pm$0.1  &  65$\pm$3  &  0.58$\pm$0.03  &  6  &  1.0    \\   
  &  0.029$\pm$0.004  &  14.9$\pm$0.6  &  -7.5$\pm$0.2  &  10.8$\pm$0.5  &  15.6$\pm$0.1  &  -7.5$\pm$0.1  &  385$\pm$18  &  3.24$\pm$0.25  &  6  &      \\   
  &  0.196$\pm$0.004  &  1.9$\pm$0.1  &  4.2$\pm$0.1  &  7.2$\pm$0.1  &  2.1$\pm$0.1  &  4.0$\pm$0.1  &  39$\pm$1  &  0.30$\pm$0.01  &  1  &      \\   
  &  1.369$\pm$0.006  &  2.2$\pm$0.1  &  6.9$\pm$0.1  &  24.1$\pm$0.1  &  2.4$\pm$0.1  &  7.0$\pm$0.1  &  39$\pm$7  &  2.37$\pm$0.43  &  2  &      \\   
  &  0.089$\pm$0.005  &  4.1$\pm$0.1  &  11.1$\pm$0.1  &  14.6$\pm$0.8  &  4.6$\pm$0.1  &  11.4$\pm$0.1  &  175$\pm$10  &  1.28$\pm$0.11  &  6  &      \\   
  &  0.054$\pm$0.006  &  1.3$\pm$0.1  &  10.4$\pm$0.1  &  \nodata  &  \nodata  &  \nodata  &  \nodata  &  \nodata  &  \nodata  &    \\  
\hline   
4C16.09  &  0.014$\pm$0.001  &  1.3$\pm$0.1  &  -9.7$\pm$0.1  &  0.3$\pm$0.0  &  1.6$\pm$0.1  &  -11.7$\pm$0.1  &  25$\pm$4  &  0.01$\pm$0.00  &  6  &  1.0    \\   
  &  0.436$\pm$0.002  &  4.2$\pm$0.1  &  -1.0$\pm$0.1  &  35.6$\pm$0.2  &  4.1$\pm$0.1  &  -1.0$\pm$0.1  &  106$\pm$3  &  3.82$\pm$0.13  &  6  &      \\   
  &  0.227$\pm$0.004  &  1.8$\pm$0.1  &  0.8$\pm$0.1  &  6.5$\pm$0.2  &  1.8$\pm$0.1  &  0.9$\pm$0.1  &  54$\pm$10  &  0.45$\pm$0.08  &  6  &      \\   
  &  0.086$\pm$0.001  &  5.4$\pm$0.1  &  5.9$\pm$0.1  &  21.4$\pm$0.2  &  5.9$\pm$0.1  &  5.7$\pm$0.1  &  260$\pm$3  &  2.42$\pm$0.05  &  6  &      \\   
  &  0.042$\pm$0.001  &  1.8$\pm$0.1  &  7.6$\pm$0.1  &  4.4$\pm$0.2  &  1.9$\pm$0.1  &  7.5$\pm$0.1  &  120$\pm$8  &  0.18$\pm$0.01  &  6  &      \\   
  &  0.013$\pm$0.001  &  4.1$\pm$0.2  &  -8.7$\pm$0.1  &  \nodata  &  \nodata  &  \nodata  &  \nodata  &  \nodata  &  \nodata  &    \\  
\hline   
3C111A  &  0.013$\pm$0.003  &  5.7$\pm$1.0  &  -28.9$\pm$0.4  &  6.3$\pm$0.5  &  6.2$\pm$0.1  &  -27.0$\pm$0.1  &  487$\pm$44  &  0.70$\pm$0.15  &  3  &  1.0    \\   
  &  0.041$\pm$0.003  &  2.6$\pm$0.1  &  -21.9$\pm$0.1  &  5.4$\pm$0.2  &  2.8$\pm$0.1  &  -20.5$\pm$0.1  &  131$\pm$6  &  0.27$\pm$0.02  &  1  &      \\   
  &  0.322$\pm$0.004  &  2.6$\pm$0.1  &  -16.8$\pm$0.1  &  8.5$\pm$0.1  &  2.9$\pm$0.1  &  -16.6$\pm$0.1  &  34$\pm$5  &  0.57$\pm$0.10  &  5  &      \\   
  &  0.172$\pm$0.005  &  8.6$\pm$0.3  &  -9.1$\pm$0.2  &  40.7$\pm$1.3  &  9.5$\pm$0.1  &  -9.0$\pm$0.1  &  248$\pm$10  &  7.33$\pm$0.47  &  12  &      \\   
  &  0.201$\pm$0.011  &  1.5$\pm$0.1  &  -4.7$\pm$0.1  &  12.2$\pm$0.7  &  1.6$\pm$0.1  &  -4.6$\pm$0.1  &  65$\pm$15  &  0.39$\pm$0.10  &  2  &      \\   
  &  0.790$\pm$0.009  &  3.9$\pm$0.1  &  -1.8$\pm$0.1  &  39.3$\pm$0.7  &  4.1$\pm$0.1  &  -1.6$\pm$0.1  &  77$\pm$4  &  4.66$\pm$0.32  &  4  &      \\   
  &  0.643$\pm$0.020  &  3.3$\pm$0.1  &  1.8$\pm$0.1  &  41.0$\pm$1.9  &  3.6$\pm$0.1  &  1.8$\pm$0.1  &  93$\pm$8  &  3.93$\pm$0.38  &  12  &      \\   
  &  0.232$\pm$0.004  &  5.2$\pm$0.1  &  6.1$\pm$0.1  &  26.8$\pm$0.4  &  5.7$\pm$0.1  &  6.0$\pm$0.1  &  129$\pm$1  &  3.08$\pm$0.10  &  12  &      \\   
  &  0.037$\pm$0.003  &  1.7$\pm$0.1  &  -32.4$\pm$0.1  &  \nodata  &  \nodata  &  \nodata  &  \nodata  &  \nodata  &  \nodata  &    \\  
  &  0.025$\pm$99.  &  0.1$\pm$99.  &  6.6$\pm$99.  &  \nodata  &  \nodata  &  \nodata  &  \nodata  &  \nodata  &  \nodata  &    \\  
  &  0.114$\pm$0.006  &  2.1$\pm$0.1  &  -8.3$\pm$0.1  &  \nodata  &  \nodata  &  \nodata  &  \nodata  &  \nodata  &  \nodata  &    \\  
  &  0.069$\pm$0.004  &  1.3$\pm$0.1  &  8.4$\pm$0.1  &  \nodata  &  \nodata  &  \nodata  &  \nodata  &  \nodata  &  \nodata  &    \\  
\hline   
3C111B  &  0.019$\pm$0.002  &  1.7$\pm$0.1  &  -54.8$\pm$0.1  &  0.5$\pm$0.0  &  1.7$\pm$0.1  &  -56.2$\pm$0.1  &  26$\pm$1  &  0.02$\pm$0.00  &  10  &  1.0    \\   
  &  0.111$\pm$0.003  &  2.7$\pm$0.1  &  -16.8$\pm$0.1  &  2.8$\pm$0.1  &  3.0$\pm$0.1  &  -16.7$\pm$0.1  &  30$\pm$5  &  0.18$\pm$0.03  &  1  &      \\   
  &  0.092$\pm$0.004  &  12.1$\pm$0.3  &  -10.8$\pm$0.3  &  35.9$\pm$1.0  &  12.7$\pm$0.1  &  -10.8$\pm$0.1  &  409$\pm$10  &  9.05$\pm$0.42  &  10  &      \\   
  &  0.400$\pm$0.005  &  2.8$\pm$0.1  &  -8.3$\pm$0.1  &  3.7$\pm$0.1  &  3.1$\pm$0.1  &  -8.3$\pm$0.1  &  14$\pm$8  &  0.33$\pm$0.20  &  10  &      \\   
  &  0.967$\pm$0.020  &  3.7$\pm$0.1  &  -1.7$\pm$0.1  &  29.2$\pm$1.1  &  3.9$\pm$0.1  &  -1.7$\pm$0.1  &  56$\pm$5  &  3.97$\pm$0.43  &  10  &      \\   
  &  0.347$\pm$0.054  &  2.7$\pm$0.1  &  2.1$\pm$0.1  &  11.1$\pm$3.1  &  2.7$\pm$0.1  &  2.1$\pm$0.1  &  58$\pm$13  &  1.09$\pm$0.31  &  2  &      \\   
  &  0.342$\pm$0.032  &  5.6$\pm$0.4  &  3.7$\pm$0.4  &  33.8$\pm$3.8  &  6.2$\pm$0.1  &  3.9$\pm$0.1  &  120$\pm$12  &  4.60$\pm$0.71  &  10  &      \\   
  &  0.181$\pm$0.006  &  0.8$\pm$0.1  &  -2.3$\pm$0.1  &  \nodata  &  \nodata  &  \nodata  &  \nodata  &  \nodata  &  \nodata  &    \\  
  &  0.015$\pm$0.002  &  2.2$\pm$0.2  &  -51.3$\pm$0.1  &  \nodata  &  \nodata  &  \nodata  &  \nodata  &  \nodata  &  \nodata  &    \\  
  &  0.212$\pm$0.005  &  2.1$\pm$0.1  &  8.0$\pm$0.1  &  \nodata  &  \nodata  &  \nodata  &  \nodata  &  \nodata  &  \nodata  &    \\  
\hline   
3C111C  &  0.027$\pm$0.004  &  2.4$\pm$0.2  &  -31.4$\pm$0.1  &  0.9$\pm$0.1  &  2.6$\pm$0.1  &  -31.4$\pm$0.1  &  35$\pm$3  &  0.05$\pm$0.01  &  11  &  1.0    \\   
  &  0.029$\pm$0.004  &  2.6$\pm$0.2  &  -27.8$\pm$0.1  &  1.1$\pm$0.1  &  2.9$\pm$0.1  &  -26.0$\pm$0.1  &  39$\pm$2  &  0.06$\pm$0.01  &  11  &      \\   
  &  0.282$\pm$0.045  &  2.1$\pm$0.1  &  -16.9$\pm$0.1  &  3.9$\pm$0.9  &  2.1$\pm$0.1  &  -17.8$\pm$0.1  &  20$\pm$4  &  0.24$\pm$0.07  &  11  &      \\   
  &  0.088$\pm$0.027  &  2.5$\pm$0.6  &  -15.3$\pm$0.5  &  8.4$\pm$2.4  &  2.8$\pm$0.1  &  -14.1$\pm$0.1  &  90$\pm$34  &  0.40$\pm$0.22  &  1  &      \\   
  &  0.182$\pm$0.011  &  11.2$\pm$0.6  &  -8.1$\pm$0.6  &  17.8$\pm$1.3  &  12.4$\pm$0.1  &  -8.1$\pm$0.1  &  118$\pm$28  &  4.79$\pm$1.20  &  11  &      \\   
  &  0.816$\pm$0.050  &  2.3$\pm$0.1  &  -2.1$\pm$0.1  &  3.5$\pm$1.0  &  2.3$\pm$0.1  &  -2.4$\pm$0.1  &  20$\pm$11  &  0.76$\pm$0.44  &  11  &      \\   
  &  0.456$\pm$0.044  &  7.3$\pm$0.3  &  1.6$\pm$0.4  &  34.6$\pm$4.7  &  8.1$\pm$0.1  &  1.6$\pm$0.1  &  108$\pm$18  &  7.15$\pm$1.41  &  11  &      \\   
  &  0.584$\pm$0.043  &  2.5$\pm$0.1  &  1.1$\pm$0.1  &  \nodata  &  \nodata  &  \nodata  &  \nodata  &  \nodata  &  \nodata  &    \\  
  &  0.416$\pm$0.008  &  1.6$\pm$0.1  &  -8.4$\pm$0.1  &  \nodata  &  \nodata  &  \nodata  &  \nodata  &  \nodata  &  \nodata  &    \\  
  &  0.211$\pm$0.014  &  1.5$\pm$0.1  &  -5.5$\pm$0.1  &  \nodata  &  \nodata  &  \nodata  &  \nodata  &  \nodata  &  \nodata  &    \\  
  &  0.058$\pm$0.007  &  1.8$\pm$0.2  &  6.3$\pm$0.1  &  \nodata  &  \nodata  &  \nodata  &  \nodata  &  \nodata  &  \nodata  &    \\  
\hline   
3C120  &  0.617$\pm$0.005  &  1.3$\pm$0.1  &  6.1$\pm$0.1  &  7.7$\pm$0.1  &  1.3$\pm$0.1  &  5.5$\pm$0.1  &  20$\pm$9  &  0.32$\pm$0.15  &  2  &  1.0    \\   
  &  1.681$\pm$0.004  &  4.4$\pm$0.1  &  7.3$\pm$0.1  &  36.6$\pm$0.2  &  4.9$\pm$0.1  &  6.6$\pm$0.1  &  44$\pm$4  &  6.49$\pm$0.60  &  1  &      \\   
  &  0.776$\pm$0.012  &  2.3$\pm$0.1  &  9.9$\pm$0.1  &  12.7$\pm$0.3  &  2.5$\pm$0.1  &  10.6$\pm$0.1  &  23$\pm$5  &  0.83$\pm$0.18  &  3  &      \\   
\hline   
3C123A  &  0.064$\pm$0.001  &  3.1$\pm$0.1  &  -19.9$\pm$0.1  &  1.5$\pm$0.0  &  3.8$\pm$0.1  &  -21.1$\pm$0.1  &  26$\pm$2  &  0.11$\pm$0.01  &  7  &  1.0    \\   
  &  0.379$\pm$0.035  &  2.2$\pm$0.1  &  1.6$\pm$0.1  &  3.2$\pm$0.7  &  2.2$\pm$0.1  &  1.9$\pm$0.1  &  19$\pm$14  &  0.32$\pm$0.23  &  7  &      \\   
  &  0.044$\pm$0.004  &  10.1$\pm$0.3  &  3.7$\pm$0.1  &  27.0$\pm$2.8  &  12.2$\pm$0.1  &  3.9$\pm$0.1  &  619$\pm$79  &  5.47$\pm$0.91  &  7  &      \\   
  &  0.810$\pm$0.071  &  4.4$\pm$0.1  &  5.3$\pm$0.2  &  23.1$\pm$4.6  &  5.3$\pm$0.1  &  5.2$\pm$0.1  &  67$\pm$25  &  4.77$\pm$1.86  &  7  &      \\   
  &  0.628$\pm$0.037  &  1.8$\pm$0.1  &  5.5$\pm$0.1  &  4.3$\pm$0.7  &  1.7$\pm$0.1  &  6.3$\pm$0.1  &  18$\pm$11  &  0.41$\pm$0.25  &  7  &      \\   
  &  0.008$\pm$0.001  &  4.0$\pm$0.3  &  20.2$\pm$0.1  &  1.0$\pm$0.1  &  4.0$\pm$0.1  &  17.9$\pm$0.1  &  124$\pm$9  &  0.08$\pm$0.01  &  7  &      \\   
  &  1.020$\pm$0.096  &  2.3$\pm$0.1  &  3.7$\pm$0.1  &  \nodata  &  \nodata  &  \nodata  &  \nodata  &  \nodata  &  \nodata  &    \\  
\hline   
3C123B  &  0.064$\pm$0.001  &  3.2$\pm$0.1  &  -19.3$\pm$0.1  &  0.9$\pm$0.0  &  3.6$\pm$0.1  &  -20.4$\pm$0.1  &  16$\pm$2  &  0.07$\pm$0.01  &  6  &  1.0    \\   
  &  1.648$\pm$0.005  &  4.3$\pm$0.1  &  4.4$\pm$0.1  &  17.6$\pm$0.2  &  4.8$\pm$0.1  &  4.4$\pm$0.1  &  36$\pm$14  &  5.19$\pm$1.98  &  3  &      \\   
  &  0.042$\pm$0.003  &  10.9$\pm$0.3  &  4.9$\pm$0.1  &  27.8$\pm$2.2  &  10.9$\pm$0.1  &  4.5$\pm$0.1  &  664$\pm$52  &  6.11$\pm$0.71  &  2  &      \\   
  &  0.253$\pm$0.007  &  1.5$\pm$0.1  &  5.5$\pm$0.1  &  5.4$\pm$0.6  &  1.6$\pm$0.1  &  6.0$\pm$0.1  &  81$\pm$43  &  0.60$\pm$0.32  &  6  &      \\   
  &  0.004$\pm$0.001  &  3.2$\pm$0.5  &  20.1$\pm$0.2  &  1.4$\pm$0.2  &  3.5$\pm$0.1  &  18.3$\pm$0.1  &  334$\pm$52  &  0.09$\pm$0.02  &  1  &      \\   
  &  0.064$\pm$0.003  &  1.7$\pm$0.1  &  8.5$\pm$0.1  &  \nodata  &  \nodata  &  \nodata  &  \nodata  &  \nodata  &  \nodata  &    \\  
\hline   
3C132  &  0.197$\pm$0.009  &  7.2$\pm$0.1  &  1.7$\pm$0.1  &  27.8$\pm$1.5  &  7.2$\pm$0.1  &  1.9$\pm$0.1  &  165$\pm$12  &  4.66$\pm$0.41  &  7  &  1.0    \\   
  &  1.472$\pm$0.005  &  2.0$\pm$0.1  &  7.8$\pm$0.9  &  7.1$\pm$0.1  &  2.0$\pm$0.1  &  7.8$\pm$0.1  &  27$\pm$19  &  1.59$\pm$1.12  &  4  &      \\   
  &  0.230$\pm$0.003  &  6.7$\pm$0.1  &  12.1$\pm$0.1  &  16.2$\pm$0.1  &  7.3$\pm$0.1  &  11.4$\pm$0.1  &  80$\pm$4  &  2.44$\pm$0.15  &  3  &      \\   
  &  0.129$\pm$0.022  &  1.8$\pm$0.1  &  1.7$\pm$2000.0  &  \nodata  &  \nodata  &  \nodata  &  \nodata  &  \nodata  &  \nodata  &    \\  
  &  0.273$\pm$0.010  &  1.5$\pm$0.1  &  9.4$\pm$0.1  &  \nodata  &  \nodata  &  \nodata  &  \nodata  &  \nodata  &  \nodata  &    \\  
  &  0.059$\pm$0.004  &  2.4$\pm$0.1  &  14.8$\pm$0.1  &  \nodata  &  \nodata  &  \nodata  &  \nodata  &  \nodata  &  \nodata  &    \\  
  &  0.042$\pm$0.017  &  3.4$\pm$0.1  &  2.2$\pm$0.1  &  \nodata  &  \nodata  &  \nodata  &  \nodata  &  \nodata  &  \nodata  &    \\  
\hline   
3C133  &  0.033$\pm$0.005  &  9.5$\pm$0.3  &  -29.8$\pm$0.1  &  17.2$\pm$0.9  &  10.4$\pm$0.1  &  -28.8$\pm$0.1  &  537$\pm$28  &  3.33$\pm$0.26  &  8  &  1.0    \\   
  &  0.134$\pm$0.007  &  2.4$\pm$0.1  &  -1.5$\pm$0.1  &  2.1$\pm$0.1  &  2.7$\pm$0.1  &  -1.6$\pm$0.1  &  26$\pm$16  &  0.17$\pm$0.11  &  2  &      \\   
  &  0.240$\pm$0.011  &  9.6$\pm$0.1  &  3.1$\pm$0.1  &  59.2$\pm$3.0  &  9.6$\pm$0.1  &  2.8$\pm$0.1  &  283$\pm$13  &  12.94$\pm$0.87  &  1  &      \\   
  &  0.706$\pm$0.010  &  2.7$\pm$0.1  &  3.7$\pm$0.1  &  2.7$\pm$0.1  &  3.0$\pm$0.1  &  3.4$\pm$0.1  &  11$\pm$15  &  0.45$\pm$0.61  &  8  &      \\   
  &  0.845$\pm$0.029  &  1.5$\pm$0.1  &  7.6$\pm$0.1  &  1.6$\pm$0.7  &  1.5$\pm$0.1  &  7.6$\pm$0.1  &  23$\pm$19  &  0.59$\pm$0.48  &  8  &      \\   
  &  0.780$\pm$0.022  &  2.6$\pm$0.1  &  8.4$\pm$0.1  &  18.0$\pm$1.0  &  2.9$\pm$0.1  &  8.7$\pm$0.1  &  46$\pm$9  &  1.86$\pm$0.37  &  8  &      \\   
  &  0.084$\pm$0.005  &  3.2$\pm$0.1  &  -27.8$\pm$0.1  &  \nodata  &  \nodata  &  \nodata  &  \nodata  &  \nodata  &  \nodata  &    \\  
  &  0.077$\pm$0.005  &  1.6$\pm$0.1  &  -4.2$\pm$0.1  &  \nodata  &  \nodata  &  \nodata  &  \nodata  &  \nodata  &  \nodata  &    \\  
\hline   
3C138  &  0.033$\pm$0.001  &  2.3$\pm$0.1  &  -22.0$\pm$0.1  &  1.9$\pm$0.1  &  2.5$\pm$0.1  &  -21.2$\pm$0.1  &  60$\pm$2  &  0.09$\pm$0.01  &  8  &  1.0    \\   
  &  0.026$\pm$0.002  &  4.3$\pm$0.2  &  -6.7$\pm$0.1  &  7.7$\pm$0.3  &  4.5$\pm$0.1  &  -6.7$\pm$0.1  &  297$\pm$11  &  0.66$\pm$0.05  &  8  &      \\   
  &  0.322$\pm$0.003  &  4.3$\pm$0.1  &  0.0$\pm$0.1  &  33.4$\pm$0.2  &  4.8$\pm$0.1  &  0.0$\pm$0.1  &  125$\pm$3  &  3.46$\pm$0.09  &  8  &      \\   
  &  0.079$\pm$0.003  &  1.2$\pm$0.1  &  1.6$\pm$0.1  &  2.0$\pm$0.2  &  1.3$\pm$0.1  &  1.6$\pm$0.1  &  47$\pm$19  &  0.09$\pm$0.04  &  8  &      \\   
  &  0.003$\pm$0.003  &  37.4$\pm$4.9  &  2.9$\pm$1.5  &  4.8$\pm$1.1  &  33.7$\pm$0.1  &  2.6$\pm$0.1  &  1427$\pm$302  &  3.59$\pm$1.23  &  8  &      \\   
  &  1.057$\pm$0.005  &  2.4$\pm$0.1  &  6.3$\pm$0.1  &  20.6$\pm$0.2  &  2.6$\pm$0.1  &  5.9$\pm$0.1  &  39$\pm$7  &  1.96$\pm$0.38  &  8  &      \\   
  &  0.432$\pm$0.003  &  3.0$\pm$0.1  &  9.0$\pm$0.1  &  23.5$\pm$0.2  &  3.3$\pm$0.1  &  9.5$\pm$0.1  &  66$\pm$1  &  1.72$\pm$0.04  &  8  &      \\   
  &  0.011$\pm$0.002  &  2.7$\pm$0.5  &  13.3$\pm$0.2  &  4.2$\pm$0.5  &  3.0$\pm$0.1  &  12.0$\pm$0.1  &  422$\pm$55  &  0.25$\pm$0.06  &  8  &      \\   
\hline   
PKS0531  &  0.001$\pm$0.001  &  12.1$\pm$1.5  &  -24.3$\pm$0.5  &  2.9$\pm$0.5  &  13.3$\pm$0.1  &  -24.3$\pm$0.1  &  2049$\pm$400  &  0.70$\pm$0.20  &  10  &  1.0    \\   
  &  0.007$\pm$0.001  &  3.5$\pm$0.2  &  -23.0$\pm$0.1  &  0.1$\pm$0.0  &  3.8$\pm$0.1  &  -20.7$\pm$0.1  &  24$\pm$4  &  0.01$\pm$0.00  &  10  &      \\   
  &  0.006$\pm$0.001  &  6.2$\pm$0.3  &  -9.0$\pm$0.1  &  0.8$\pm$0.0  &  6.8$\pm$0.1  &  -9.0$\pm$0.1  &  240$\pm$130  &  0.17$\pm$0.09  &  1  &      \\   
  &  0.403$\pm$0.009  &  1.9$\pm$0.1  &  1.6$\pm$0.1  &  3.7$\pm$0.3  &  1.9$\pm$0.1  &  1.5$\pm$0.1  &  28$\pm$26  &  0.45$\pm$0.40  &  3  &      \\   
  &  0.141$\pm$0.005  &  9.7$\pm$0.1  &  4.6$\pm$0.1  &  69.4$\pm$2.7  &  10.3$\pm$0.1  &  4.3$\pm$0.1  &  542$\pm$25  &  14.79$\pm$0.89  &  2  &      \\   
  &  0.190$\pm$0.003  &  2.4$\pm$0.1  &  9.5$\pm$0.1  &  14.9$\pm$0.3  &  2.6$\pm$0.1  &  10.0$\pm$0.1  &  100$\pm$10  &  0.90$\pm$0.09  &  10  &      \\   
  &  0.019$\pm$0.001  &  2.3$\pm$0.2  &  12.1$\pm$0.1  &  3.0$\pm$0.2  &  2.5$\pm$0.1  &  12.1$\pm$0.1  &  132$\pm$31  &  0.11$\pm$0.03  &  10  &      \\   
  &  0.005$\pm$0.001  &  4.1$\pm$0.2  &  21.0$\pm$0.1  &  2.0$\pm$0.1  &  4.5$\pm$0.1  &  20.9$\pm$0.1  &  380$\pm$14  &  0.17$\pm$0.01  &  10  &      \\   
  &  0.075$\pm$0.005  &  2.7$\pm$0.1  &  6.0$\pm$0.1  &  \nodata  &  \nodata  &  \nodata  &  \nodata  &  \nodata  &  \nodata  &    \\  
  &  0.088$\pm$0.007  &  1.9$\pm$0.1  &  3.2$\pm$0.1  &  \nodata  &  \nodata  &  \nodata  &  \nodata  &  \nodata  &  \nodata  &    \\  
\hline   
3C147  &  0.013$\pm$0.001  &  2.0$\pm$0.1  &  -19.4$\pm$0.1  &  0.8$\pm$0.0  &  2.2$\pm$0.1  &  -19.4$\pm$0.1  &  65$\pm$5  &  0.03$\pm$0.00  &  9  &  1.0    \\   
  &  0.054$\pm$0.009  &  1.9$\pm$0.1  &  -13.7$\pm$0.1  &  2.0$\pm$0.3  &  2.0$\pm$0.1  &  -15.0$\pm$0.1  &  36$\pm$8  &  0.07$\pm$0.02  &  9  &      \\   
  &  0.120$\pm$0.021  &  6.4$\pm$0.3  &  -11.1$\pm$0.1  &  23.9$\pm$4.5  &  6.6$\pm$0.1  &  -11.2$\pm$0.1  &  218$\pm$37  &  3.32$\pm$0.82  &  9  &      \\   
  &  0.165$\pm$0.020  &  2.3$\pm$0.1  &  -10.7$\pm$0.1  &  5.0$\pm$0.7  &  2.1$\pm$0.1  &  -9.9$\pm$0.1  &  37$\pm$6  &  0.28$\pm$0.06  &  9  &      \\   
  &  0.278$\pm$0.004  &  5.9$\pm$0.1  &  0.5$\pm$0.1  &  23.6$\pm$0.4  &  6.5$\pm$0.1  &  0.6$\pm$0.1  &  109$\pm$9  &  3.56$\pm$0.33  &  9  &      \\   
  &  0.035$\pm$0.001  &  1.6$\pm$0.1  &  5.1$\pm$0.1  &  0.9$\pm$0.0  &  1.8$\pm$0.1  &  5.6$\pm$0.1  &  25$\pm$3  &  0.03$\pm$0.00  &  9  &      \\   
  &  0.725$\pm$0.006  &  1.7$\pm$0.1  &  -8.0$\pm$0.1  &  \nodata  &  \nodata  &  \nodata  &  \nodata  &  \nodata  &  \nodata  &    \\  
  &  0.208$\pm$0.004  &  2.0$\pm$0.1  &  -0.1$\pm$0.1  &  \nodata  &  \nodata  &  \nodata  &  \nodata  &  \nodata  &  \nodata  &    \\  
  &  0.042$\pm$0.003  &  1.5$\pm$0.1  &  1.8$\pm$0.1  &  \nodata  &  \nodata  &  \nodata  &  \nodata  &  \nodata  &  \nodata  &    \\  
\hline   
3C154  &  0.061$\pm$0.005  &  3.1$\pm$0.2  &  -23.2$\pm$0.1  &  1.6$\pm$0.1  &  3.4$\pm$0.1  &  -23.1$\pm$0.1  &  25$\pm$3  &  0.10$\pm$0.02  &  5  &  1.0    \\   
  &  0.012$\pm$0.005  &  7.4$\pm$1.4  &  -22.9$\pm$0.3  &  7.2$\pm$3.0  &  7.4$\pm$0.1  &  -22.4$\pm$0.1  &  587$\pm$240  &  1.08$\pm$0.65  &  6  &      \\   
  &  0.123$\pm$0.020  &  14.0$\pm$0.5  &  0.6$\pm$0.2  &  87.5$\pm$14.8  &  15.0$\pm$0.1  &  0.7$\pm$0.1  &  762$\pm$130  &  26.09$\pm$6.15  &  2  &      \\   
  &  1.300$\pm$0.015  &  2.3$\pm$0.1  &  2.0$\pm$0.1  &  4.2$\pm$0.2  &  2.2$\pm$0.1  &  2.0$\pm$0.1  &  10$\pm$15  &  0.63$\pm$0.91  &  4  &      \\   
  &  0.733$\pm$0.016  &  3.9$\pm$0.1  &  4.8$\pm$0.1  &  7.0$\pm$0.3  &  3.9$\pm$0.1  &  4.8$\pm$0.1  &  19$\pm$17  &  1.13$\pm$0.99  &  1  &      \\   
  &  0.521$\pm$0.004  &  2.2$\pm$0.1  &  10.8$\pm$0.1  &  7.0$\pm$0.1  &  2.5$\pm$0.1  &  11.8$\pm$0.1  &  19$\pm$3  &  0.45$\pm$0.08  &  3  &      \\   
  &  0.297$\pm$0.008  &  1.1$\pm$0.1  &  -2.2$\pm$0.1  &  \nodata  &  \nodata  &  \nodata  &  \nodata  &  \nodata  &  \nodata  &    \\  
  &  0.975$\pm$0.015  &  4.6$\pm$0.1  &  -2.6$\pm$0.1  &  \nodata  &  \nodata  &  \nodata  &  \nodata  &  \nodata  &  \nodata  &    \\  
\hline   
PKS0742  &  0.009$\pm$0.001  &  3.2$\pm$0.1  &  1.5$\pm$0.1  &  1.5$\pm$0.1  &  3.5$\pm$0.2  &  1.6$\pm$0.1  &  164$\pm$9  &  0.10$\pm$0.01  &  1  &  0.0    \\   
\hline   
3C225A  &  0.043$\pm$0.002  &  1.8$\pm$0.1  &  -40.2$\pm$0.1  &  0.9$\pm$0.0  &  1.9$\pm$0.1  &  -39.6$\pm$0.1  &  22$\pm$0  &  0.03$\pm$0.00  &  5  &  1.0    \\   
  &  0.020$\pm$0.002  &  4.8$\pm$0.3  &  -37.4$\pm$0.2  &  1.2$\pm$0.0  &  4.7$\pm$0.1  &  -38.3$\pm$0.1  &  60$\pm$1  &  0.11$\pm$0.01  &  5  &      \\   
  &  0.048$\pm$0.002  &  2.5$\pm$0.1  &  -27.2$\pm$0.1  &  1.0$\pm$0.0  &  2.8$\pm$0.1  &  -27.7$\pm$0.1  &  22$\pm$0  &  0.05$\pm$0.00  &  5  &      \\   
  &  0.013$\pm$0.002  &  7.7$\pm$0.3  &  -5.2$\pm$0.1  &  4.3$\pm$0.1  &  8.1$\pm$0.1  &  -4.9$\pm$0.1  &  327$\pm$11  &  0.66$\pm$0.04  &  5  &      \\   
  &  0.805$\pm$0.002  &  1.3$\pm$0.1  &  4.0$\pm$0.1  &  5.7$\pm$0.0  &  1.4$\pm$0.1  &  3.6$\pm$0.1  &  11$\pm$2  &  0.23$\pm$0.04  &  5  &      \\   
\hline   
3C225B  &  0.044$\pm$0.003  &  2.0$\pm$0.1  &  -40.3$\pm$0.1  &  0.8$\pm$0.0  &  1.9$\pm$0.1  &  -39.8$\pm$0.1  &  18$\pm$0  &  0.03$\pm$0.00  &  5  &  1.0    \\   
  &  0.023$\pm$0.003  &  4.0$\pm$0.3  &  -37.2$\pm$0.1  &  3.3$\pm$0.1  &  4.0$\pm$0.1  &  -37.4$\pm$0.1  &  145$\pm$4  &  0.27$\pm$0.02  &  5  &      \\   
  &  0.053$\pm$0.003  &  2.4$\pm$0.1  &  -27.3$\pm$0.1  &  1.0$\pm$0.0  &  2.9$\pm$0.1  &  -27.8$\pm$0.1  &  20$\pm$0  &  0.05$\pm$0.00  &  5  &      \\   
  &  0.013$\pm$0.003  &  8.3$\pm$0.4  &  -5.6$\pm$0.2  &  6.1$\pm$0.2  &  8.3$\pm$0.1  &  -4.5$\pm$0.1  &  458$\pm$17  &  1.02$\pm$0.07  &  5  &      \\   
  &  0.774$\pm$0.003  &  1.3$\pm$0.1  &  4.0$\pm$0.1  &  7.7$\pm$0.0  &  1.5$\pm$0.1  &  3.5$\pm$0.1  &  14$\pm$0  &  0.29$\pm$0.01  &  5  &      \\   
\hline   
3C236  &  \nodata  &  \nodata  &  \nodata  &  \nodata  &  \nodata  &  \nodata  &  \nodata  &  \nodata  &  \nodata  &      \\   
\hline   
3C237  &  0.006$\pm$0.001  &  15.0$\pm$0.6  &  -4.0$\pm$0.3  &  2.1$\pm$0.1  &  14.8$\pm$0.1  &  -4.0$\pm$0.1  &  382$\pm$17  &  0.64$\pm$0.05  &  3  &  1.0    \\   
  &  0.415$\pm$0.001  &  1.2$\pm$0.1  &  2.3$\pm$0.1  &  4.3$\pm$0.0  &  1.2$\pm$0.1  &  2.1$\pm$0.1  &  13$\pm$0  &  0.13$\pm$0.00  &  3  &      \\   
  &  0.005$\pm$0.001  &  1.9$\pm$0.2  &  -2.4$\pm$0.1  &  \nodata  &  \nodata  &  \nodata  &  \nodata  &  \nodata  &  \nodata  &    \\  
\hline   
3C245A  &  0.010$\pm$0.002  &  5.3$\pm$0.3  &  -9.1$\pm$0.1  &  3.9$\pm$0.3  &  5.3$\pm$0.1  &  -8.8$\pm$0.1  &  385$\pm$30  &  0.41$\pm$0.05  &  2  &  0.5    \\   
  &  0.006$\pm$0.002  &  1.6$\pm$0.3  &  -9.8$\pm$0.1  &  \nodata  &  \nodata  &  \nodata  &  \nodata  &  \nodata  &  \nodata  &    \\  
\hline   
3C245B  &  \nodata  &  \nodata  &  \nodata  &  \nodata  &  \nodata  &  \nodata  &  \nodata  &  \nodata  &  \nodata  &      \\   
\hline   
1055+018  &  0.006$\pm$0.001  &  7.1$\pm$0.3  &  -7.3$\pm$0.1  &  6.0$\pm$0.2  &  7.0$\pm$0.1  &  -6.8$\pm$0.1  &  941$\pm$30  &  0.85$\pm$0.05  &  1  &  1.0    \\   
\hline   
3C263.1  &  0.007$\pm$0.001  &  1.3$\pm$0.1  &  -68.9$\pm$0.1  &  0.1$\pm$0.0  &  1.4$\pm$0.1  &  -65.4$\pm$0.1  &  15$\pm$1  &  0.00$\pm$0.00  &  3  &  1.0    \\   
  &  0.020$\pm$0.001  &  2.0$\pm$0.1  &  -52.8$\pm$0.1  &  0.7$\pm$0.0  &  2.1$\pm$0.1  &  -54.3$\pm$0.1  &  35$\pm$0  &  0.03$\pm$0.00  &  3  &      \\   
  &  0.007$\pm$0.001  &  0.6$\pm$0.1  &  15.6$\pm$0.1  &  0.3$\pm$0.0  &  0.6$\pm$0.1  &  16.9$\pm$0.1  &  42$\pm$5  &  0.00$\pm$0.00  &  3  &      \\   
\hline   
3C273  &  0.019$\pm$0.001  &  2.3$\pm$0.1  &  -6.3$\pm$0.1  &  0.3$\pm$0.0  &  2.5$\pm$0.1  &  -5.7$\pm$0.1  &  17$\pm$0  &  0.02$\pm$0.00  &  3  &  1.0    \\   
  &  0.005$\pm$0.001  &  6.4$\pm$0.3  &  -5.8$\pm$0.1  &  2.4$\pm$0.2  &  7.0$\pm$0.1  &  -5.7$\pm$0.1  &  455$\pm$47  &  0.30$\pm$0.04  &  3  &      \\   
  &  0.003$\pm$0.001  &  2.0$\pm$0.2  &  31.6$\pm$0.1  &  \nodata  &  \nodata  &  \nodata  &  \nodata  &  \nodata  &  \nodata  &    \\  
\hline   
4C32.44  &  0.018$\pm$0.001  &  2.8$\pm$0.1  &  -16.2$\pm$0.1  &  2.0$\pm$0.0  &  3.1$\pm$0.1  &  -16.5$\pm$0.1  &  112$\pm$2  &  0.12$\pm$0.00  &  2  &  1.0    \\   
  &  0.004$\pm$0.001  &  3.7$\pm$0.3  &  -4.2$\pm$0.1  &  0.9$\pm$0.1  &  3.9$\pm$0.1  &  -4.6$\pm$0.1  &  255$\pm$17  &  0.07$\pm$0.01  &  2  &      \\   
\hline   
4C25.43  &  \nodata  &  \nodata  &  \nodata  &  \nodata  &  \nodata  &  \nodata  &  \nodata  &  \nodata  &  \nodata  &      \\   
\hline   
3C286  &  0.006$\pm$0.001  &  2.4$\pm$0.1  &  -28.5$\pm$0.1  &  0.5$\pm$0.0  &  2.6$\pm$0.1  &  -28.8$\pm$0.1  &  76$\pm$2  &  0.02$\pm$0.00  &  3  &  0.0    \\   
  &  0.005$\pm$0.001  &  3.2$\pm$0.2  &  -14.2$\pm$0.1  &  0.3$\pm$0.0  &  3.5$\pm$0.1  &  -13.6$\pm$0.1  &  60$\pm$2  &  0.02$\pm$0.00  &  3  &      \\   
  &  0.007$\pm$0.001  &  4.3$\pm$0.1  &  -7.3$\pm$0.1  &  0.6$\pm$0.0  &  4.8$\pm$0.1  &  -6.6$\pm$0.1  &  78$\pm$2  &  0.05$\pm$0.00  &  3  &      \\   
\hline   
4C12.50  &  0.016$\pm$0.001  &  6.6$\pm$0.1  &  -2.5$\pm$0.1  &  5.1$\pm$0.2  &  6.5$\pm$0.1  &  -2.5$\pm$0.1  &  318$\pm$11  &  0.67$\pm$0.04  &  2  &  1.0    \\   
  &  0.077$\pm$0.002  &  2.2$\pm$0.1  &  -1.0$\pm$0.1  &  7.2$\pm$0.1  &  2.4$\pm$0.1  &  -1.0$\pm$0.1  &  97$\pm$1  &  0.33$\pm$0.01  &  1  &      \\   
\hline   
3C298  &  0.019$\pm$0.001  &  3.6$\pm$0.1  &  -1.1$\pm$0.1  &  2.1$\pm$0.2  &  4.3$\pm$0.4  &  -1.1$\pm$0.2  &  112$\pm$10  &  0.15$\pm$0.01  &  1  &  0.5    \\   
\hline   
UGC09799  &  0.058$\pm$0.010  &  2.7$\pm$0.1  &  -4.2$\pm$0.1  &  3.1$\pm$0.2  &  2.9$\pm$0.1  &  -4.2$\pm$0.1  &  58$\pm$4  &  0.18$\pm$0.02  &  1  &  1.0    \\   
\hline   
4C04.51  &  0.002$\pm$0.001  &  14.0$\pm$1.5  &  -9.2$\pm$0.8  &  0.3$\pm$0.0  &  15.5$\pm$0.1  &  -9.1$\pm$0.1  &  114$\pm$8  &  0.07$\pm$0.01  &  3  &  0.0    \\   
  &  0.025$\pm$0.001  &  2.4$\pm$0.1  &  -4.0$\pm$0.1  &  5.2$\pm$0.1  &  2.4$\pm$0.1  &  -4.0$\pm$0.1  &  207$\pm$5  &  0.24$\pm$0.01  &  3  &      \\   
  &  0.067$\pm$0.001  &  3.4$\pm$0.1  &  -0.6$\pm$0.1  &  12.1$\pm$0.1  &  3.8$\pm$0.1  &  -0.6$\pm$0.1  &  183$\pm$2  &  0.84$\pm$0.02  &  3  &      \\   
\hline   
3C327.1A  &  0.126$\pm$0.006  &  3.4$\pm$0.1  &  -2.7$\pm$0.1  &  16.3$\pm$0.3  &  3.7$\pm$0.1  &  -2.9$\pm$0.1  &  140$\pm$3  &  1.18$\pm$0.06  &  1  &  1.0    \\   
  &  0.425$\pm$0.010  &  1.9$\pm$0.1  &  -0.0$\pm$0.1  &  19.8$\pm$0.6  &  2.0$\pm$0.1  &  -0.0$\pm$0.1  &  63$\pm$4  &  1.02$\pm$0.08  &  2  &      \\   
  &  0.401$\pm$0.008  &  2.2$\pm$0.1  &  2.0$\pm$0.1  &  22.2$\pm$0.4  &  2.4$\pm$0.1  &  2.0$\pm$0.1  &  69$\pm$3  &  1.21$\pm$0.06  &  3  &      \\   
\hline   
3C327.1B  &  0.118$\pm$0.006  &  3.2$\pm$0.1  &  -2.7$\pm$0.1  &  15.6$\pm$0.3  &  3.6$\pm$0.1  &  -3.0$\pm$0.1  &  142$\pm$3  &  1.07$\pm$0.06  &  4  &  1.0    \\   
  &  0.359$\pm$0.011  &  1.9$\pm$0.1  &  -0.0$\pm$0.1  &  20.3$\pm$0.8  &  2.1$\pm$0.1  &  -0.0$\pm$0.1  &  74$\pm$4  &  1.03$\pm$0.08  &  4  &      \\   
  &  0.419$\pm$0.008  &  2.2$\pm$0.1  &  1.9$\pm$0.1  &  22.2$\pm$0.6  &  2.4$\pm$0.1  &  1.9$\pm$0.1  &  68$\pm$3  &  1.23$\pm$0.07  &  4  &      \\   
  &  0.013$\pm$0.003  &  0.7$\pm$0.1  &  46.1$\pm$0.1  &  \nodata  &  \nodata  &  \nodata  &  \nodata  &  \nodata  &  \nodata  &    \\  
\hline   
PKS1607  &  0.128$\pm$0.001  &  2.0$\pm$0.1  &  -10.3$\pm$0.1  &  3.0$\pm$0.0  &  2.0$\pm$0.1  &  -10.7$\pm$0.1  &  26$\pm$1  &  0.13$\pm$0.01  &  1  &  1.0    \\   
  &  0.013$\pm$0.001  &  6.5$\pm$0.3  &  -7.8$\pm$0.2  &  1.4$\pm$0.0  &  7.2$\pm$0.1  &  -7.8$\pm$0.1  &  114$\pm$10  &  0.20$\pm$0.02  &  2  &      \\   
  &  0.064$\pm$0.004  &  3.9$\pm$0.1  &  -1.2$\pm$0.1  &  7.0$\pm$0.4  &  4.3$\pm$0.1  &  -1.1$\pm$0.1  &  115$\pm$10  &  0.58$\pm$0.06  &  4  &      \\   
  &  0.123$\pm$0.004  &  2.2$\pm$0.1  &  -2.3$\pm$0.1  &  \nodata  &  \nodata  &  \nodata  &  \nodata  &  \nodata  &  \nodata  &    \\  
\hline   
J1613  &  \nodata  &  \nodata  &  \nodata  &  \nodata  &  \nodata  &  \nodata  &  \nodata  &  \nodata  &  \nodata  &      \\   
\hline   
3C345  &  0.009$\pm$0.001  &  0.6$\pm$0.1  &  -5.2$\pm$0.1  &  \nodata  &  \nodata  &  \nodata  &  \nodata  &  \nodata  &  \nodata  &  0.0    \\  
\hline   
3C346  &  0.035$\pm$0.003  &  4.5$\pm$0.1  &  -6.4$\pm$0.1  &  4.5$\pm$0.1  &  4.9$\pm$0.1  &  -6.4$\pm$0.1  &  133$\pm$3  &  0.41$\pm$0.02  &  3  &  1.0    \\   
  &  0.279$\pm$0.004  &  2.0$\pm$0.1  &  -0.6$\pm$0.1  &  13.2$\pm$0.1  &  2.1$\pm$0.1  &  -0.6$\pm$0.1  &  58$\pm$4  &  0.63$\pm$0.04  &  3  &      \\   
  &  0.197$\pm$0.004  &  1.9$\pm$0.1  &  2.1$\pm$0.1  &  9.8$\pm$0.1  &  2.1$\pm$0.1  &  2.1$\pm$0.1  &  59$\pm$4  &  0.44$\pm$0.03  &  3  &      \\   
\hline   
3C390  &  0.103$\pm$0.006  &  2.2$\pm$0.1  &  1.0$\pm$0.1  &  14.6$\pm$0.8  &  2.5$\pm$0.1  &  0.9$\pm$0.1  &  143$\pm$11  &  0.65$\pm$0.06  &  10  &  0.5    \\   
  &  0.146$\pm$0.003  &  3.7$\pm$0.2  &  3.7$\pm$0.1  &  30.5$\pm$0.6  &  4.0$\pm$0.1  &  3.4$\pm$0.1  &  221$\pm$7  &  2.36$\pm$0.14  &  3  &      \\   
  &  0.082$\pm$0.003  &  5.7$\pm$0.2  &  8.8$\pm$0.2  &  26.2$\pm$0.4  &  6.3$\pm$0.1  &  8.8$\pm$0.1  &  331$\pm$7  &  3.08$\pm$0.13  &  10  &      \\   
  &  0.043$\pm$0.003  &  2.0$\pm$0.1  &  11.2$\pm$0.1  &  1.1$\pm$0.1  &  2.2$\pm$0.1  &  12.3$\pm$0.1  &  27$\pm$4  &  0.05$\pm$0.01  &  10  &      \\   
  &  0.095$\pm$0.003  &  9.8$\pm$0.1  &  24.5$\pm$0.1  &  37.7$\pm$0.5  &  10.7$\pm$0.1  &  25.5$\pm$0.1  &  415$\pm$6  &  7.67$\pm$0.17  &  1  &      \\   
  &  0.067$\pm$0.003  &  1.4$\pm$0.1  &  28.7$\pm$0.1  &  2.0$\pm$0.0  &  1.5$\pm$0.1  &  30.8$\pm$0.1  &  30$\pm$4  &  0.06$\pm$0.01  &  10  &      \\   
  &  0.007$\pm$0.002  &  9.2$\pm$1.1  &  42.2$\pm$0.3  &  18.1$\pm$1.1  &  10.2$\pm$0.1  &  42.1$\pm$0.1  &  2613$\pm$177  &  3.31$\pm$0.48  &  2  &      \\   
  &  0.022$\pm$0.002  &  1.7$\pm$0.1  &  35.8$\pm$0.1  &  \nodata  &  \nodata  &  \nodata  &  \nodata  &  \nodata  &  \nodata  &    \\  
  &  0.033$\pm$0.003  &  2.0$\pm$0.1  &  24.0$\pm$0.1  &  \nodata  &  \nodata  &  \nodata  &  \nodata  &  \nodata  &  \nodata  &    \\  
  &  0.006$\pm$0.002  &  0.6$\pm$0.2  &  17.1$\pm$0.1  &  \nodata  &  \nodata  &  \nodata  &  \nodata  &  \nodata  &  \nodata  &    \\  
\hline   
4C33.48  &  0.170$\pm$0.006  &  2.2$\pm$0.1  &  3.5$\pm$0.1  &  3.6$\pm$0.1  &  2.4$\pm$0.1  &  3.3$\pm$0.1  &  26$\pm$5  &  0.20$\pm$0.04  &  4  &  1.0    \\   
  &  0.139$\pm$0.006  &  6.8$\pm$0.1  &  25.4$\pm$0.1  &  8.5$\pm$0.3  &  6.8$\pm$0.1  &  23.5$\pm$0.1  &  75$\pm$13  &  1.42$\pm$0.25  &  4  &      \\   
  &  0.277$\pm$0.006  &  2.3$\pm$0.1  &  23.8$\pm$0.1  &  \nodata  &  \nodata  &  \nodata  &  \nodata  &  \nodata  &  \nodata  &    \\  
  &  0.128$\pm$0.005  &  1.7$\pm$0.1  &  28.8$\pm$0.1  &  \nodata  &  \nodata  &  \nodata  &  \nodata  &  \nodata  &  \nodata  &    \\  
\hline   
3C409A  &  0.443$\pm$0.004  &  3.2$\pm$0.1  &  4.2$\pm$0.1  &  23.6$\pm$0.1  &  3.5$\pm$0.1  &  4.2$\pm$0.1  &  64$\pm$9  &  1.81$\pm$0.27  &  1  &  0.5    \\   
  &  0.332$\pm$0.005  &  3.0$\pm$0.1  &  7.9$\pm$0.1  &  34.0$\pm$0.2  &  3.0$\pm$0.1  &  7.9$\pm$0.1  &  122$\pm$11  &  2.41$\pm$0.22  &  2  &      \\   
  &  0.732$\pm$0.007  &  2.1$\pm$0.1  &  13.8$\pm$0.1  &  5.8$\pm$0.1  &  2.1$\pm$0.1  &  12.5$\pm$0.1  &  13$\pm$5  &  0.42$\pm$0.16  &  7  &      \\   
  &  0.440$\pm$0.008  &  6.5$\pm$0.1  &  14.6$\pm$0.1  &  49.2$\pm$1.2  &  6.8$\pm$0.1  &  14.5$\pm$0.1  &  145$\pm$20  &  8.28$\pm$1.18  &  7  &      \\   
  &  0.735$\pm$0.006  &  1.7$\pm$0.1  &  15.9$\pm$0.1  &  0.6$\pm$0.1  &  1.6$\pm$0.1  &  15.9$\pm$0.1  &  12$\pm$18  &  0.30$\pm$0.45  &  7  &      \\   
  &  0.020$\pm$0.003  &  4.3$\pm$0.2  &  24.3$\pm$0.1  &  7.1$\pm$0.3  &  4.8$\pm$0.1  &  24.6$\pm$0.1  &  366$\pm$39  &  0.64$\pm$0.08  &  7  &      \\   
  &  0.004$\pm$0.002  &  0.6$\pm$0.2  &  -53.8$\pm$0.1  &  \nodata  &  \nodata  &  \nodata  &  \nodata  &  \nodata  &  \nodata  &    \\  
\hline   
3C409B  &  0.429$\pm$0.003  &  2.9$\pm$0.1  &  4.0$\pm$0.1  &  14.0$\pm$0.1  &  3.2$\pm$0.1  &  3.6$\pm$0.1  &  49$\pm$9  &  1.20$\pm$0.24  &  1  &  1.0    \\   
  &  0.280$\pm$0.007  &  3.0$\pm$0.1  &  8.0$\pm$0.1  &  24.9$\pm$0.8  &  3.1$\pm$0.1  &  8.0$\pm$0.1  &  120$\pm$13  &  2.02$\pm$0.23  &  3  &      \\   
  &  0.631$\pm$0.081  &  1.9$\pm$0.1  &  13.6$\pm$0.1  &  7.2$\pm$5.2  &  1.8$\pm$0.1  &  13.9$\pm$0.1  &  66$\pm$22  &  1.54$\pm$0.55  &  6  &      \\   
  &  0.106$\pm$0.004  &  13.1$\pm$0.2  &  14.5$\pm$0.2  &  28.5$\pm$1.1  &  14.4$\pm$0.1  &  15.9$\pm$0.1  &  285$\pm$12  &  7.83$\pm$0.46  &  6  &      \\   
  &  0.890$\pm$0.065  &  3.0$\pm$0.1  &  15.3$\pm$0.1  &  16.7$\pm$3.0  &  3.0$\pm$0.1  &  15.6$\pm$0.1  &  47$\pm$13  &  2.50$\pm$0.73  &  2  &      \\   
  &  0.193$\pm$0.053  &  3.6$\pm$1.2  &  12.3$\pm$0.9  &  \nodata  &  \nodata  &  \nodata  &  \nodata  &  \nodata  &  \nodata  &    \\  
\hline   
3C410A  &  0.014$\pm$0.002  &  1.5$\pm$0.1  &  -30.2$\pm$0.1  &  0.2$\pm$0.0  &  1.5$\pm$0.1  &  -30.2$\pm$0.1  &  15$\pm$3  &  0.01$\pm$0.00  &  13  &  1.0    \\   
  &  0.020$\pm$0.003  &  4.0$\pm$0.2  &  -22.7$\pm$0.1  &  12.6$\pm$0.6  &  4.0$\pm$0.1  &  -23.4$\pm$0.1  &  647$\pm$31  &  1.01$\pm$0.09  &  13  &      \\   
  &  0.613$\pm$0.018  &  3.6$\pm$0.1  &  -0.2$\pm$0.1  &  17.5$\pm$0.9  &  3.9$\pm$0.1  &  -0.2$\pm$0.1  &  48$\pm$12  &  2.11$\pm$0.55  &  13  &      \\   
  &  0.648$\pm$0.023  &  3.2$\pm$0.1  &  2.7$\pm$0.1  &  17.9$\pm$1.1  &  3.6$\pm$0.1  &  2.7$\pm$0.1  &  49$\pm$14  &  2.06$\pm$0.60  &  13  &      \\   
  &  1.693$\pm$0.150  &  3.2$\pm$0.1  &  7.4$\pm$0.1  &  47.0$\pm$9.9  &  3.5$\pm$0.1  &  6.7$\pm$0.1  &  70$\pm$13  &  7.49$\pm$1.64  &  13  &      \\   
  &  1.864$\pm$0.125  &  1.5$\pm$0.1  &  8.1$\pm$0.1  &  11.6$\pm$12.2  &  1.6$\pm$0.1  &  8.1$\pm$0.1  &  104$\pm$60  &  5.91$\pm$3.46  &  13  &      \\   
  &  0.149$\pm$0.081  &  1.6$\pm$0.3  &  11.1$\pm$0.1  &  8.5$\pm$8.3  &  1.8$\pm$0.1  &  10.6$\pm$0.1  &  106$\pm$45  &  0.51$\pm$0.37  &  13  &      \\   
  &  0.575$\pm$0.096  &  4.4$\pm$0.4  &  11.3$\pm$0.4  &  43.6$\pm$10.2  &  4.4$\pm$0.1  &  11.6$\pm$0.1  &  111$\pm$21  &  5.58$\pm$1.48  &  13  &      \\   
  &  0.186$\pm$0.003  &  3.3$\pm$0.1  &  17.9$\pm$0.1  &  10.9$\pm$0.1  &  3.3$\pm$0.1  &  16.5$\pm$0.1  &  71$\pm$8  &  0.88$\pm$0.11  &  13  &      \\   
  &  0.060$\pm$0.003  &  5.2$\pm$0.1  &  25.3$\pm$0.1  &  3.5$\pm$0.2  &  5.8$\pm$0.1  &  25.3$\pm$0.1  &  69$\pm$14  &  0.44$\pm$0.09  &  13  &      \\   
  &  0.055$\pm$0.002  &  1.9$\pm$0.1  &  -46.5$\pm$0.1  &  \nodata  &  \nodata  &  \nodata  &  \nodata  &  \nodata  &  \nodata  &    \\  
  &  0.048$\pm$0.003  &  3.0$\pm$0.2  &  -4.8$\pm$0.1  &  \nodata  &  \nodata  &  \nodata  &  \nodata  &  \nodata  &  \nodata  &    \\  
  &  0.076$\pm$0.003  &  1.8$\pm$0.1  &  24.6$\pm$0.1  &  \nodata  &  \nodata  &  \nodata  &  \nodata  &  \nodata  &  \nodata  &    \\  
\hline   
3C410B  &  0.021$\pm$0.002  &  5.3$\pm$0.4  &  -47.6$\pm$0.2  &  0.8$\pm$0.1  &  5.8$\pm$0.1  &  -52.0$\pm$0.1  &  39$\pm$3  &  0.09$\pm$0.01  &  12  &  0.0    \\   
  &  0.019$\pm$0.002  &  1.8$\pm$0.3  &  -30.2$\pm$0.1  &  0.4$\pm$0.0  &  2.0$\pm$0.1  &  -30.1$\pm$0.1  &  18$\pm$3  &  0.01$\pm$0.00  &  12  &      \\   
  &  0.476$\pm$0.007  &  2.7$\pm$0.1  &  -0.2$\pm$0.1  &  12.2$\pm$0.1  &  2.9$\pm$0.1  &  -0.2$\pm$0.1  &  28$\pm$2  &  0.72$\pm$0.07  &  12  &      \\   
  &  0.344$\pm$0.023  &  11.4$\pm$0.5  &  5.1$\pm$0.3  &  46.3$\pm$3.5  &  12.6$\pm$0.1  &  4.6$\pm$0.1  &  153$\pm$12  &  11.96$\pm$1.36  &  1  &      \\   
  &  0.292$\pm$0.011  &  1.5$\pm$0.1  &  5.9$\pm$0.1  &  7.5$\pm$0.3  &  1.4$\pm$0.1  &  6.0$\pm$0.1  &  31$\pm$3  &  0.27$\pm$0.03  &  12  &      \\   
  &  2.798$\pm$0.014  &  2.2$\pm$0.1  &  8.0$\pm$0.1  &  17.2$\pm$0.2  &  2.4$\pm$0.1  &  8.0$\pm$0.1  &  18$\pm$1  &  2.19$\pm$0.24  &  12  &      \\   
  &  0.430$\pm$0.137  &  1.9$\pm$0.1  &  10.9$\pm$0.1  &  22.0$\pm$8.4  &  2.0$\pm$0.1  &  10.2$\pm$0.1  &  65$\pm$21  &  1.08$\pm$0.49  &  12  &      \\   
  &  0.041$\pm$0.017  &  19.2$\pm$2.0  &  12.7$\pm$3.1  &  40.7$\pm$16.7  &  19.1$\pm$0.1  &  11.4$\pm$0.1  &  1006$\pm$403  &  15.85$\pm$9.16  &  12  &      \\   
  &  0.112$\pm$0.004  &  2.5$\pm$0.1  &  25.2$\pm$0.1  &  3.9$\pm$0.1  &  2.8$\pm$0.1  &  25.1$\pm$0.1  &  32$\pm$3  &  0.19$\pm$0.02  &  12  &      \\   
  &  0.426$\pm$0.010  &  2.3$\pm$0.1  &  2.5$\pm$0.1  &  \nodata  &  \nodata  &  \nodata  &  \nodata  &  \nodata  &  \nodata  &    \\  
  &  0.119$\pm$0.005  &  2.5$\pm$0.1  &  18.4$\pm$0.1  &  \nodata  &  \nodata  &  \nodata  &  \nodata  &  \nodata  &  \nodata  &    \\  
  &  0.347$\pm$0.087  &  2.6$\pm$0.3  &  12.2$\pm$0.4  &  \nodata  &  \nodata  &  \nodata  &  \nodata  &  \nodata  &  \nodata  &    \\  
\hline   
B2050  &  0.007$\pm$0.001  &  3.4$\pm$0.2  &  -21.3$\pm$0.1  &  1.0$\pm$0.1  &  3.7$\pm$0.1  &  -21.2$\pm$0.1  &  141$\pm$9  &  0.07$\pm$0.01  &  9  &  1.0    \\   
  &  0.011$\pm$0.001  &  2.4$\pm$0.1  &  -11.5$\pm$0.1  &  0.8$\pm$0.0  &  2.7$\pm$0.1  &  -12.5$\pm$0.1  &  78$\pm$12  &  0.04$\pm$0.01  &  9  &      \\   
  &  0.043$\pm$0.002  &  10.0$\pm$0.3  &  -0.1$\pm$0.2  &  18.2$\pm$1.0  &  10.0$\pm$0.1  &  -0.1$\pm$0.1  &  453$\pm$29  &  3.87$\pm$0.34  &  9  &      \\   
  &  0.039$\pm$0.002  &  2.0$\pm$0.1  &  2.5$\pm$0.1  &  0.6$\pm$0.0  &  2.0$\pm$0.1  &  2.7$\pm$0.1  &  13$\pm$5  &  0.02$\pm$0.01  &  9  &      \\   
  &  0.150$\pm$0.002  &  1.7$\pm$0.1  &  9.0$\pm$0.1  &  0.4$\pm$0.0  &  1.7$\pm$0.1  &  8.2$\pm$0.1  &  10$\pm$6  &  0.05$\pm$0.03  &  2  &      \\   
  &  0.190$\pm$0.002  &  6.0$\pm$0.1  &  10.2$\pm$0.1  &  21.5$\pm$0.3  &  6.0$\pm$0.1  &  9.4$\pm$0.1  &  142$\pm$15  &  3.23$\pm$0.35  &  1  &      \\   
  &  0.074$\pm$0.002  &  3.8$\pm$0.1  &  15.9$\pm$0.1  &  1.7$\pm$0.0  &  4.1$\pm$0.1  &  16.2$\pm$0.1  &  28$\pm$6  &  0.16$\pm$0.04  &  9  &      \\   
  &  0.086$\pm$0.002  &  2.8$\pm$0.1  &  -0.8$\pm$0.1  &  \nodata  &  \nodata  &  \nodata  &  \nodata  &  \nodata  &  \nodata  &    \\  
  &  0.014$\pm$0.002  &  1.2$\pm$0.1  &  -6.1$\pm$0.1  &  \nodata  &  \nodata  &  \nodata  &  \nodata  &  \nodata  &  \nodata  &    \\  
\hline   
3C433  &  0.181$\pm$0.010  &  1.5$\pm$0.1  &  2.6$\pm$0.1  &  2.2$\pm$0.2  &  1.6$\pm$0.1  &  2.4$\pm$0.1  &  21$\pm$9  &  0.12$\pm$0.05  &  2  &  1.0    \\   
  &  0.304$\pm$0.009  &  3.9$\pm$0.1  &  3.5$\pm$0.1  &  19.1$\pm$0.7  &  4.1$\pm$0.1  &  3.5$\pm$0.1  &  78$\pm$5  &  1.86$\pm$0.15  &  1  &      \\   
  &  0.080$\pm$0.005  &  2.1$\pm$0.1  &  7.3$\pm$0.1  &  1.8$\pm$0.1  &  2.1$\pm$0.1  &  7.3$\pm$0.1  &  29$\pm$5  &  0.10$\pm$0.02  &  4  &      \\   
  &  0.059$\pm$0.004  &  2.5$\pm$0.1  &  16.5$\pm$0.1  &  1.6$\pm$0.0  &  2.7$\pm$0.1  &  15.7$\pm$0.1  &  29$\pm$2  &  0.09$\pm$0.01  &  4  &      \\   
\hline   
PKS2127  &  0.073$\pm$0.001  &  2.7$\pm$0.1  &  -0.9$\pm$0.1  &  7.6$\pm$0.1  &  2.7$\pm$0.1  &  -0.9$\pm$0.1  &  105$\pm$4  &  0.40$\pm$0.02  &  1  &  1.0    \\   
  &  0.013$\pm$0.001  &  8.3$\pm$0.2  &  1.1$\pm$0.1  &  9.7$\pm$0.6  &  8.3$\pm$0.1  &  1.2$\pm$0.1  &  769$\pm$43  &  1.62$\pm$0.14  &  2  &      \\   
  &  0.107$\pm$0.001  &  2.1$\pm$0.1  &  1.5$\pm$0.1  &  5.8$\pm$0.1  &  2.0$\pm$0.1  &  1.5$\pm$0.1  &  61$\pm$5  &  0.28$\pm$0.02  &  3  &      \\   
\hline   
J2136  &  0.096$\pm$0.002  &  2.2$\pm$0.1  &  1.1$\pm$0.1  &  3.1$\pm$0.0  &  2.4$\pm$0.1  &  1.0$\pm$0.1  &  28$\pm$3  &  0.12$\pm$0.02  &  1  &  0.0    \\   
  &  0.062$\pm$0.002  &  7.0$\pm$0.1  &  3.4$\pm$0.1  &  2.4$\pm$0.0  &  7.7$\pm$0.1  &  3.1$\pm$0.1  &  22$\pm$13  &  0.20$\pm$0.12  &  2  &      \\   
  &  0.079$\pm$0.002  &  2.5$\pm$0.1  &  5.8$\pm$0.1  &  7.7$\pm$0.1  &  2.7$\pm$0.1  &  6.3$\pm$0.1  &  98$\pm$4  &  0.38$\pm$0.02  &  3  &      \\   
\hline   
J2232  &  0.066$\pm$0.001  &  2.3$\pm$0.1  &  -14.3$\pm$0.1  &  4.2$\pm$0.0  &  2.5$\pm$0.1  &  -15.0$\pm$0.1  &  65$\pm$0  &  0.19$\pm$0.00  &  5  &  0.5    \\   
  &  0.056$\pm$0.002  &  2.0$\pm$0.1  &  -7.6$\pm$0.1  &  3.1$\pm$0.1  &  2.0$\pm$0.1  &  -7.6$\pm$0.1  &  58$\pm$3  &  0.13$\pm$0.01  &  5  &      \\   
  &  0.064$\pm$0.002  &  5.7$\pm$0.1  &  -5.0$\pm$0.1  &  6.6$\pm$0.2  &  5.7$\pm$0.1  &  -5.0$\pm$0.1  &  105$\pm$4  &  0.77$\pm$0.04  &  5  &      \\   
  &  0.099$\pm$0.002  &  2.0$\pm$0.1  &  -3.6$\pm$0.1  &  2.9$\pm$0.1  &  2.1$\pm$0.1  &  -3.6$\pm$0.1  &  32$\pm$4  &  0.13$\pm$0.02  &  5  &      \\   
  &  0.046$\pm$0.002  &  3.3$\pm$0.1  &  2.1$\pm$0.1  &  4.3$\pm$0.1  &  3.6$\pm$0.1  &  2.3$\pm$0.1  &  96$\pm$3  &  0.29$\pm$0.01  &  1  &      \\   
\hline   
3C454.3  &  0.313$\pm$0.024  &  2.5$\pm$0.1  &  -10.3$\pm$0.1  &  5.0$\pm$0.5  &  2.8$\pm$0.1  &  -10.6$\pm$0.1  &  20$\pm$2  &  0.32$\pm$0.04  &  1  &  1.0    \\   
  &  0.025$\pm$0.013  &  3.7$\pm$1.2  &  -8.5$\pm$1.3  &  4.0$\pm$2.1  &  3.7$\pm$0.1  &  -8.1$\pm$0.1  &  159$\pm$90  &  0.28$\pm$0.24  &  8  &      \\   
  &  0.096$\pm$0.002  &  3.6$\pm$0.1  &  -2.1$\pm$0.1  &  12.9$\pm$0.1  &  3.6$\pm$0.1  &  -2.1$\pm$0.1  &  142$\pm$2  &  0.97$\pm$0.02  &  8  &      \\   
  &  0.081$\pm$0.002  &  1.8$\pm$0.1  &  0.7$\pm$0.1  &  4.6$\pm$0.1  &  1.9$\pm$0.1  &  0.6$\pm$0.1  &  61$\pm$3  &  0.17$\pm$0.01  &  8  &      \\   
  &  0.026$\pm$0.002  &  4.7$\pm$0.3  &  3.0$\pm$0.2  &  2.7$\pm$0.1  &  5.1$\pm$0.1  &  3.3$\pm$0.1  &  109$\pm$3  &  0.26$\pm$0.02  &  8  &      \\   
  &  0.047$\pm$0.002  &  2.1$\pm$0.1  &  -30.5$\pm$0.1  &  \nodata  &  \nodata  &  \nodata  &  \nodata  &  \nodata  &  \nodata  &    \\  
  &  0.016$\pm$0.002  &  5.9$\pm$0.2  &  -16.4$\pm$0.1  &  \nodata  &  \nodata  &  \nodata  &  \nodata  &  \nodata  &  \nodata  &    \\  
  &  0.010$\pm$0.002  &  3.5$\pm$0.2  &  -35.1$\pm$0.1  &  \nodata  &  \nodata  &  \nodata  &  \nodata  &  \nodata  &  \nodata  &    \\  
\hline   
3C459  &  0.010$\pm$0.001  &  2.8$\pm$0.1  &  -13.2$\pm$0.1  &  0.7$\pm$0.0  &  3.1$\pm$0.1  &  -12.4$\pm$0.1  &  72$\pm$2  &  0.04$\pm$0.00  &  2  &  1.0    \\   
  &  0.039$\pm$0.002  &  5.2$\pm$0.3  &  -6.2$\pm$0.2  &  14.4$\pm$0.8  &  5.5$\pm$0.1  &  -6.7$\pm$0.1  &  384$\pm$22  &  1.51$\pm$0.14  &  1  &      \\   
  &  0.057$\pm$0.002  &  1.9$\pm$0.1  &  0.4$\pm$0.1  &  2.7$\pm$0.1  &  2.0$\pm$0.1  &  0.4$\pm$0.1  &  37$\pm$8  &  0.08$\pm$0.02  &  7  &      \\   
  &  0.039$\pm$0.003  &  7.7$\pm$0.6  &  0.8$\pm$0.3  &  18.3$\pm$1.4  &  7.7$\pm$0.1  &  0.7$\pm$0.1  &  478$\pm$35  &  2.85$\pm$0.37  &  7  &      \\   
  &  0.088$\pm$0.003  &  2.2$\pm$0.1  &  2.9$\pm$0.1  &  3.5$\pm$0.1  &  2.2$\pm$0.1  &  2.9$\pm$0.1  &  46$\pm$6  &  0.18$\pm$0.03  &  7  &      \\   
  &  0.016$\pm$0.001  &  1.4$\pm$0.1  &  7.7$\pm$0.1  &  0.9$\pm$0.0  &  1.6$\pm$0.1  &  7.7$\pm$0.1  &  56$\pm$2  &  0.03$\pm$0.00  &  7  &      \\   
  &  0.102$\pm$0.003  &  2.2$\pm$0.1  &  -7.3$\pm$0.1  &  \nodata  &  \nodata  &  \nodata  &  \nodata  &  \nodata  &  \nodata  &    \\  
\hline

\hline   
 
\enddata
\tablecomments{Cols. (2-4): Gaussian parameters fit to \hi\ absorption (Equation~\ref{e:tau}). Cols. (5-7): Gaussian parameters fit to \hi\ emission (Equations~\ref{e:tb}). Col. (8): Average spin temperature from all permutations of components with overlap along the LOS (Equations~\ref{e:order},~\ref{e:fval}). Col. (9): Column density computed from fitted parameters (Equation~\ref{e:nhi}). Col. (10): Order of components along the LOS corresponding to the smallest model residuals. Components whose position along the LOS is extremely uncertain or unaffected by order permutations are assumed to lie behind all others (i.e., $\mathscr{O} = N$, for $N$ total components). Col (11): Fraction of WNM (emission-detected only) components which lie in front of all absorption-detected components, allowed to be 1.0 or 0.0 for all emission-detected components. Fit parameters for components with $T_s\leq3\rm\,K$ are omitted, as these are either spurious AGD detections or were not recovered in the fit to $T_{B, \rm exp}(v)$ due to strong line blending.}
\end{deluxetable*}

}

\end{document}